\documentclass[preprint,12pt]{elsarticle}

\usepackage{amssymb}

\usepackage{amssymb}
\usepackage{amsmath}
\allowdisplaybreaks[4]
\usepackage{bm}

\usepackage{algorithm}
\usepackage{algorithmic}
\usepackage{subfigure}

\begin{document}

\begin{frontmatter}



\title{Self-optimization wavelet-learning method for predicting nonlinear thermal conductivity of highly heterogeneous materials with randomly hierarchical configurations}


\author[author1,author2]{Jiale Linghu}
\author[author1]{Hao Dong\corref{cor1}}\ead{donghao@mail.nwpu.edu.cn}
\cortext[cor1]{Corresponding author.}
\author[author1,author2]{Weifeng Gao}
\author[author3]{Yufeng Nie}

\address[author1]{School of Mathematics and Statistics, Xidian University, Xi'an 710071, PR China}
\address[author2]{Key Laboratory of Collaborative Intelligence Systems, Ministry of Education, Xidian University, Xi'an 710071, China}
\address[author3]{School of Mathematics and Statistics, Northwestern Polytechnical University, Xi'an 710129, PR China}

\begin{abstract}
In the present work, we propose a self-optimization wavelet-learning method (SO-W-LM) with high accuracy and efficiency to compute the equivalent nonlinear thermal conductivity of highly heterogeneous materials with randomly hierarchical configurations. The randomly structural heterogeneity, temperature-dependent nonlinearity and material property uncertainty of heterogeneous materials are considered within the proposed self-optimization wavelet-learning framework. Firstly, meso- and micro-structural modeling of random heterogeneous materials are achieved by the proposed computer representation method, whose simulated hierarchical configurations have relatively high volume ratio of material inclusions. Moreover, temperature-dependent nonlinearity and material property uncertainties of random heterogeneous materials are modeled by a polynomial nonlinear model and Weibull probabilistic model, which can closely resemble actual material properties of heterogeneous materials. Secondly, an innovative stochastic three-scale homogenized method (STSHM) is developed to compute the macroscopic nonlinear thermal conductivity of random heterogeneous materials. Background meshing and filling techniques are devised to extract geometry and material features of random heterogeneous materials for establishing material databases. Thirdly, high-dimensional and highly nonlinear material features of material databases are preprocessed and reduced by wavelet decomposition technique. The neural networks are further employed to excavate the predictive models from dimension-reduced low-dimensional data. At the same time, advanced intelligent optimization algorithms are utilized to self-search the optimal network structure and learning rate for obtaining the optimal predictive models. Finally, the computational accuracy and efficiency of the presented approach are validated via various numerical experiments on realistic random composites.
\end{abstract}


\end{frontmatter}


\section{Introduction}
\label{sec:1}
Since the beginning of the industrial era, the research and design of materials have been occupied the innovative midst of many disruptive technology. Especially in aerospace, aviation, mechanical and civil engineering, heterogeneous materials with outstanding physical properties have been widely used to meet practical demands within severe constraints. In recent years, random heterogeneous materials have been sustained and rapid development, such as concrete materials, functional gradient materials, and alloy materials, etc. These random composites possess flexible design-ability, which leads to complicated structural configurations in multiple spatial scales. According to previous studies, when serving under complex thermal environments, the thermal conductivity of materials changes significantly with temperature, and there is a significant nonlinear performance in the thermal conductivity of materials \cite{R1,R2}. Therefore, extensively engineering demands strongly prompt to effectively analyze and accurately predict the nonlinear thermal conductivity of random heterogeneous materials.

Considerable success has been also reported to predict the physical performances of heterogeneous materials. The predictive approaches of the physical properties of composites are mainly composed of analytical and numerical methods. Since 1960s, numerous analytical methods were developed, such as Hashin-Shtrikman method \cite{R3}, self-consistent method \cite{R4} and Mori-Tanaka method \cite{R5}, etc. However, most of these analytical methods can only acquire better prediction for heterogeneous materials with regular or simple geometric configurations since their theoretical foundation is established on some simplified assumptions. In practical engineering applications, inhomogeneous materials often have considerably complicated morphology and each phase of these inhomogeneous materials has obvious interaction effect. Hence, extensive numerical methods are widely employed to calculate the effective performance of inhomogeneous materials, such as finite element method (FEM) \cite{R6,R7}, boundary element method (BEM) \cite{R8}, fast fourier transforms (FFT) \cite{R9} and finite volume method (FVM) \cite{R10}, etc. In recent years, multi-scale methods are extensively utilized to establish simulated tool and predictive model in material science field since they can establish the bridge among different spatial scales of inhomogeneous materials. In reference \cite{R11}, Yu and Cui et al. proposed a two-order two-scale method for predicting the thermal conduction properties of random inhomogeneous materials. A theoretical method and its numerical implementation was developed for predicting the effective transport coefficients in periodic heterogeneous media by Mathieu-Potvin in \cite{R12}. Furthermore, the simulation of heat conduction and prediction of effective material performance inside a heterogeneous material with multiple spatial scales are presented in references \cite{R13,R14,R15}. In references \cite{R16,R17}, scholars developed stochastic multi-scale computational models and their numerical algorithms for predicting mechanical properties and yield strength of random heterogeneous materials with multi-level spatial configurations. With the enlargement of application areas, heterogeneous materials exhibit nonlinear temperature-dependent properties when serving in extreme high-temperature circumstances \cite{R18,R19,R20,R21,R22,R23}. Few studies have investigated the nonlinear temperature-dependent behaviors of inhomogeneous materials using the multi-scale approaches in recent years. Li et al. \cite{R18} proposed a new multi-scale analysis model that not only accurately measures the thermal shock response, but also precisely reflects the effects of diverse microstructures of porous ceramics. Zhang et al. \cite{R19} presented a multi-scale finite element method to predict the nonlinear thermoelastic behaviors of heterogeneous multiphase medias with temperature-dependent performances. Dong and Cui et al. \cite{R20} first presented a second-order two-scale analysis model and its numerical algorithm for time-dependent nonlinear thermo-mechanical problems of cylindrical periodic composites. In reference \cite{R21}, Iwama et al. proposed and experimentally validated a multi-scale model for structural concrete materials at high temperatures. Li et al. \cite{R22} developed a new multi-scale computational model to evaluate the thermal protection performance of three-dimensional reinforced weave fabrics and multifunctional zones. Zhou et al. \cite{R23} established a stochastic multi-scale fracture model to simulate the nonlinear mechanical performances of 2D woven composites in uniaxial tension experiments and predicted the fracture strength influenced by fibrous volume fraction and temperature. Nowadays, stochastic multi-scale modeling and prediction for nonlinear temperature-dependent properties of random heterogeneous materials is still a challenging problem, involving the macro-, meso- and micro-scales of inhomogeneous materials.

Along the development of computer technology, machine learning algorithms are rapidly developed and applied in many fields. In the field of materials, Liu et al. \cite{R24} developed a data-driven multi-scale model and a two-step approach to estimate the equivalent thermal conductivity of 2D woven composites. In reference \cite{R25}, Wei et al. employed the Gaussian process regression, support vector regression and convolution neural network to calculate the equivalent thermal conductivities of composites. A new wavelet-based machine learning-assisted multi-scale analysis method was developed by Dong \cite{R26,R27,R31}, and has been employed to estimate the equivalent thermal conductivity of particulate composites, braided composites and hybrid composites with complicated microstructures. Rong et al. \cite{R28} used convolutional neural networks to estimate the equivalent thermal conductivity of three-dimensional inhomogeneous materials. The predictive data features were originated from two-dimensional cross-sectional images. Ankel et al. \cite{R29} developed a convolutional neural network to automatically extract features from thermal tomography images. In reference \cite{R30}, Ye et al. proposed a machine learning approach to compute the equivalent mechanical properties of composite materials with any arbitrary shape and distribution of inclusions. Rao et al. \cite{R32} presented a three-dimensional convolutional neural network to predict effective material parameters for composites with random inclusions and the advantages of the three-dimensional convolutional neural network over traditional numerical methods are discussed at length. A novel deep material network is developed by Liu et al. \cite{R33,R34} and further enhanced by advanced model compression algorithms. The proposed deep material network in data-driven multi-scale mechanics can tackle general three-dimensional problems with arbitrary material and geometric nonlinearities. Tao et al. \cite{R35} utilized the versatile finite element analysis capabilities of Abaqus software and the powerful machine learning capabilities of deep neural networks to integrate the Abaqus-DNN system to learn the constitutive law of composites. Guo et al. \cite{R36} developed a machine learning method that innovatively combined micromechanics to predict the ductility, compressive and tensile strength of high-performance fiber-reinforced cementitious composites for the first time. Huang et al. \cite{R37} employed a machine learning method to calculate the mechanical performances of carbon nanotube-reinforced cement composites, which has better generalization and prediction performance compared with classical response surface methods. Yang et al. \cite{R38} developed a cGAN method for strain and stress tensor prediction, which provided a powerful tool for the design of multifunctional composites and optimization of layered structures. Lyngdoh et al. \cite{R39} developed an interpretable machine learning-based model to predict the seamless performance of auxetic cellular cementitious composites (ACCCs) and can accelerate the acceptance and widespread utilization of ACCCs. Avci Karatas \cite{R40} introduced regression methods to estimate the shear load capacity of headed steel studs via utilizing the minimax probability machine regression and extreme machine learning. Bhaduri et al. \cite{R41} predicted the cross-sectional stress field of composites with a fixed number of fibers and random distribution by means of a convolutional neural network and U-network architecture. Liu et al. \cite{R42} developed a stochastic multi-scale method to estimate the thermal conductivity of carbon nanotube reinforced polymeric composites utilizing integrated machine learning. To sum up, most of existing studies on machine learning prediction have merely considered the random geometric features of inhomogeneous materials, without considering the randomness and nonlinearity of material parameters. Moreover, their studies mainly focus on inhomogeneous materials with two-scale spatial structures. Meanwhile, it is difficult to determine the optimal predictive model due to a large number of hyperparameters of machine learning algorithms. Therefore, there are extensive demands in practical applications to deeply investigate machine learning approaches that can precisely estimate the nonlinear physical properties of highly heterogeneous materials with randomly hierarchical configurations.

In this study, a self-optimization wavelet-learning approach is developed to estimate the equivalent nonlinear thermal conductivity of highly heterogeneous materials with randomly hierarchical configurations. The random heterogeneous materials studied in this paper have temperature-dependent nonlinearity and material property uncertainties. In addition, fibrous or particulate inclusions at the meso- or micro-scale are randomly distributed. Thereupon, it is extremely challenging to establish an accurate prediction model of the macroscopic effective nonlinear thermal conductivity of random heterogeneous materials. The primary contributions of this study are presented as follows. Firstly, a computer generation algorithm for three-dimensional fibrous composites with high volume ratio inclusions is proposed, and the maximum volume ratio of fibrous inclusions can reach 26.7\%. Secondly, a stochastic three-scale homogenized method is developed to calculate the macroscopic nonlinear thermal conductivity of random heterogeneous materials, and the proposed stochastic three-scale homogenized method can accurately describe spatial structural heterogeneities, material property uncertainties and temperature-dependent nonlinearity of heterogeneous materials. Thirdly, the material parameters and geometric structure information are extracted using the background meshing and filling techniques, which overcomes the difficulty of missing material data of mesoscopic matrix. The proposed two techniques can also be applied to the digital image processing of real multi-scale composites. Fourthly, a wavelet-learning approach for predicting effective nonlinear thermal conductivity is obtained by integrating the superiorities of wavelet transform in data compression and feature extraction with the excellent learning ability of artificial neural network. The preprocessing data by wavelet transform can prominently reduce the data scale in input layer of neural networks, diminish over-fitting phenomenon, and improve the training efficiency of artificial neural network. Finally, advanced intelligent optimization algorithms are used to self-search the optimal network structure and learning rate of the wavelet-learning approach. The above-mentioned technical contributions constitute the self-optimization wavelet-learning approach, which is obtained to estimate the macroscopic effective nonlinear thermal conductivity of highly heterogeneous materials with randomly hierarchical configurations.

The article is outlined as follows. Section \ref{sec:2} presents the computer modeling of random composites with hierarchical configurations in detail, especially for fibrous composites with high volume ratio inclusions. Then, a stochastic three-scale homogenized method is established to calculate the macroscopic effective nonlinear thermal conductivity of random heterogeneous materials. The stochastic three-scale homogenized method takes into account not only
the random structural heterogeneities at meso- and micro-scales but also the nonlinearity and uncertainties of the material performance. Next, the background meshing and filling techniques are introduced to extract the geometry and material features at meso- and micro-scales to build the material database. Moreover, it should be noted that the filling technique is presented to tackle the missing data issue of mesoscopic matrix material. In Section \ref{sec:3}, the self-optimization wavelet-learning framework for estimating the macroscopic equivalent nonlinear thermal conductivity of random heterogeneous materials is first established using the respective merits of neural networks, wavelet transform and intelligent optimization algorithms. In the presented self-optimization wavelet-learning framework, the intelligent optimization algorithms are employed to self-optimize the network structure and learning rate of the wavelet-learning predictive models. Section \ref{sec:4} conducts a systematic verification study of the self-optimization wavelet-learning approach by extensive numerical experiments on the basis of realistic random composites. The meaningful conclusions and future research prospects are summarized in Section \ref{sec:5}.

In this present work, Einstein summation convention is applied to simplify repetitious indices for readability.
\section{Multi-scale modeling for nonlinear heterogeneous materials with randomly hierarchical configurations}
\label{sec:2}
This section implements multi-scale modeling for random structural heterogeneity, temperature-dependent nonlinearity and material property uncertainty of heterogeneous materials. Firstly, an improved algorithm for representing the random structural heterogeneities of the investigated composites is presented. Then, a polynomial nonlinear model and Weibull probabilistic model are employed to characterize the temperature-dependent nonlinearity and random uncertainties of the investigated composites. Furthermore, a stochastic three-scale homogenized approach is developed for computing macroscopic nonlinear thermal conductivity of random heterogeneous materials. Eventually, a background mesh method and filling technique are presented to extract the hierarchical geometry and material characteristics of random heterogeneous materials, so the material database of random heterogeneous materials for self-optimization wavelet-learning is established.
\subsection{Geometric model of random heterogeneous materials}
\begin{figure}
	\centering
	\begin{minipage}[c]{0.35\textwidth}
		\centering
		\includegraphics[width=42mm]{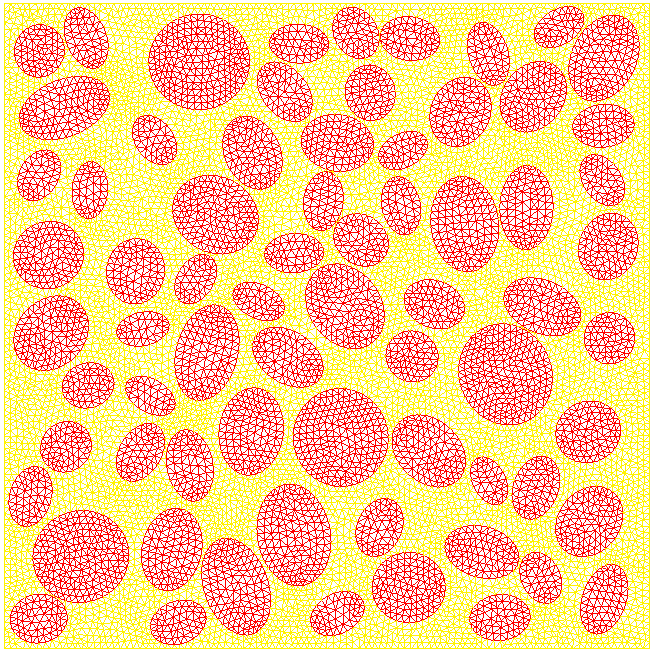}\\
		(a)
	\end{minipage}
	\begin{minipage}[c]{0.35\textwidth}
		\centering
		\includegraphics[width=42mm]{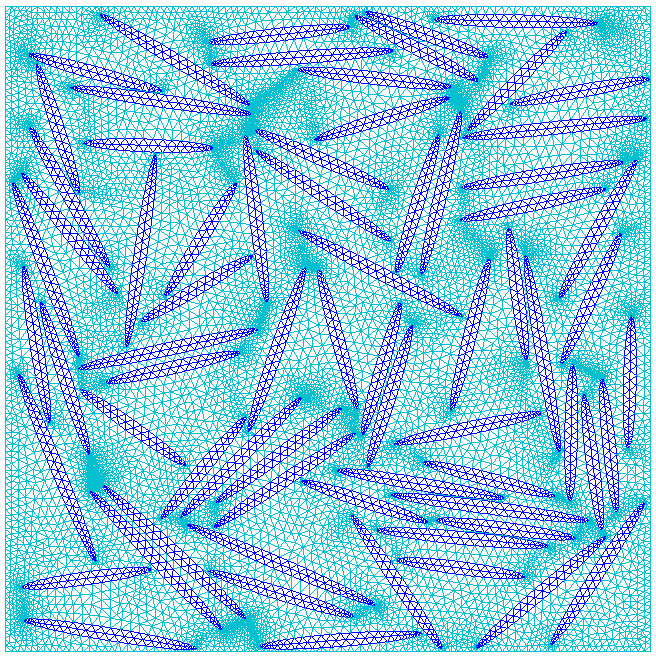}\\
		(b)
	\end{minipage}
	\begin{minipage}[c]{0.35\textwidth}
		\centering
		\includegraphics[width=40mm]{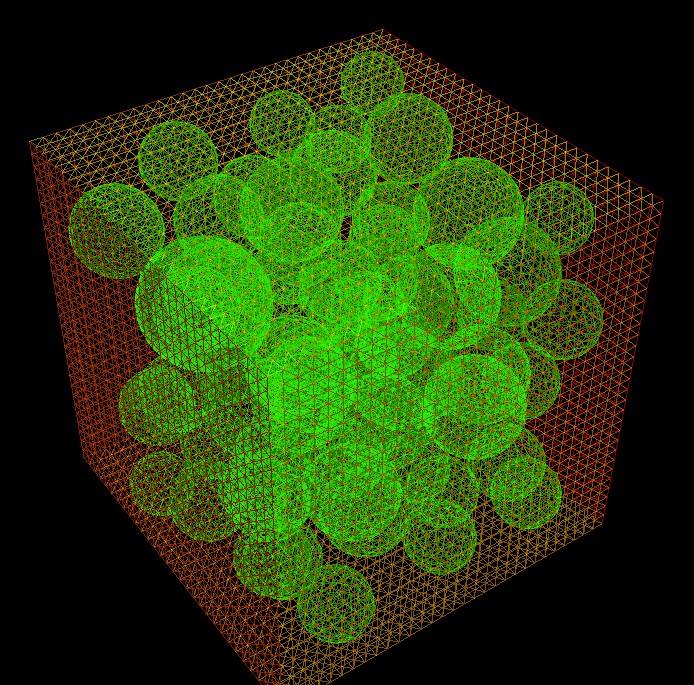}\\
		(c)
	\end{minipage}
	\begin{minipage}[c]{0.35\textwidth}
		\centering
		\includegraphics[width=40mm]{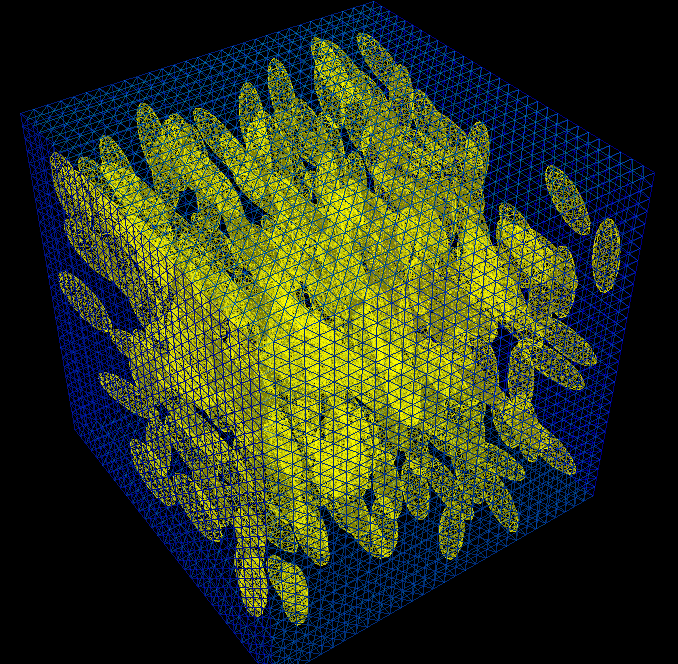}\\
		(d)
	\end{minipage}
	\caption{The illustrations of random composites with hierarchical configurations: (a) 2D microscopic configuration at statistical screen; (b) 2D mesoscopic configuration at statistical screen; (c) 3D microscopic configuration at statistical screen; (d) 3D mesoscopic configuration at statistical screen.}\label{fig:1}
\end{figure}

The spatial structural heterogeneities in composites significantly affect the physical properties of composites. Accurate computer modeling for random heterogeneous materials is essential to predict their macroscopic physical properties. In this present study, we develop the computer algorithm to represent the random structural heterogeneities of the investigated inhomogeneous materials with 2D and 3D hierarchical configurations. The investigated inhomogeneous materials are considered for both particulate inclusions and matrix in the micro-scale, as well as fibrous inclusions and matrix in the meso-scale, as illustrated in Fig.\hspace{1mm}\ref{fig:1}.

In order to better simulate random composites with a large number of particulate or fibrous inclusions, an improved algorithm is developed on the basis of the computer representation method in \cite{R43}. Firstly, randomly generate geometric parameters $\left({x_1},{x_2},{x_3},\alpha, \beta, \gamma\right)$ of ellipsoidal inclusions, set the long, middle and short axes of ellipsoidal inclusions as $a$, $b$ and $c$. Whereupon, judge whether newly generated inclusions are completely situated within the statistical screen of representative volume element (RVE). Furthermore, the points on the center of the previously generated inclusions and the surface of the newly generated inclusions are connected, and the existence of an intersection between the connected line segment and the surface of the previously generated inclusions is used as the judgment criterion. In this paper, ellipsoidal particles with relatively large ratio of long and short axes are employed to approximate the fibrous inclusions. The details of the generation algorithm are presented in Algorithm 1 as below.
\begin{algorithm}[!h]
	\caption{Computer generation algorithm}
	\label{alg:AOA}
	\begin{algorithmic}[1]
		\STATE Define the number $N$ of inclusions, and the long, middle and short axes $a$, $b$ and $c$. Then define $i=0$
		\WHILE{$i<N$}
		\STATE Randomly generate geometric parameters $({{x_1},{x_2},{x_3},\alpha ,\beta ,\gamma })$
		\STATE Uniformly take points on ellipsoidal inclusions in the global coordinate system
		\STATE The global coordinates of the obtained points are transformed into local coordinates by using the rotation matrix
		\IF{all point coordinates are located within the RVE}
		\IF{the line between all points and the center of the previously generated inclusions has an intersection with the ellipsoidal surface}
		\STATE Store all geometric parameters of ellipsoidal inclusions
		\STATE $i=i+1$
		\ENDIF
		\ENDIF
		\ENDWHILE
	\end{algorithmic}
\end{algorithm}

\begin{figure}
	\centering
	\subfigure[]{
		\label{fig:subfig:a} 
		\includegraphics[width=5in]{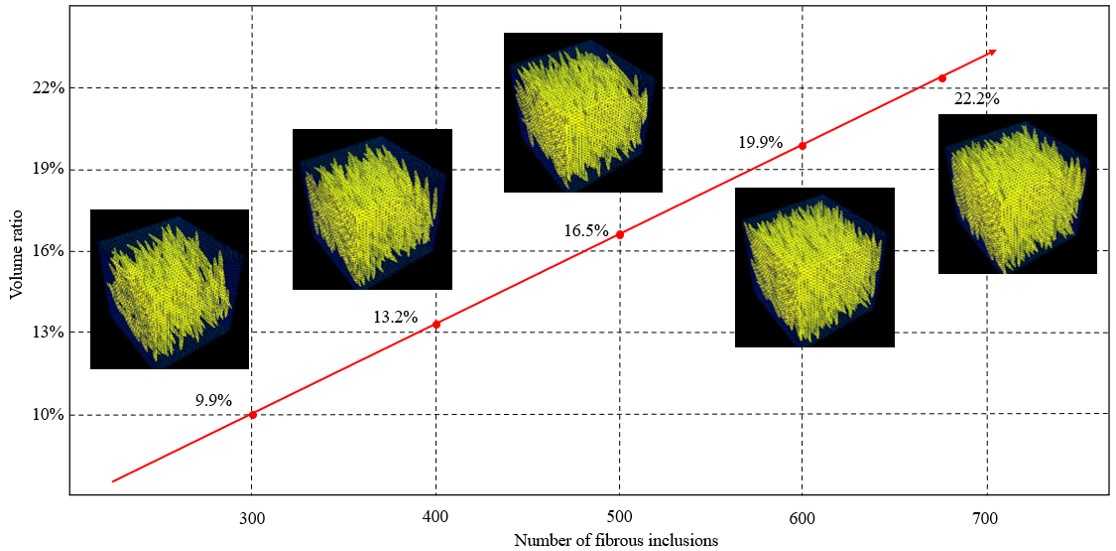}}
	\hspace{1in}
	\subfigure[]{
		\label{fig:subfig:b} 
		\includegraphics[width=5in]{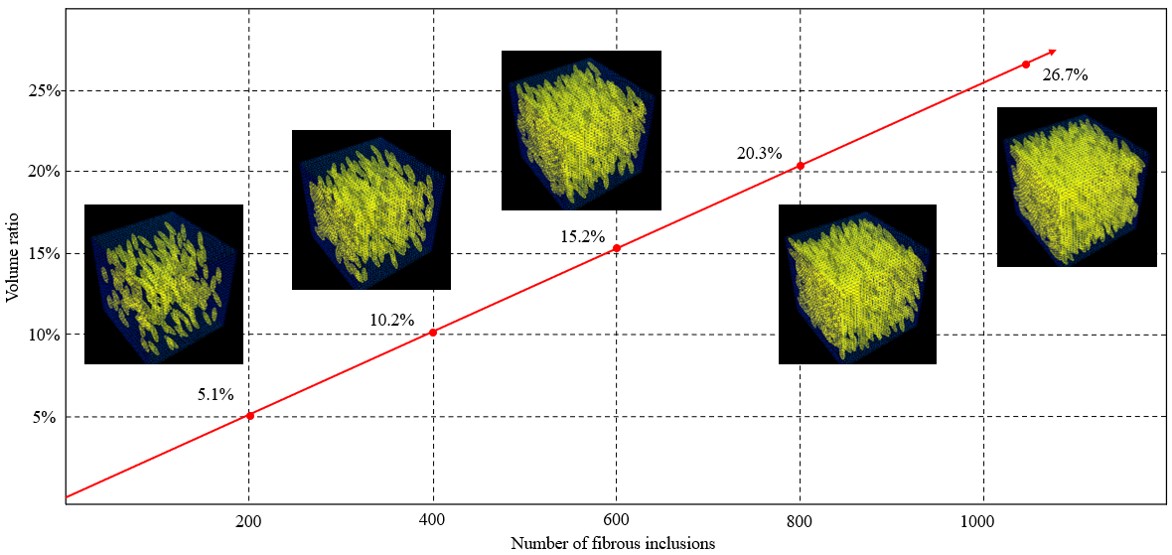}}
	\caption{The illustrations of computer modeling results: (a) $a = 20mm$, $b = c = 2mm$; (b) $a = 10mm$, $b = c = 2.5mm$.}\label{fig:2}
\end{figure}

By experimental simulation, the spatial size of 3D statistical screen is set as $100mm \times 100mm \times 100mm$. When the long, middle and short axes of ellipsoidal particles are set as $20 mm$, $2mm$ and $2mm$, the fraction ratio of fibrous inclusions can be up to 22.2\% in the statistical screen. When the long, middle and short axes of ellipsoidal particles are set as $10mm$, $2.5mm$ and $2.5mm$, the fraction ratio of fibrous inclusions can be up to 26.7\%. The detailed computer modeling results are shown in Fig.\hspace{1mm}\ref{fig:2}.

\textbf{Remark} It should be illustrated that the computer modeling of all composites quoted above is achieved via the secondary development and free programming of open-source Freefem++ software, that is developed by the researchers of Universit$\rm\acute{e}$ Paris VI.
\subsection{Material model of random heterogeneous materials}
According to physical experiments, when the service temperature changes, the thermal conductivity of the materials will change, which leads to nonlinear material properties. In this paper, the temperature-dependent nonlinearity of random heterogeneous materials is modeled by a so-called polynomial nonlinear model, which is widely utilized to approximate the authentic nonlinear material properties of heterogeneous materials. The detailed polynomial nonlinear model is given by Eq.\hspace{1mm}(\ref{eq:29})
\begin{equation}
\kappa = {\kappa_a} + {\kappa_b}\theta  + {\kappa_c}{\theta ^2} +  \cdots ,
\label{eq:29}
\end{equation}
where $\kappa$ denotes the thermal conductivity coefficient, $\theta$ represents temperature variable, $\kappa_a$, $\kappa_b$ and $\kappa_c$ are the coefficients of the polynomial model.

In this work, three kinds of random composites are investigated, including particle-reinforced ZrO$_{2}$/Ti-6Al-4V, fiber-reinforced C/SiC and multilevel concrete material. The thermal transfer properties of these investigated composites possess random uncertainties. For ZrO$_{2}$/Ti-6Al-4V and C/SiC, we assume that the thermal conductivity $\kappa_a$ of the component materials is scattered in the form of a normal distribution \cite{R44}. This work assumes that the expectation of the thermal conductivity of the material is ${\kappa_a}\left( {W/\left( {m \bullet K} \right)} \right)$, the variance is 1, and the range is limited to $\left[{0.9{\kappa_a},1.1{\kappa_a}} \right]$. In the research field of concrete materials, Weibull probabilistic distribution is extensively applied to represent stochastic uncertainties of thermal conductivity of concrete materials at meso- and micro-scales. In the present study, we utilize Weibull probabilistic distribution to characterize the random thermal conductivity of concrete materials, which is specifically given as
\begin{equation}
f\left( \lambda  \right) = \frac{m}{{{\lambda _0}}}{\left( {\frac{\lambda }{{{\lambda _0}}}} \right)^{m - 1}}{e^{ - {{\left( {\frac{\lambda }{{{\lambda _0}}}} \right)}^m}}}.
\label{eq:30}
\end{equation}
Among formula (\ref{eq:30}), $\lambda$ is defined as the thermal conductivity of the investigated material, $\lambda _0$ and $m$ represent the scale parameter and shape parameter separately. To obtain the value of the thermal conductivity that follows the Weibull distribution law, the Monte Carlo-based method \cite{R45} is used to generate a random number $\omega$ from 0 to 1, and the thermal conductivity of the material can be obtained as follows
\begin{equation}
\lambda  = {\lambda _0}{\left[ {\ln \frac{1}{{1 - \omega }}} \right]^{\frac{1}{m}}}.
\label{eq:31}
\end{equation}
In this study, we presume that the thermal conductivity $\kappa_a$ of all component materials of concrete composites containing cement matrix, aggregate particles and fibrous inclusions ranges from $\left[ {0.9{\kappa_a},1.1{\kappa_a}} \right]$ with Weibull distribution parameters ${\lambda _0} = {\kappa_a}$ and $m=10$.
\subsection{Stochastic three-scale homogenized method for nonlinear thermal problems of random heterogeneous materials}
Consider heterogeneous materials with randomly hierarchical configurations shown in Fig.\hspace{1mm}\ref{fig:1} and utilize the classical heat transfer theory, the governing differential equation for describing their nonlinear thermal transfer behaviors in static case is reported as follows.
\begin{equation}
\left\{ {\begin{aligned}
&\displaystyle {- \frac{\partial }{{\partial {x_i}}}\left[ {\kappa_{ij}^{{\varepsilon _1}{\varepsilon _2}}({\bm{x}},\bm{\omega},{\theta ^{{\varepsilon _1}{\varepsilon _2}}})\frac{{\partial {\theta ^{{\varepsilon _1}{\varepsilon _2}}}({\bm{x}},\bm{\omega})}}{{\partial {x_j}}}} \right] = h({\bm{x}}),\text { in } \Omega,}\\
&\displaystyle{\theta ^{{\varepsilon _1}{\varepsilon _2}}}({\bm{x}},\bm{\omega}) = \hat \theta ({\bm{x}}),\text { on } \partial \Omega_\theta,\\
&\displaystyle \kappa_{ij}^{{\varepsilon _1}{\varepsilon _2}}({\bm{x}},\bm{\omega},{\theta ^{{\varepsilon _1}{\varepsilon _2}}})\frac{{\partial {\theta ^{{\varepsilon _1}{\varepsilon _2}}}({\bm{x}},\bm{\omega})}}{{\partial {x_j}}}{n_i} = \bar q({\bm{x}}),\text { on } \partial \Omega_q,
\end{aligned}} \right.\label{eq:1}
\end{equation}
in which $\Omega$ is a bounded convex domain in ${\mathbb{R}^\mathcal{N}}(\mathcal{N} = 2,3)$ with a outer boundary $\partial \Omega=\partial \Omega_\theta \cup \partial \Omega_q$. The spatial sizes of mesoscopic RVE $Y_{\bm{\omega}}$ and microscopic RVE $Z_{\bm{\omega}}$ are ${\varepsilon _1}$ and ${\varepsilon _2}$ respectively. $\hat \theta ({\bm{x}})$ is the temperature data specified on the boundary $\partial {{\rm{\Omega }}_\theta }$, and $\bar q({\bm{x}})$ is the heat flux supplied on the boundary $\partial {{\rm{\Omega }}_q}$ with the normal vector ${n_i}$; $h({\bm{x}})$ is the internal heat source. ${\theta ^{{\varepsilon _1}{\varepsilon _2}}}$ is the unresolved temperature fields. $\bm{\omega}$ represents the random variable, which characterizes the double uncertainties of spatial structures and material properties. $\kappa_{ij}^{{\varepsilon _1}{\varepsilon _2}}$ is the second-order heat conduction tensor with nonlinear temperature-dependent performance \cite{R46,R47,R53}. The detailed expression of $\kappa_{ij}^{{\varepsilon _1}{\varepsilon _2}}$ with double uncertainties can be presented as below.
\begin{equation}
\kappa_{ij}^{{\varepsilon _1}{\varepsilon _2}}({\bm{x}},\bm{\omega},{\theta ^{{\varepsilon _1}{\varepsilon _2}}})=\left\{ {\begin{aligned}
&\kappa_{ij}^{1}({\omega_m^{1}},{\theta ^{{\varepsilon _1}{\varepsilon _2}}}),\;\;{\bm{x}}\in \Omega_1^{\varepsilon_1}(\omega_g^{1}),\\
&\kappa_{ij}^{2}({\omega_m^{2}},{\theta ^{{\varepsilon _1}{\varepsilon _2}}}),\;\;{\bm{x}}\in \Omega_2^{\varepsilon_2}(\omega_g^{2}),\\
&\kappa_{ij}^{3}({\omega_m^{3}},{\theta ^{{\varepsilon _1}{\varepsilon _2}}}),\;\;{\bm{x}}\in \Omega_3^{\varepsilon_2}(\omega_g^{2}),
\end{aligned}} \right.\label{eq:2}
\end{equation}
where $\omega_m^{1}$, $\omega_m^{2}$ and $\omega_m^{3}$ are denoted as random variables for characterizing the material uncertainty, $\omega_g^{1}$ and $\omega_g^{2}$ are defined as random variables for for characterizing the geometric uncertainty of inclusion materials in meso- and micro-scales respectively.

Let us set $\displaystyle{\bm{y}} = {{\bm{x}}}/{{{\varepsilon _1}}}$ and $\displaystyle{\bm{z}} = {{{\varepsilon _1}{\bm{y}}}}/{{{\varepsilon _2}}} = {{\bm{x}}}/{{{\varepsilon _2}}}$, $\bm{x}$ is macro-scale coordinate of macroscopic structure $\Omega$, $\bm{y}$ is meso-scale coordinate in mesoscopic RVE $Y_{\bm{\omega}}$ and $\bm{z}$ is micro-scale coordinate in microscopic RVE $Z_{\bm{\omega}}$. Then the chain rules on multi-variable derivatives can be presented as below.
\begin{equation}
\begin{aligned}
\text{Macro-meso}:\;\;&\frac{\partial }{{\partial {x_i}}} \to \frac{\partial }{{\partial {x_i}}} + \frac{1}{{{\varepsilon _1}}}\frac{\partial }{{\partial {y_i}}},\\
\text{Meso-micro}:\;\;&\frac{\partial }{{\partial {y_i}}} \to \frac{\partial }{{\partial {y_i}}} + \frac{{{\varepsilon _1}}}{{{\varepsilon _2}}}\frac{\partial }{{\partial {z_i}}},\\
\text{Macro-meso-micro}:\;\;&\frac{\partial }{{\partial {x_i}}} \to \frac{\partial }{{\partial {x_i}}} + \frac{1}{{{\varepsilon _1}}}\frac{\partial }{{\partial {y_i}}} + \frac{1}{{{\varepsilon _2}}}\frac{\partial }{{\partial {z_i}}}.\label{eq:3}
\end{aligned}
\end{equation}

Enlightened by references \cite{R46,R47}, we suppose that $\theta ^{{\varepsilon _1}{\varepsilon _2}}$ has the following power series form of the asymptotic expansion in a multi-scale framework
\begin{equation}
\begin{aligned}
{\theta ^{{\varepsilon _1}{\varepsilon _2}}}({\bm{x}},\bm{\omega}) &= {\theta_{0}}({\bm{x}},{\bm{y}},{\bm{z}},\bm{\omega}) + {\varepsilon _1}{\theta_{1}}({\bm{x}},{\bm{y}},{\bm{z}},\bm{\omega})+ {\varepsilon _2}{\theta _{2}}({\bm{x}},{\bm{y}},{\bm{z}},\bm{\omega}) + O(\varepsilon _2).\label{eq:4}
\end{aligned}
\end{equation}
In the above formula (\ref{eq:4}), $\theta_{0}$, $\theta_{1}$ and $\theta_{2}$ are the specific terms corresponding to the coefficients 1, $\varepsilon _1$ and $\varepsilon _2$ of the asymptotic expansion, respectively.

Furthermore, the Taylor's formula is introduced and further written by multi-index notation as follow
\begin{equation}
\begin{aligned}
f({x_0},{y_0},{z_0},w_0 + \delta h)& = f({x_0},{y_0},{z_0},w_0) + {f_w}({x_0},{y_0},{z_0},w_0)\delta h\\
& + \frac{1}{2}{f_{ww}}({x_0},{y_0},{z_0},w_0){\delta h^2} + O({\delta h^3})\\
&= f({x_0},{y_0},{z_0},w_0) + {{\bf{D}}^{(0,0,0,1)}}f({x_0},{y_0},{z_0,w_0})\delta h \\
&+ \frac{1}{2}{{\bf{D}}^{(0,0,0,2)}}f({x_0},{y_0},{z_0},w_0){\delta h^2} + O({\delta h^3}).\label{eq:5}
\end{aligned}
\end{equation}
With the help of formula (\ref{eq:5}), the temperature-dependent material parameter $\kappa_{ij}^{{\varepsilon _1}{\varepsilon _2}}({\bm{x}},\bm{\omega},{\theta ^{{\varepsilon _1}{\varepsilon _2}}})$ can be expanded as
\begin{equation}
\begin{aligned}
&\kappa_{ij}^{{\varepsilon _1}{\varepsilon _2}}({\bm{x}},\bm{\omega},{\theta ^{{\varepsilon _1}{\varepsilon _2}}})= {\kappa_{ij}}\big({\bm{y}},{\bm{z}},\bm{\omega},{\theta_0} + {\varepsilon _1}{\theta_1} + {\varepsilon _2}{\theta_2} + O(\varepsilon _2)\big)\\
&= {\kappa_{ij}}({\bm{y}},{\bm{z}},\bm{\omega},{\theta_0}) + {{\bf{D}}^{( {0,0,0,1} )}}{\kappa_{ij}}({\bm{y}},{\bm{z}},\bm{\omega},{\theta_0})\big[ {{\varepsilon _1}{\theta_1} + {\varepsilon _2}{\theta_2} + O(\varepsilon _2)} \big]\\
&+ \frac{1}{2}{{\bf{D}}^{( {0,0,0,2} )}}{\kappa_{ij}}({\bm{y}},{\bm{z}},\bm{\omega},{\theta_0}){\big[ {{\varepsilon _1}{\theta_1} + {\varepsilon _2}{\theta_2} + O(\varepsilon _2)} \big]^2} +  \cdots \\
&= {\kappa_{ij}}({\bm{y}},{\bm{z}},\bm{\omega},{\theta_0}) + {\varepsilon _1}{{\bf{D}}^{( {0,0,0,1} )}}{\kappa_{ij}}({\bm{y}},{\bm{z}},\bm{\omega},{\theta_0}){\theta_1}\\
&+ {\varepsilon _2}{{\bf{D}}^{( {0,0,0,1} )}}{\kappa_{ij}}({\bm{y}},{\bm{z}},\bm{\omega},{\theta_0}){\theta_2} + O(\varepsilon _2)\\
&= \kappa_{ij}^{(0)} + {\varepsilon _1}\kappa_{ij}^{(1)} + {\varepsilon _2}\kappa_{ij}^{(2)} + O(\varepsilon _2).
\end{aligned}\label{eq:6}
\end{equation}

Then straightforward substitution of formulas (\ref{eq:4}) and (\ref{eq:6}) into the multi-scale nonlinear problem (\ref{eq:1}) leads to the following equations by virtue of the chain rule (\ref{eq:3}).
\begin{equation}
\begin{aligned}
& - \frac{\partial }{{\partial {x_i}}}\left[ {\left( {\kappa_{ij}^{(0)} + {\varepsilon _1}\kappa_{ij}^{(1)} + {\varepsilon _2}\kappa_{ij}^{(2)} } \right)\frac{\partial }{{\partial {x_j}}}\left( {{\theta_0} + {\varepsilon _1}{\theta_1} + {\varepsilon _2}{\theta_2} } \right)} \right]\\
&- \varepsilon _1^{ - 1}\frac{\partial }{{\partial {x_i}}}\left[ {\left( {\kappa_{ij}^{(0)} + {\varepsilon _1}\kappa_{ij}^{(1)} + {\varepsilon _2}\kappa_{ij}^{(2)} } \right)\frac{\partial }{{\partial {y_j}}}\left({{\theta_0} + {\varepsilon _1}{\theta_1} + {\varepsilon _2}{\theta_2} }\right)} \right]\\
&- \varepsilon _2^{ - 1}\frac{\partial }{{\partial {x_i}}}\left[ {\left( {\kappa_{ij}^{(0)} + {\varepsilon _1}\kappa_{ij}^{(1)} + {\varepsilon _2}\kappa_{ij}^{(2)}} \right)\frac{\partial }{{\partial {z_j}}}\left({{\theta_0} + {\varepsilon _1}{\theta_1} + {\varepsilon _2}{\theta_2} } \right)} \right]\\
&- \varepsilon _1^{ - 1}\frac{\partial }{{\partial {y_i}}}\left[ {\left( {\kappa_{ij}^{(0)} + {\varepsilon _1}\kappa_{ij}^{(1)} + {\varepsilon _2}\kappa_{ij}^{(2)} } \right)\frac{\partial }{{\partial {x_j}}}\left( {{\theta_0} + {\varepsilon _1}{\theta_1} + {\varepsilon _2}{\theta_2} } \right)} \right]\\
&- \varepsilon _1^{ - 2}\frac{\partial }{{\partial {y_i}}}\left[ {\left( {\kappa_{ij}^{(0)} + {\varepsilon _1}\kappa_{ij}^{(1)} + {\varepsilon _2}\kappa_{ij}^{(2)} } \right)\frac{\partial }{{\partial {y_j}}}\left( {{\theta_0} + {\varepsilon _1}{\theta_1} + {\varepsilon _2}{\theta_2} } \right)} \right]\\
&- \varepsilon _1^{ - 1}\varepsilon _2^{ - 1}\frac{\partial }{{\partial {y_i}}}\left[ {\left( {\kappa_{ij}^{(0)} + {\varepsilon _1}\kappa_{ij}^{(1)} + {\varepsilon _2}\kappa_{ij}^{(2)} } \right)\frac{\partial }{{\partial {z_j}}}\left( {{\theta_0} + {\varepsilon _1}{\theta_1} + {\varepsilon _2}{\theta_2}} \right)} \right]\\
&- \varepsilon _2^{ - 1}\frac{\partial }{{\partial {z_i}}}\left[ {\left( {\kappa_{ij}^{(0)} + {\varepsilon _1}\kappa_{ij}^{(1)} + {\varepsilon _2}\kappa_{ij}^{(2)} } \right)\frac{\partial }{{\partial {x_j}}}\left( {{\theta_0} + {\varepsilon _1}{\theta_1} + {\varepsilon _2}{\theta_2} } \right)} \right]\\
&- \varepsilon _1^{ - 1}\varepsilon _2^{ - 1}\frac{\partial }{{\partial {z_i}}}\left[ {\left( {\kappa_{ij}^{(0)} + {\varepsilon _1}\kappa_{ij}^{(1)} + {\varepsilon _2}\kappa_{ij}^{(2)} } \right)\frac{\partial }{{\partial {y_j}}}\left( {{\theta_0} + {\varepsilon _1}{\theta_1} + {\varepsilon _2}{\theta_2} } \right)} \right]\\
&- \varepsilon _2^{ - 2}\frac{\partial }{{\partial {z_i}}}\left[ {\left( {\kappa_{ij}^{(0)} + {\varepsilon _1}\kappa_{ij}^{(1)} + {\varepsilon _2}\kappa_{ij}^{(2)}} \right)\frac{\partial }{{\partial {z_j}}}\left( {{\theta_0} + {\varepsilon _1}{\theta_1} + {\varepsilon _2}{\theta_2} } \right)} \right]\\
&= h({\bm{x}})+O(1).
\end{aligned}
\label{eq:7}
\end{equation}

For the sake of simplifying the multi-scale analysis procedures, we shall define the following differential operators for multi-scale nonlinear problem.
\begin{equation}
\begin{aligned}
&{L_0}\left( {\mu ,\nu } \right) =  - \frac{\partial }{{\partial {x_i}}}\left[ {\mu \frac{{\partial \nu }}{{\partial {x_j}}}} \right],\;\;{L_1}\left( {\mu ,\nu } \right) =  - \frac{\partial }{{\partial {x_i}}}\left[ {\mu \frac{{\partial \nu }}{{\partial {y_j}}}} \right] - \frac{\partial }{{\partial {y_i}}}\left[ {\mu \frac{{\partial \nu }}{{\partial {x_j}}}} \right],\\
&{L_2}\left( {\mu ,\nu } \right) =  - \frac{\partial }{{\partial {x_i}}}\left[ {\mu \frac{{\partial \nu }}{{\partial {z_j}}}} \right] - \frac{\partial }{{\partial {z_i}}}\left[ {\mu \frac{{\partial \nu }}{{\partial {x_j}}}} \right],\;\;{L_3}\left( {\mu ,\nu } \right) =  - \frac{\partial }{{\partial {y_i}}}\left[ {\mu \frac{{\partial \nu }}{{\partial {y_j}}}} \right],\\
&{L_4}\left( {\mu ,\nu } \right) =  - \frac{\partial }{{\partial {y_i}}}\left[ {\mu \frac{{\partial \nu }}{{\partial {z_j}}}} \right] - \frac{\partial }{{\partial {z_i}}}\left[ {\mu \frac{{\partial \nu }}{{\partial {y_j}}}} \right],\;\;{L_5}\left( {\mu ,\nu } \right) =  - \frac{\partial }{{\partial {z_i}}}\left[ {\mu \frac{{\partial \nu }}{{\partial {z_j}}}} \right].
\end{aligned}
\label{eq:8}
\end{equation}

By virtue of the defined operators in (\ref{eq:8}), we acquire the following equilibrium equations with different power series $\varepsilon _1$ and $\varepsilon _2$ from (\ref{eq:7}).
\begin{equation}
O\left( {\varepsilon _2^{ - 2}} \right):{L_5}\big( {\kappa_{ij}^{(0)},{\theta_0}} \big) = 0,
\label{eq:9}
\end{equation}
\begin{equation}
O\left( {\varepsilon _1^{ - 1}\varepsilon _2^{ - 1}} \right):{L_4}\big( {\kappa_{ij}^{(0)},{\theta_0}} \big) = 0,
\label{eq:10}
\end{equation}
\begin{equation}
O\left( {\varepsilon _1^{ - 2}} \right):{L_3}\big( {\kappa_{ij}^{(0)},{\theta_0}} \big) = 0,
\label{eq:11}
\end{equation}
\begin{equation}
O\left( {{\varepsilon _1}\varepsilon _2^{ - 2}} \right):{L_5}\big( {\kappa_{ij}^{(1)},{\theta_0}} \big) + {L_5}\big( {\kappa_{ij}^{(0)},{\theta_1}} \big) = 0,
\label{eq:12}
\end{equation}
\begin{equation}
\begin{aligned}
O\left( {\varepsilon _2^{ - 1}} \right):&{L_5}\big( {\kappa_{ij}^{(2)},{\theta_0}} \big) + {L_4}\big( {\kappa_{ij}^{(1)},{\theta_0}} \big) + {L_2}\big( {\kappa_{ij}^{(0)},{\theta_0}} \big)\\
&+ {L_4}\big( {\kappa_{ij}^{(0)},{\theta_1}} \big) + {L_5}\big( {\kappa_{ij}^{(0)},{\theta_2}} \big) = 0,
\end{aligned}
\label{eq:13}
\end{equation}
\begin{equation}
\begin{aligned}
O\left( {\varepsilon _1^{ - 1}} \right):&{L_4}\big( {\kappa_{ij}^{(2)},{\theta_0}} \big) + {L_3}\big( {\kappa_{ij}^{(1)},{\theta_0}} \big) + {L_1}\big( {\kappa_{ij}^{(0)},{\theta_0}} \big)\\
&+ {L_3}\big( {\kappa_{ij}^{(0)},{\theta _1}} \big) + {L_4}\big( {\kappa_{ij}^{(0)},{\theta_2}} \big) = 0,
\end{aligned}
\label{eq:14}
\end{equation}
\begin{equation}
\begin{aligned}
O\left( {{\varepsilon _1}\varepsilon _2^{ - 1}} \right):&{L_5}\big( {\kappa_{ij}^{(2)},{\theta_1}} \big) + {L_2}\big( {\kappa_{ij}^{(1)},{\theta_0}} \big) + {L_4}\big( {\kappa_{ij}^{(1)},{\theta_1}} \big)\\
&+ {L_5}\big( {\kappa_{ij}^{(1)},{\theta_2}} \big) = 0,
\end{aligned}
\label{eq:15}
\end{equation}
\begin{equation}
O\left( {{\varepsilon _1^{2}}\varepsilon _2^{ - 1}} \right):{L_5}\big( {\kappa_{ij}^{(1)},{\theta_1}} \big) = 0,
\label{eq:16}
\end{equation}
\begin{equation}
O\left( {{\varepsilon _1^{0}}\varepsilon _2^{0}} \right):{L_0}\big( {\kappa_{ij}^{(0)},{\theta_0}} \big) + {L_1}\big( {\kappa_{ij}^{(0)},{\theta_1}} \big) + {L_2}\big( {\kappa_{ij}^{(0)},{\theta_2}} \big) = h({\bm{x}}).
\label{eq:17}
\end{equation}

From (\ref{eq:9}), (\ref{eq:10}) and (\ref{eq:11}), it is easy to acquire that $\theta_0$ is independent of mesoscopic variable $\bm{y}$ and microscopic variable $\bm{z}$.
\begin{equation}
{\theta_0}({\bm{x}},{\bm{y}},{\bm{z}},\bm{\omega}) = {\theta_0}({\bm{x}},\bm{\omega}).
\label{eq:18}
\end{equation}

Similarly, from (\ref{eq:12}), it follows that
\begin{equation}
{\theta_1}({\bm{x}},{\bm{y}},{\bm{z}},\bm{\omega}) = {\theta_1}({\bm{x}},{\bm{y}},\bm{\omega}).
\label{eq:19}
\end{equation}

According to (\ref{eq:19}), we assume that $\theta_1$ has the following expression
\begin{equation}
{\theta_1}({\bm{x}},{\bm{y}},\bm{\omega}) = {M_{{\alpha _1}}}({\theta_0},{\bm{y}},\bm{\omega})\frac{{\partial {\theta _0}}}{{\partial {x_{{\alpha _1}}}}},
\label{eq:20}
\end{equation}
where ${M_{{\alpha _1}}}({\theta_0},{\bm{y}},\bm{\omega})$ is so-called reference cell functions defined in mesoscopic RVE $Y_{\bm{\omega}}$.

Further, taking (\ref{eq:20}) into (\ref{eq:13}), the following equation can be obtained after computation and simplification
\begin{equation}
\frac{\partial }{{\partial {z_i}}}\left[ {\kappa_{ij}^{(0)}\frac{{\partial {\theta_2}}}{{\partial {z_j}}}} \right] =  - \frac{{\partial \kappa_{i{\alpha _1}}^{(0)}}}{{\partial {z_i}}}\frac{{\partial {\theta_0}}}{{\partial {x_{{\alpha _1}}}}} - \frac{{\partial \kappa_{i{\alpha _2}}^{(0)}}}{{\partial {z_i}}}\frac{{\partial {M_{{\alpha _1}}}}}{{\partial {y_{{\alpha _2}}}}}\frac{{\partial {\theta_0}}}{{\partial {x_{{\alpha _1}}}}}.
\label{eq:21}
\end{equation}
Based on (\ref{eq:21}), we assume that $\theta_2$ has the following expression
\begin{equation}
{\theta_2}({\bm{x}},{\bm{y}},{\bm{z}},\bm{\omega}) = {H_{{\alpha _1}}}({\theta_0},{\bm{y}},{\bm{z}},\bm{\omega})\frac{{\partial {\theta_0}}}{{\partial {x_{{\alpha _1}}}}} + {H_{{\alpha _2}}}({\theta_0},{\bm{y}},{\bm{z}},\bm{\omega})\frac{{\partial {M_{{\alpha _1}}}({\theta_0},{\bm{y}},\bm{\omega})}}{{\partial {y_{{\alpha _2}}}}}\frac{{\partial {\theta_0}}}{{\partial {x_{{\alpha _1}}}}},
\label{eq:22}
\end{equation}
where ${H_{{\alpha _1}}}({\theta_0},{\bm{y}},{\bm{z}},\bm{\omega})$ is so-called reference cell functions defined in microscopic RVE $Z_{\bm{\omega}}$. Inserting (\ref{eq:22}) into (\ref{eq:21}), auxiliary cell problems imposing with homogeneous Dirichlet boundary condition are gained for computing ${H_{{\alpha _1}}}({\theta_0},{\bm{y}},{\bm{z}},\bm{\omega})$ in microscopic RVE $Z_{\bm{\omega}}$ after computation and simplification
\begin{equation}
\left\{{\begin{aligned}
&{\frac{\partial }{{\partial {z_i}}}\left[ {\kappa_{ij}^{(0)}\frac{{\partial {H_{{\alpha _1}}}({\theta_0},{\bm{y}},{\bm{z}},\bm{\omega})}}{{\partial {z_j}}}} \right] =  - \frac{{\partial \kappa_{i{\alpha _1}}^{(0)}}}{{\partial {z_i}}},\quad {\bm{z}} \in Z_{\bm{\omega}},}\\
&{{H_{{\alpha _1}}}({\theta_0},{\bm{y}},{\bm{z}},\bm{\omega}) = 0,\quad {\bm{z}} \in \partial Z_{\bm{\omega}}.}
\end{aligned}}\right.
\label{eq:23}
\end{equation}

Subsequently, replacing $\theta_0$, $\theta_1$ and $\theta_2$ in (\ref{eq:14}) with their detailed expressions (\ref{eq:18}), (\ref{eq:20}) and (\ref{eq:22}), and making integration on both sides of (\ref{eq:14}) with respect to microscopic variable $\bm{z}$ on the RVE $Z_{\bm{\omega}}$, the following equation can be gained
\begin{equation}
- \frac{\partial }{{\partial {y_i}}}\left[ {\kappa_{ij}^ * ({\bm{y}},\bm{\omega},{\theta_0})\frac{{\partial {M_{{\alpha _1}}}({\theta _0},{\bm{y}},\bm{\omega})}}{{\partial {y_j}}}} \right]\frac{{\partial {\theta_0}}}{{\partial {x_{{\alpha _1}}}}} - \frac{{\partial \kappa_{i{\alpha _1}}^ * ({\bm{y}},\bm{\omega},{\theta_0})}}{{\partial {y_i}}}\frac{{\partial {\theta_0}}}{{\partial {x_{{\alpha _1}}}}} = 0,
\label{eq:24}
\end{equation}
where $\kappa_{ij}^*({\bm{y}},\bm{\omega},{\theta_0})$ is mesoscopic equivalent thermal conductivity and can be computed by the following formula
\begin{equation}
\kappa_{ij}^*({\bm{y}},\bm{\omega},{\theta_0}) = {\left\langle {\kappa_{ij}^{(0)} + \kappa_{i{\alpha _1}}^{(0)}\frac{{\partial {H_j}({\theta_0},{\bm{y}},{\bm{z}},\bm{\omega})}}{{\partial {z_{{\alpha _1}}}}}} \right\rangle _{Z_{\bm{\omega}}}.}
\label{eq:25}
\end{equation}
In which the volume average operator $\left\langle\bullet\right\rangle_{Z_{\bm{\omega}}}=\displaystyle\frac{1}{|Z_{\bm{\omega}}|}{\int_{Z_{\bm{\omega}}}}\bullet dZ_{\bm{\omega}}$. By means of making integration of microscopic variable $\bm{z}$, Eq.\hspace{1mm}(\ref{eq:14}) converted to the Eq.\hspace{1mm}(\ref{eq:24}), which is only related to the mesoscopic variable $\bm{y}$ and independent of the microscopic variable $\bm{z}$. By the aforementioned up-scaling method from micro-scale to meso-scale, the equivalent thermal conductivity can be computed on meso-scale by the established formula (\ref{eq:25}). Currently, it can easily establish the following auxiliary cell problems attached with homogeneous Dirichlet boundary condition in mesoscopic RVE $Y_{\bm{\omega}}$ from equality (\ref{eq:24}).
\begin{equation}
\left\{ {\begin{aligned}
&{\frac{\partial }{{\partial {y_i}}}\left[ {\kappa_{ij}^*({\bm{y}},\bm{\omega},{\theta_0})\frac{{\partial {M_{{\alpha _1}}}({\theta_0},{\bm{y}},\bm{\omega})}}{{\partial {y_j}}}} \right] =  - \frac{{\partial \kappa_{i{\alpha _1}}^*({\bm{y}},\bm{\omega},{\theta_0})}}{{\partial {y_i}}},\quad {\bm{y}} \in Y_{\bm{\omega}}},\\
&{{M_{{\alpha _1}}}({\theta_0},{\bm{y}},\bm{\omega}) = 0,\quad {\bm{y}} \in \partial Y_{\bm{\omega}}}.
\end{aligned}} \right.
\label{eq:26}
\end{equation}

Finally, we integrate variable $\bm{z}$ and $\bm{y}$ on both sides of equation (\ref{eq:17}) in sequence. Moreover, the macroscopic homogenized thermal equation for multi-scale nonlinear problem (\ref{eq:1}) can be established as follows
\begin{equation}
\left\{\begin{aligned}
&-\frac{\partial}{\partial x_i}\left[\kappa_{i j}^{**}\left({\bm{\omega},{\theta_0}} \right) \frac{\partial \theta_0(\bm{x},\bm{\omega})}{\partial x_j}\right]=h(\bm{x}), \text { in } \Omega, \\
&\theta_0(\bm{x},\bm{\omega})=\hat{\theta}(\bm{x}), \text { on } \partial \Omega_\theta, \\
&\kappa_{i j}^{**}\left({\bm{\omega},{\theta_0}} \right) \frac{\partial \theta_0(\bm{x},\bm{\omega})}{\partial x_j} n_i=\bar{q}(\bm{x}), \text { on } \partial \Omega_q,
\end{aligned}\right.
\label{eq:27}
\end{equation}
where $\kappa_{i j}^{**}\left({\bm{\omega},{\theta_0}} \right)$ is macroscopic equivalent parameter and can be computed by the following formula.
\begin{equation}
\begin{aligned}
\kappa_{ij}^{**}\left( {\bm{\omega},{\theta_0}} \right)& = \left\langle \left\langle \left[\kappa_{ij}^{(0)} + \kappa_{i{\alpha _1}}^{(0)}\frac{\partial {H_j}({\theta_0},{\bm{y}},{\bm{z}},\bm{\omega})}{\partial z_{\alpha _1}} \right]\right.\right.\\
&\left.\left.+ \left[ {\kappa_{i{\alpha _2}}^{(0)} + \kappa_{i{\alpha _1}}^{(0)}\frac{{\partial {H_{{\alpha _2}}}({\theta_0},{\bm{y}},{\bm{z}},\bm{\omega})}}{{\partial {z_{{\alpha _1}}}}}} \right]\frac{{\partial {M_j}({\theta_0},{\bm{y}},\bm{\omega})}}{\partial y_{\alpha _2}} \right\rangle_{Z_{\bm{\omega}}} \right\rangle _{Y_{\bm{\omega}}}\\
&= \left\langle \kappa_{ij}^*({\bm{y}},\bm{\omega},{\theta_0}) + \kappa_{i{\alpha _2}}^*({\bm{y}},\bm{\omega},{\theta_0})\frac{{\partial {M_j}({\theta_0},{\bm{y}},\bm{\omega})}}{{\partial {y_{{\alpha _2}}}}} \right\rangle _{Y_{\bm{\omega}}}.
\end{aligned}
\label{eq:28}
\end{equation}
In the above formula (\ref{eq:28}), the volume average operator $\left\langle\bullet\right\rangle_{Y_{\bm{\omega}}}$ is defined as  $\left\langle\bullet\right\rangle_{Y_{\bm{\omega}}}=\displaystyle\frac{1}{|Y_{\bm{\omega}}|}{\int_{Y_{\bm{\omega}}}}\bullet dY_{\bm{\omega}}$. Through making integration of variable $\bm{z}$ on microscopic RVE $Z_{\bm{\omega}}$ and then making integration of variable $\bm{y}$ on mesoscopic RVE $Y_{\bm{\omega}}$, Eq.\hspace{1mm}(\ref{eq:17}) is transformed into the Eq.\hspace{1mm}(\ref{eq:27}), that is only related to macro-scale variable $\bm{x}$ and is irrelevant to micro-scale variable $\bm{z}$ and meso-scale variable $\bm{y}$. By this double up-scaling method from microscopic scale to mesoscopic scale and then to macroscopic scale, the equivalent material parameters can be calculated on macro-scale by formula (\ref{eq:28}).
\subsection{Material database of random heterogeneous materials}
Random heterogeneous materials possess complicated spatial configurations, nonlinear and uncertain material properties, and the establishment of an effective database of random composites is essential for successful machine learning predictions.
There exist two technical challenges for establishing material database for random composites with hierarchical configurations. One challenge is to determine the data labels of supervised learning. We established a stochastic three-scale homogenized method to calculate the effective nonlinear thermal conductivity as data labels. The other challenge is to sufficiently extract microscopic and mesoscopic data features for supervised learning of random heterogeneous materials. The principal difficulty is that material parameters of mesoscopic matrix material are unknown and necessary to have innovative treatment.

In this study, novel background meshing and filling techniques are developed to extract geometry and material features of random heterogeneous materials for establishing material databases. For 2D statistical configurations, a $99\times99$ equispaced rectangular mesh is chosen to pick up data features of 2D materials. For 3D statistical configurations, a $29\times29\times29$ equispaced hexahedral mesh is established to extract the data features of 3D materials. For highly heterogeneous materials with random hierarchical configurations, we denote $\kappa_{{z_1}{z_2}}^{1 \sim m_{i,1}}$ and $\kappa_{{z_1}{z_2}}^{1 \sim m_{i,2}}$ as the thermal conductivity of each voxel $({z_1},{z_2})$ assigned by matrix and inclusion components of 2D composites at micro-scale respectively, while $\kappa_{{z_1}{z_2}{z_3}}^{1 \sim m_{i,1}}$ and $\kappa_{{z_1}{z_2}{z_3}}^{1 \sim m_{i,2}}$ as the thermal conductivity of each voxel $({z_1},{z_2},{z_3})$ assigned by matrix and inclusion components of 3D composites at micro-scale respectively. To solve the issue of missing material parameters for matrix materials at meso-scale, we utilize a virtual material as filler matrix material with material parameter $K$. For random heterogeneous materials with hierarchical configurations, we further denote $K_{{y_1}{y_2}}^{1 \sim m_{i,4}}$ and $\kappa_{{y_1}{y_2}}^{1 \sim m_{i,3}}$ as the thermal conductivity of each voxel $({y_1},{y_2})$ assigned by matrix and inclusion constituents of 2D composites at meso-scale respectively, while $K_{{y_1}{y_2}{y_3}}^{1 \sim m_{i,4}}$ and $\kappa_{{y_1}{y_2}{y_3}}^{1 \sim m_{i,3}}$ as the thermal conductivity of each voxel $({y_1},{y_2},{y_3})$ assigned by matrix and inclusion constituents of 3D composites at meso-scale respectively. In this paper, all filling material parameters are assigned as value 0. For the two-scale composite materials, we can also apply the proposed method to establish material databases, which is similar as the previous work \cite{R26,R27,R54}. Moreover, it should be emphasized that the proposed approach can extract both geometric and material characteristics.

\begin{figure}
	\centering
	\includegraphics[width=140mm]{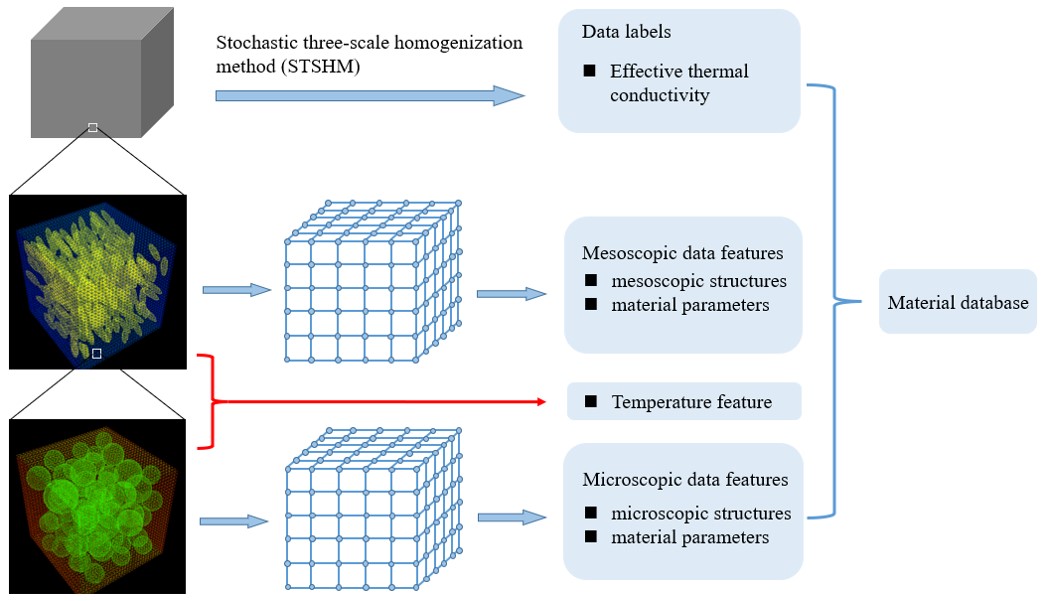}
	\caption{The generation diagram of three-scale material database.}\label{fig:3}
\end{figure}

According to the aforementioned background meshing and filling techniques, the geometric and material information at micro- and meso-scales are assigned as data features, which together with the service temperature $\theta$ make up the data features of the material database. The macroscopic effective nonlinear thermal conductivity is regarded as data labels for the material database. The detailed three-scale process of obtaining data features and establishing material databases is exhibited in Fig.\hspace{1mm}\ref{fig:3}. The material databases of 2D and 3D three-scale composites are presented in Tables \ref{tab:1} and \ref{tab:2}, respectively.

\begin{table}[]
	\footnotesize
	\centering
	\caption{2D three-scale material database.}
	\begin{tabular}{ccccccc}
		\hline
		Sample number & \multicolumn{2}{c}{\begin{tabular}[c]{@{}c@{}}Microscopic data\\ features\end{tabular}} & \multicolumn{2}{c}{\begin{tabular}[c]{@{}c@{}}Mesoscopic data\\ features\end{tabular}} & Temperature & Data labels \\ \hline
		Sample 1 & $\kappa_{{z_1}{z_2}}^{1 \sim m_{1,1}}$ & $\kappa_{{z_1}{z_2}}^{1 \sim m_{1,2}}$   & $K_{{y_1}{y_2}}^{1 \sim m_{1,4}}$    &  $\kappa_{{y_1}{y_2}}^{1 \sim m_{1,3}}$  & $\theta$ & $\kappa_{ij}^{ *  * }$ \\
		...    &  ...       & ...    & ...    &   ...      &  ...       & ...    \\
		Sample n & $\kappa_{{z_1}{z_2}}^{1 \sim m_{n,1}}$ & $\kappa_{{z_1}{z_2}}^{1 \sim m_{n,2}}$   & $K_{{y_1}{y_2}}^{1 \sim m_{n,4}}$    &  $\kappa_{{y_1}{y_2}}^{1 \sim m_{n,3}}$  & $\theta$ & $\kappa_{ij}^{ *  * }$  \\ \hline
	\end{tabular}\label{tab:1}
\end{table}

\begin{table}[]
	\footnotesize
	\centering
	\caption{3D three-scale material database.}
	\begin{tabular}{ccccccc}
		\hline
		Sample number & \multicolumn{2}{c}{\begin{tabular}[c]{@{}c@{}}Microscopic data\\ features\end{tabular}} & \multicolumn{2}{c}{\begin{tabular}[c]{@{}c@{}}Mesoscopic data\\ features\end{tabular}} & Temperature & Data labels \\ \hline
		Sample 1 & $\kappa_{{z_1}{z_2}{z_3}}^{1 \sim m_{1,1}}$ & $\kappa_{{z_1}{z_2}{z_3}}^{1 \sim m_{1,2}}$   & $K_{{y_1}{y_2}{y_3}}^{1 \sim m_{1,4}}$    &  $\kappa_{{y_1}{y_2}{y_3}}^{1 \sim m_{1,3}}$  & $\theta$ & $\kappa_{ij}^{ *  * }$ \\
		...    &  ...       & ...    & ...    &   ...      &  ...       & ...    \\
		Sample n & $\kappa_{{z_1}{z_2}{z_3}}^{1 \sim m_{n,1}}$ & $\kappa_{{z_1}{z_2}{z_3}}^{1 \sim m_{n,2}}$   & $K_{{y_1}{y_2}{y_3}}^{1 \sim m_{n,4}}$    &  $\kappa_{{y_1}{y_2}{y_3}}^{1 \sim m_{n,3}}$  & $\theta$ & $\kappa_{ij}^{ *  * }$  \\ \hline
	\end{tabular}\label{tab:2}
\end{table}
\section{Self-optimization wavelet-learning method}
\label{sec:3}
\begin{figure}
	\centering
	\includegraphics[width=140mm]{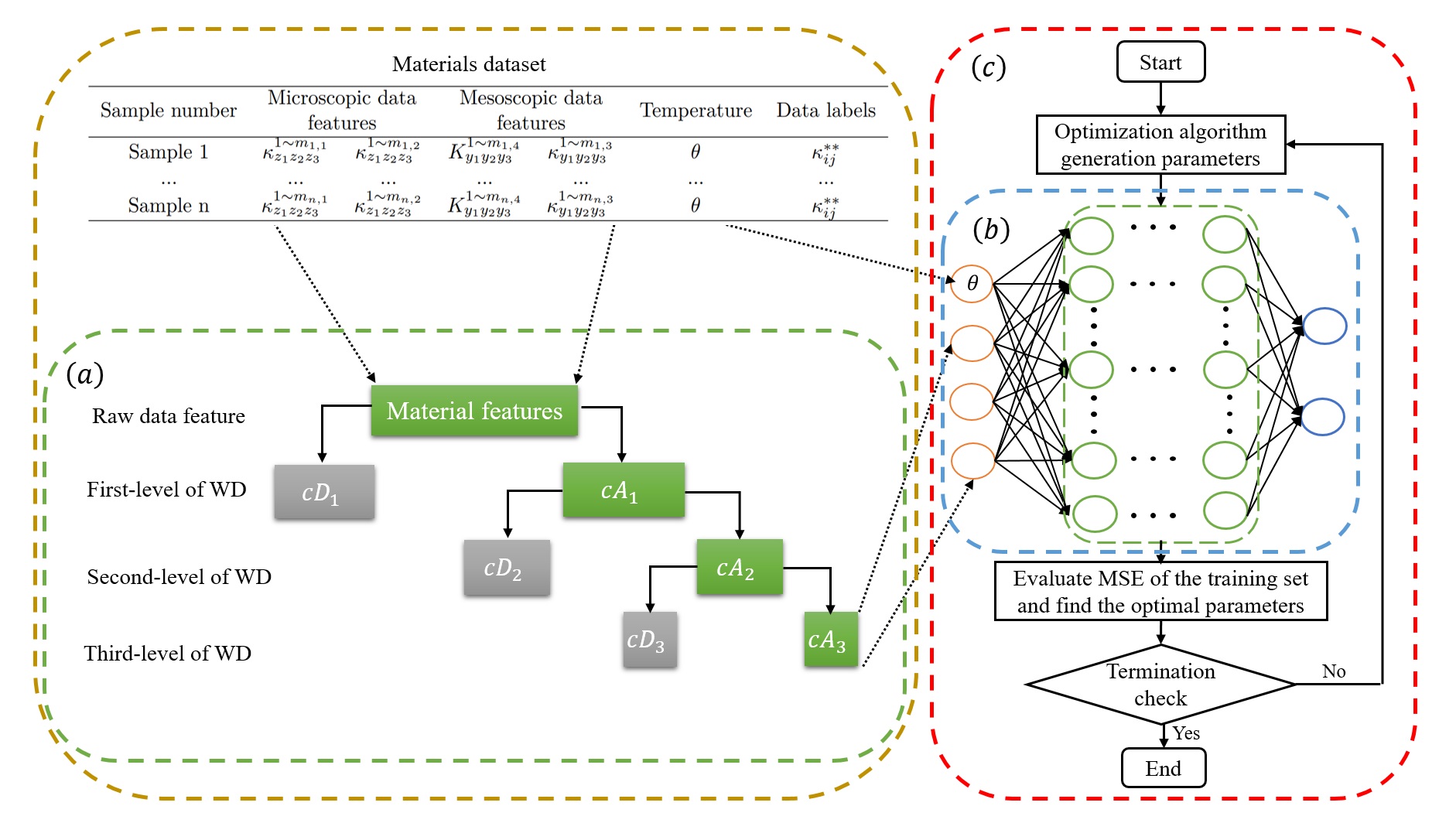}
	\caption{The diagram of self-optimization wavelet-learning framework.}\label{fig:4}
\end{figure}

This section establishes a new self-optimization wavelet-learning method in details. This novel framework is comprised of wavelet preprocess, machine learning prediction and intelligent self-optimization mechanism. It should be underlined that wavelet preprocess can prominently reduce the input data scale of predictive models and improve the training efficiency of supervised Learning. Moreover, intelligent self-optimization mechanism is beneficial to optimize the network structure and learning rate of the predictive models. The detailed structure of self-optimization wavelet-learning method is illustrated in Fig.\hspace{1mm}\ref{fig:4}.
\subsection{Wavelet-learning method}
Inspired by biology and neurology, artificial neural network (ANN) is devised by scientists, which can simulate the operating mechanism of biological neural networks by abstracting the biological neural networks, constructing artificial neurons, and establishing connections between artificial neurons according to a definite topological structure.

The goal of a neural network is to learn a certain mapping functional $\varphi$ such that a given input $\textbf{x}$ yields an output $\textbf{y}$. General neural networks are comprised of input, hidden and output layers. Further, their hidden layer is a kind of multi-layer structure, and the detailed structure is shown in Fig.\hspace{1mm}\ref{fig:4}(b). In neural network, the neuron is its basic unit. Assume that a neuron receives the input vector $\textbf{x} = \left[ {{x_1};{x_2}; \cdots ;{x_d}} \right]$ with $d$ dimension, $z$ denotes the weighted sum of input vector $\textbf{x}$ acquired from a neuron \cite{R48}.
\begin{equation}
z = \sum\limits_{i = 1}^d {{w_i}} {x_i} + b = {\textbf{w}^{\rm{T}}}\textbf{x} + b,
\label{eq:32}
\end{equation}
where $\textbf{w} = \left[ {{w_1};{w_2}; \cdots ;{w_d}} \right]$ denotes the weight vector and $b$ is the bias.

The neuron activity value is obtained via the activation function $f(z)$. The frequently-used activation functions are Tanh$(z) = ({{{e^z} - {e^{ - z}}}})/({{{e^z} + {e^{ - z}}}})$, Sigmoid$(z) = {1}/({{1 + e^{ - z}}})$ and ReLU$(z) = \max (0,z)$. It should be mentioned that the activation function used in this paper is the ReLU function. Assuming that the neural network has $n$ layers in total and entire neural network can be generally characterized as a composite function $\varphi$, this function has the mathematical expression as
\begin{equation}
\varphi (\textbf{x}) = {f_n}\left( { \cdots {f_2}\left( {{f_1}\left( {{\textbf{W}^1}\textbf{x} + {\textbf{b}^1}} \right)} \right)} \right),
\label{eq:33}
\end{equation}
where $\textbf{W}^i$ and $\textbf{b}^i$ are presented as weights and biases of layer $i$. When the structure of the neural network model is given, the main objective is to determine the parameters $\boldsymbol{\eta}=\left( {{\eta^1},{\eta^2}, \cdots ,{\eta^n}} \right)$ of the model with ${\eta^i} = \left( {{\textbf{W}^i},{\textbf{b}^i}} \right)$. The objective function is minimized by the gradient descent algorithm to compute parameters $\boldsymbol{\eta}$.
\begin{equation}
\mathop {\min }\limits_{\boldsymbol{\eta}}  {\cal L}(\varphi (\textbf{x}),\textbf{y}) + \lambda \sum\limits_{i = 1}^n \psi  \left( {{\eta^i}} \right).
\label{eq:34}
\end{equation}
In ANN model (\ref{eq:34}), ${\cal L}$ stands for the loss function, and the frequently adopted loss functions are primarily mean square error (MSE), mean absolute error (MAE) and binary cross-entropy loss function. $\psi$ stands for the regularization function, and some widely adopted regularization are $L^2$ regularization, $L^1$ regularization, Dropout regularization and Batch Normalization regularization. Moreover, commonly used gradient descent algorithms for solving ANN model are stochastic gradient descent (SGD), Momentum optimization, Adaptive moment estimation (Adam) and Nesterov Accelerated Gradient (NAG). In this paper, MSE loss function, $L^{2}$ regularization function, and Adam gradient descent algorithm are utilized in establishing predictive models. In order to diminish over-fitting phenomenon, stopping iteration are employed in the training process.

It is commonly recognized that a large amount of input data on the training set affects the training results of the predictive models and consumes a tremendous amount of time. Wavelet transform owns the capability of data compression and feature extraction from raw data, which is extensively employed in the research field of image and signal processing. For the sake of improving the training efficiency of the neural network, we innovatively introduce wavelet transform to preprocess the data features of material database. By using the wavelet transform, the original data is decomposed into multiple layers of wavelet coefficients. The specific process of three-layer wavelet decomposition is depicted in Fig.\hspace{1mm}\ref{fig:4}(a). By the first layer of wavelet decomposition, the raw data is decomposed into approximate coefficients $cA_{1}$ and detail coefficients $cD_{1}$. Subsequently, by applying wavelet transform again to $cA_{1}$, $cA_{1}$ is divided into two components: approximate coefficients $cA_{2}$ and detail coefficients $cD_{2}$. In the third layer of wavelet decomposition, wavelet decomposition is applied to $cA_{2}$ to obtain approximate coefficients $cA_{3}$ and detail coefficients $cD_{3}$. The raw data features are decomposed by three-layer wavelet decomposition to obtain the approximation coefficients $cA_{3}$, which form the input data of the predictive models with the temperature feature $\theta$. Processing the data after multi-layer wavelet decomposition greatly reduces the input data size of the neural network, and ensures that the neural network can be trained successfully and improves the training efficiency of the predictive models.

\subsection{Intelligent optimization algorithms}
The accuracy of the wavelet-learning model is mainly influenced by the network structure and learning rate, therefore it is momentous to determine the appropriate network structure and learning rate. In order to build a high-precision wavelet-learning approach, the network structure and learning rate of the predictive model will be self-search in combination with the intelligent optimization algorithm. Intelligent optimization algorithms are classified as ant colony algorithm (AC), simulated annealing algorithm (SA), particle swarm optimization algorithm (PSO) and artificial bee colony algorithm (ABC). In this paper, PSO and ABC are adopted to optimize the undetermined parameters of the wavelet-learning predictive models.

The PSO algorithm is a typical intelligence optimization algorithm. The PSO algorithm employs $N$ particles to establish a particle swarm. And then, PSO algorithm searches for the optimal solution repeatedly and iteratively in $D$-dimensional space. The particles move in space with two characteristics: position and velocity. In each iteration, the velocity $v_{i,j}$ and position $x_{i,j}$ of particle $i$ in the $j$-th dimension are updated on the basis of the following equation.
\begin{equation}
{v_{i,j}}(t + 1) = {v_{i,j}}(t) + {c_1} * {r_{1,j}}(t) * \left( {{p_{i,j}} - {x_{i,j}}(t)} \right) + {c_2} * {r_{2,j}}(t) * \left( {{p_{g,j}} - {x_{i,j}}(t)} \right),
\label{eq:35}
\end{equation}
\begin{equation}
{x_{i,j}}(t + 1) = {x_{i,j}}(t) + {v_{i,j}}(t + 1),
\label{eq:36}
\end{equation}
where $t$ is the current number of iterations, $i \in \{ 1,2, \ldots ,N\}$, $N$ is the number of particles, $j \in \{ 1,2, \ldots ,D\}$, $D$ is the dimension of the solution. $c_{1}$ and $c_{2}$ are acceleration coefficients that represent the degree to which the particles are influenced by individual perceptions and social knowledge. $r_{1,j}$ and $r_{2,j}$ are random number in $\left[ {0,1} \right]$. Furthermore, we denote position ${\bm{x}_i} = \left( {{x_{i,1}},{x_{i,2}}, \cdots ,{x_{i,D}}} \right)$ and velocity ${\bm{v}_i} = \left( {{v_{i,1}},{v_{i,2}}, \cdots ,{v_{i,D}}} \right)$. Particle $i$ can record the optimal position it has experienced, namely the individual most optimal position ${\bm{p}_i} = \left( {{p_{i,1}},{p_{i,2}}, \cdots ,{p_{i,D}}} \right)$, and the optimal position experienced by all particles, namely the global optimal position ${\bm{p}_g} = \left( {{p_{g,1}},{p_{g,2}}, \cdots ,{p_{g,D}}} \right)$. To avoid particle motion beyond the solution space, the particle velocity $v_{i,j}$ is restricted to $\left[ {-v_{\max,j},v_{\max,j}} \right]$, and the particle position $x_{i,j}$ is restricted to the corresponding search range $\left[ {x_{\min,j},x_{\max,j}} \right]$. In this paper, setting $T=1000$, $N=10$, $c_{1}=2$, $c_{2}=2$, $D$ is determined by specific computational examples. The PSO algorithm steps are presented in Algorithm 2 as follows.
\begin{algorithm}[!h]
	\caption{PSO algorithm}
	\label{alg:AOA}
	\begin{algorithmic}[1]
		\STATE Set parameters: $T$, $N$, $D$, $c_1$, $c_2$, $n=0$
		\STATE Randomly initialize the position and velocity of the particle and calculate the objective function value $f(\bm{x})$ of the particle, initialize the optimal position $\bm{p}_{i}$ of each particle and the global optimal position $\bm{p}_{g}$
		\WHILE{$n<T$}
		\FOR{$i=1$ to $N$}
		\FOR{$j=1$ to $D$}
		\STATE Calculate velocity $v_{i,j}$ by Eq.\hspace{1mm}(\ref{eq:35})
		\STATE Restrict $v_{i,j}$ in $[{-{v_{\max,j }},{v_{\max ,j}}}]$
		\STATE Calculate position $x_{i,j}$ by Eq.\hspace{1mm}(\ref{eq:36})
		\STATE Restrict $x_{i,j}$ in $[{{x_{\min ,j}},{x_{\max ,j}}}]$
		\ENDFOR
		\STATE Calculate $f(\bm{x}_i)$
		\STATE Update $\bm{p}_i$
		\ENDFOR
		\STATE Update $\bm{p}_g$
		\STATE $n=n+1$
		\ENDWHILE
	\end{algorithmic}
\end{algorithm}

The ABC algorithm is comprised of three main types of bees: employed bees, onlooker bees and scout bees \cite{R49,R56}. Randomly initialize $SN$ initial solutions in population $P$, where $SN$ is the number of nectar sources and also equals to the number of employed bees, with each solution ${\bm{x}_i}$ $(i = 1,2,...,SN)$ being a vector of $D$ dimensions. After initialization, based on the initial solutions, all employed, onlooker and scout bees start a cyclic search. The employed bees generate a new nectar location based on the nectar information in the memory and calculate the amount of nectar at the new location, utilizing a greedy selection mechanism to choose original location and new nectar location. After all employed bees accomplish their search task, the information from memory is shared with the onlooker bees. On the basis of the obtained information, the onlooker bees select a nectar location with a certain probability and generate a new nectar location and calculate the amount of nectar in the new location just like the employed bees, which also use the greedy selection mechanism to select the original location and the new nectar location. When the nectar source still has not improved after $limit$ cycles, the source is abandoned. The employed bee at that location becomes the scout bee and produces the nectar source according to the method of initializing the nectar source.

Nectar sources are randomly selected by initialized populations according to Eq.\hspace{1mm}(\ref{eq:37}) and the nectar quantity of the nectar sources is evaluated.
\begin{equation}
{x_{i,j}} = {x_{\min ,j}} + {\mathop{\rm rand}\nolimits} (0,1) * \left( {{x_{\max ,j}} - {x_{\min ,j}}} \right),
\label{eq:37}
\end{equation}
where $i \in \{ 1,2, \ldots ,SN\}$, $j \in \{ 1,2, \ldots ,D\}$, $x_{\max,j}$ and $x_{\min,j}$ are the upper and lower bounds of $x_{i,j}$.

The probability ${\tilde p_i}$ of nectar source selection by onlooker bees is related to nectar quality, and the probability ${\tilde p_i}$ is expressed as
\begin{equation}
{\tilde p_i} = \frac{{fi{t_i}}}{{\sum\limits_{j = 1}^{SN} f i{t_j}}},
\label{eq:38}
\end{equation}
in which $fi{t_i}$ is the fitness value of nectar source $i$. The current position $x_{i,j}$ is known, and in order to find a better nectar source, a new position is generated by
\begin{equation}
{v_{i,j}} = {x_{i,j}} + {\phi _{i,j}} * \left( {{x_{i,j}} - {x_{k,j}}} \right),
\label{eq:39}
\end{equation}
where $k \in \{ 1,2, \ldots ,SN\}$ with $k \ne i$ and $j \in \{ 1,2, \ldots ,D\}$ are randomly chosen subscripts, ${\phi _{i,j}}$ is random number in $\left[ { - 1,1} \right]$. In this paper, setting $T=1000$, $SN=5$, $limit=30$, $D$ is determined by specific computational examples. The primary procedures of ABC are presented in Algorithm 3 as follows. The training MSE of the wavelet-learning approach is considered as the objective function of the PSO and ABC algorithms to establish a self-optimization wavelet-learning approach, which can find a high-precision network structure and learning rate. The specific flow chart of self-optimization mechanism is shown in Fig.\hspace{1mm}\ref{fig:4}(c).
\begin{algorithm}[!h]
	\caption{ABC algorithm}
	\label{alg:AOA}
	
	\begin{algorithmic}[1]
		
		\STATE Set parameters: $T$, $SN$, $D$, $limit$, $t=1$, $trial_i=1$
		\STATE Initialize the population of nectar sources by using Eq.\hspace{1mm}(\ref{eq:37}) and calculate their objective function values
		\WHILE{$t<T$}
		\STATE  /* Employed bee phase */
		\FOR{$i=1$ to $SN$}
		\STATE Generate new nectar source $\bm{v}_i$ by using Eq.\hspace{1mm}(\ref{eq:39}) and calculate its objective function value
		\STATE If {$f(\bm{v}_i)<f(\bm{x}_i)$}, set $\bm{x}_i=\bm{v}_i$, $trial_i=1$, otherwise, set $trial_i=trial_i+1$
		\STATE $i=i+1$
		\ENDFOR
		
		\STATE Calculate the probability value $\tilde p_i$ by Eq.\hspace{1mm}(\ref{eq:38})
		\STATE /* Onlooker bee phase*/
		\STATE Set $i=1$, $n=0$
		\WHILE {$n<SN$}
		\IF {$rand(0,1)<\tilde p_i$}
		\STATE Generate new nectar source $\bm{v}_i$ via using Eq.\hspace{1mm}(\ref{eq:39}) and calculate its objective function value
		\STATE If {$f(\bm{v}_i)<f(\bm{x}_i)$}, set $\bm{x}_i=\bm{v}_i$, $trial_i=1$, otherwise, set $trial_i=trial_i+1$
		\STATE $n=n+1$
		\ENDIF
		\STATE $i=i+1$, if $i=SN$, set $i=1$
		\ENDWHILE
		
		\STATE /* Scout bee phase*/
		\IF{ $trial_i>limit$ }
		\STATE Replace $\bm{x}_i$ with a new randomly generated nectar source via using Eq.\hspace{1mm}(37) and calculate its objective function value
		\ENDIF
		
		\STATE Preservation of optimal individuals and optimal solutions
		\STATE $t=t+1$
		
		\ENDWHILE
		
	\end{algorithmic}
\end{algorithm}

\section{Numerical experiments and discussions}
\label{sec:4}
To confirm the computational accuracy and effectiveness of the proposed self-optimization wavelet-learning method, extensive numerical experiments are implemented on realistic random composites. In these numerical experiments, raw data set is randomly divided into the training set and test set with the ratio 8:2. Also average relative absolute errors are set as the training and test errors. Moreover, we assume that both component inclusions and matrix of the investigated random heterogeneous materials are isotropic. Our predictive models are trained on a platform equipped with NVIDIA T600 GPU and Intel Core i7-10700 CPU@2.90 GHz.
\subsection{Example 1. nonlinear particulate composite ZrO$_{2}$/Ti-6Al-4V with two-scale random configurations}
\begin{figure}[htbp]
	\centering
	\subfigure{
		\begin{minipage}[t]{0.25\linewidth}
			\centering
			\includegraphics[width=30mm]{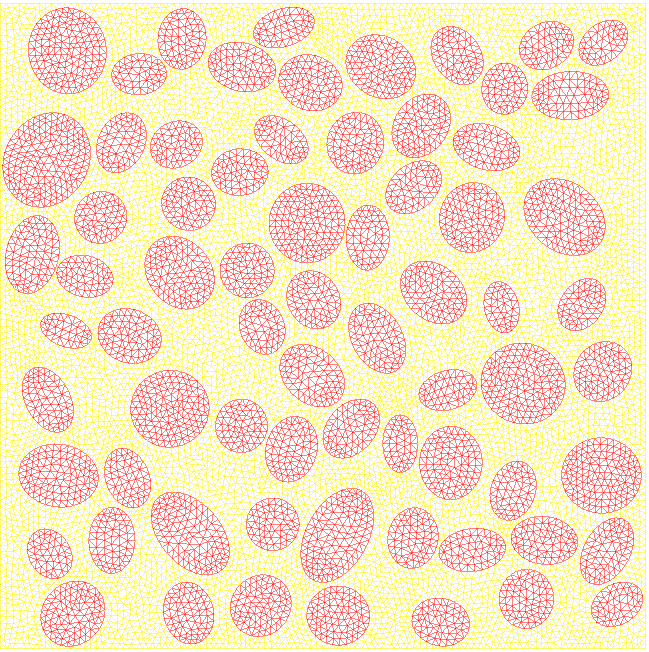}
			(a)
		\end{minipage}%
	}%
	\subfigure{
		\begin{minipage}[t]{0.25\linewidth}
			\centering
			\includegraphics[width=30mm]{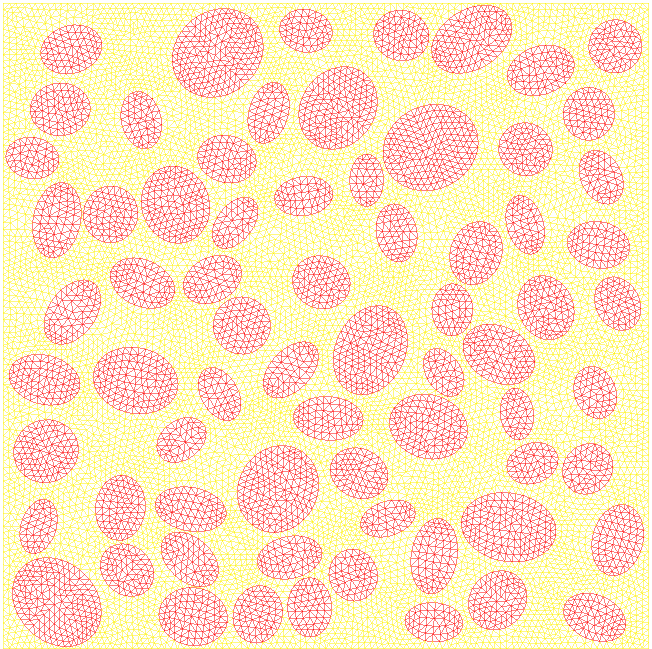}
			(b)
		\end{minipage}%
	}%
	\subfigure{
		\begin{minipage}[t]{0.25\linewidth}
			\centering
			\includegraphics[width=31mm]{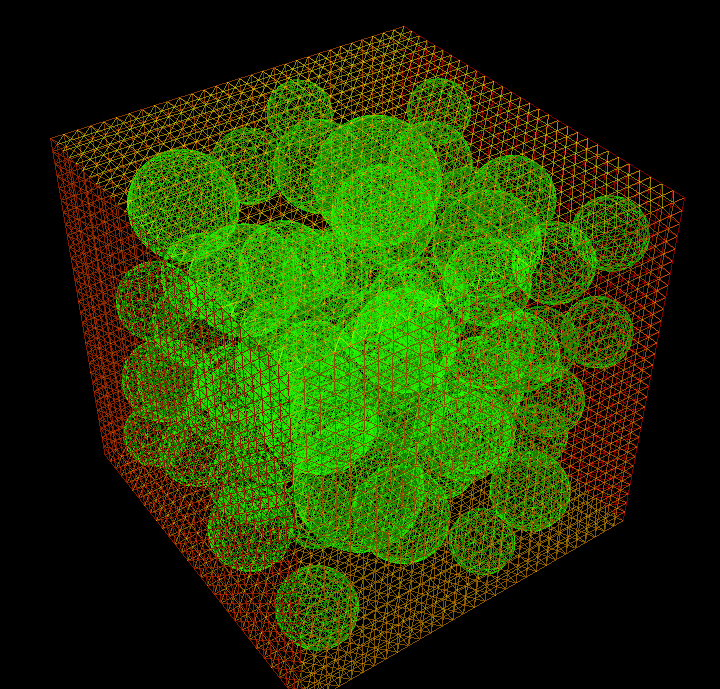}
			(c)
		\end{minipage}
	}%
	\subfigure{
		\begin{minipage}[t]{0.25\linewidth}
			\centering
			\includegraphics[width=31mm]{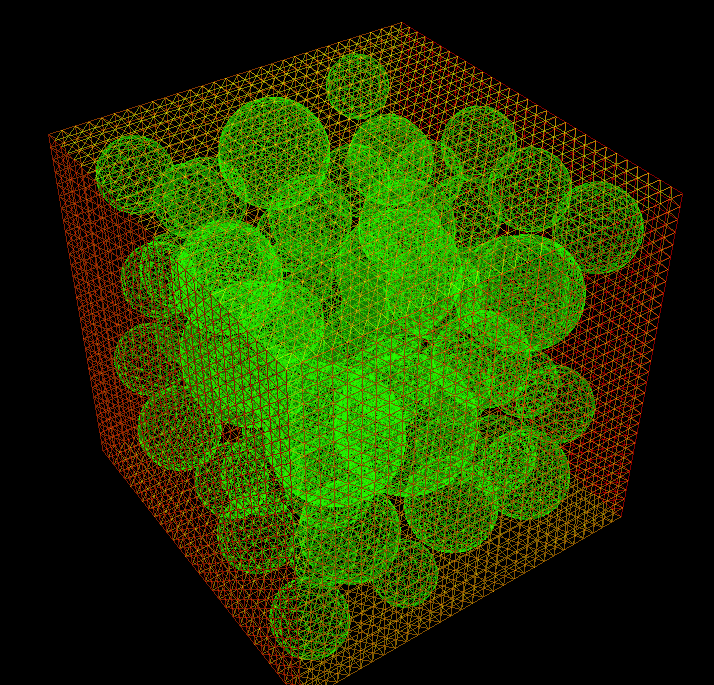}
			(d)
		\end{minipage}
	}%
	\centering
	\caption{Several microscopic RVEs of 2D and 3D ZrO$_{2}$/Ti-6Al-4V.}
	\label{fig:5}
\end{figure}

The composite material ZrO$_{2}$/Ti-6Al-4V is a typical two-scale heterogeneous material with particle-reinforced phase ZrO$_{2}$ and metal matrix Ti-6Al-4V. Some microscopic RVEs of these composites in 2D and 3D cases are exhibited in Fig.\hspace{1mm}\ref{fig:5}. The nonlinear thermal conductivities with temperature-dependent properties for components ZrO$_{2}$ and Ti-6Al-4V are given as ${\kappa_1} = 2.072 - 3.656 \times {10^{ - 4}}\theta  + 4.347 \times {10^{ - 7}}{\theta ^2}$ and ${\kappa_2} = 1.1 + 0.017\theta $, respectively \cite{R50}.

\begin{figure}[htbp]
	\centering
	\subfigure[]{
		\begin{minipage}[t]{0.5\linewidth}
			\centering
			\includegraphics[width=60mm]{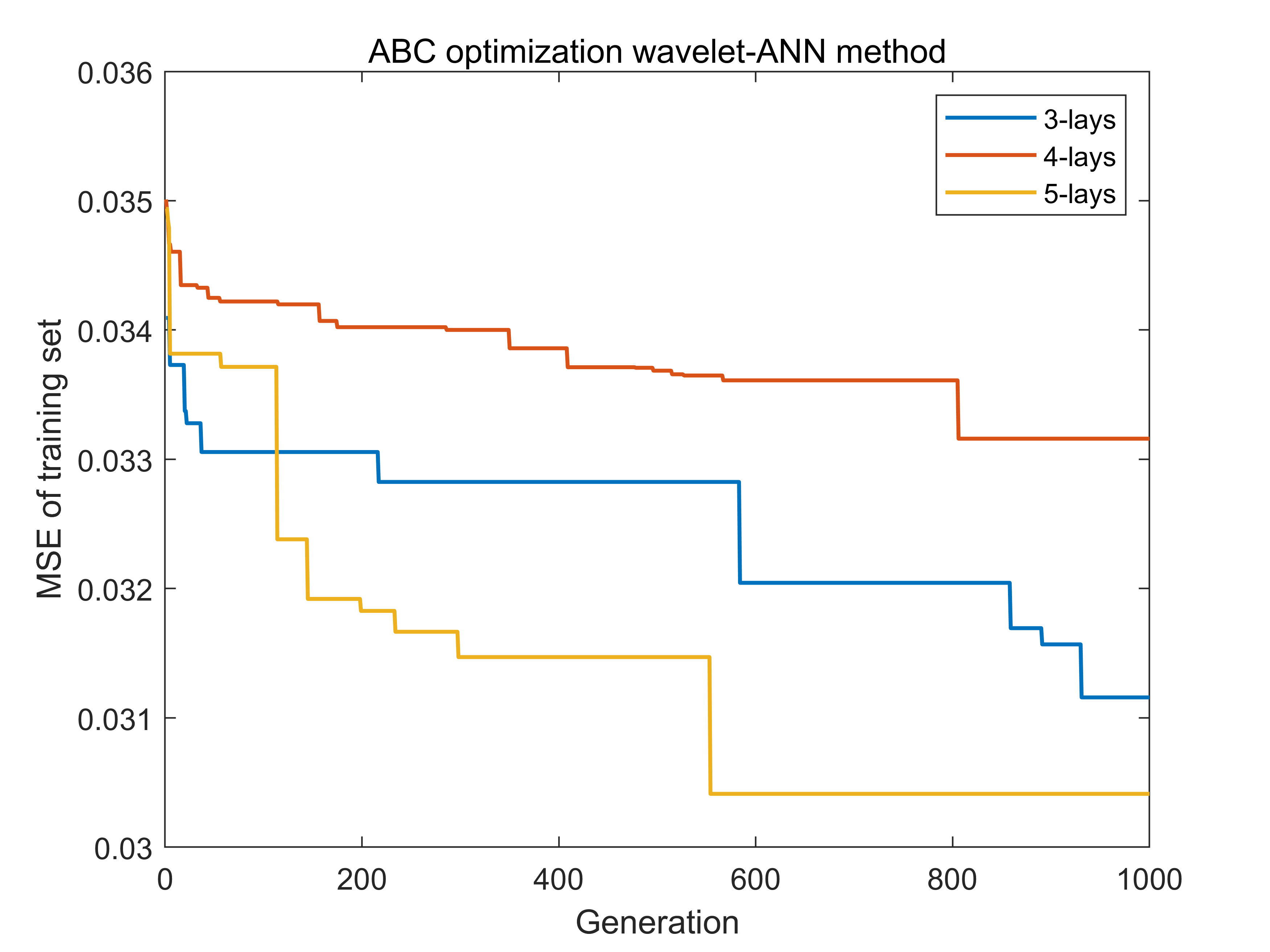}
		\end{minipage}%
	}%
	\subfigure[]{
		\begin{minipage}[t]{0.5\linewidth}
			\centering
			\includegraphics[width=60mm]{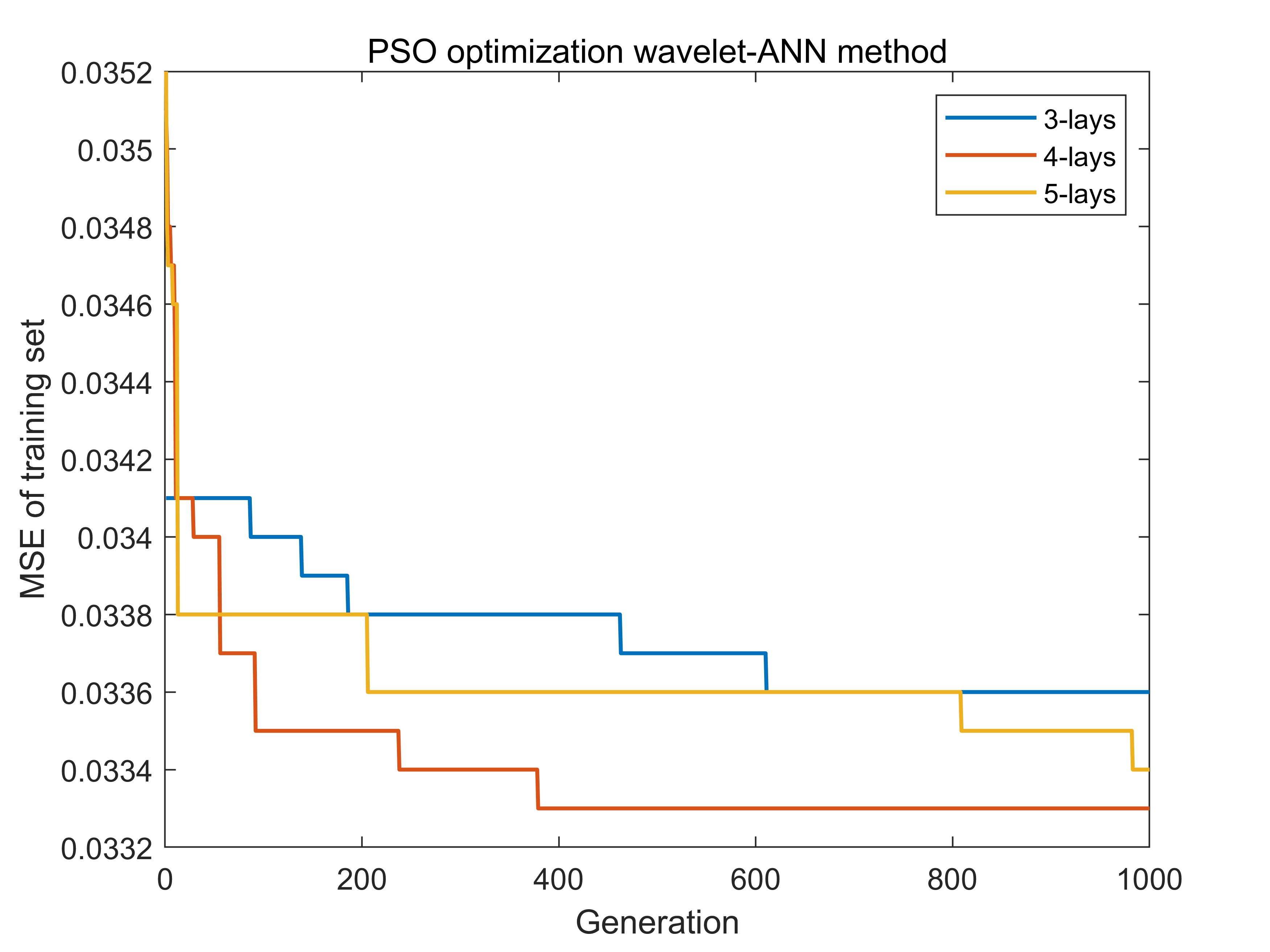}
		\end{minipage}%
	}
	\centering
	\caption{Decreasing curves of loss function of wavelet-learning predictive models: (a) ABC optimization; (b) PSO optimization.}
	\label{fig:6}
\end{figure}

In this example, the temperature changes of component materials are considered in the range of [500,1000)$K$. To ensure the diversity of material samples, 50 samples are randomly generated per 5$K$ for a total of 5000 samples. From these 5000 samples, 80\% are randomly selected as training set and the remaining as test set. For 2D composites, material features are decomposed by using three-layer wavelet decomposition, and the scale of material features is reduced from 10000 to 1250. The $cA_3$ and temperature $\theta$ are used as inputs, $\kappa_{11}^*$ and $\kappa_{22}^*$ were calculated by the stochastic three-scale homogenized method \cite{R26,R53} as outputs to construct the self-optimization wavelet-learning predictive models. In addition, self-optimization mechanism is introduced to search a optimal predictive model to accurately predict the macroscopic nonlinear thermal conductivity of heterogeneous particulate material. Next, we employed ABC and PSO algorithms to optimize the number of neurons in each hidden layer, the learning rate, and the depth of hidden layer of the wavelet-learning predictive models. The number of neurons in each hidden layer is searched in the range of 1-500, the learning rate is searched in the range of $\left[ {1 \times {{10}^{ - 6}},5 \times {{10}^{ - 4}}} \right]$, and the depth of hidden layer is searched in the range of 3-5 layers. Furthermore, the wavelet-learning approach is maximum iterated 500 epochs. Then using ABC and PSO algorithms to dynamically optimize wavelet-ANN model, the descent results of loss function are exhibited in Fig.\hspace{1mm}\ref{fig:6}.

\begin{figure}[htbp]
	\centering
	\subfigure[]{
		\begin{minipage}[t]{0.5\linewidth}
			\centering
			\includegraphics[width=60mm]{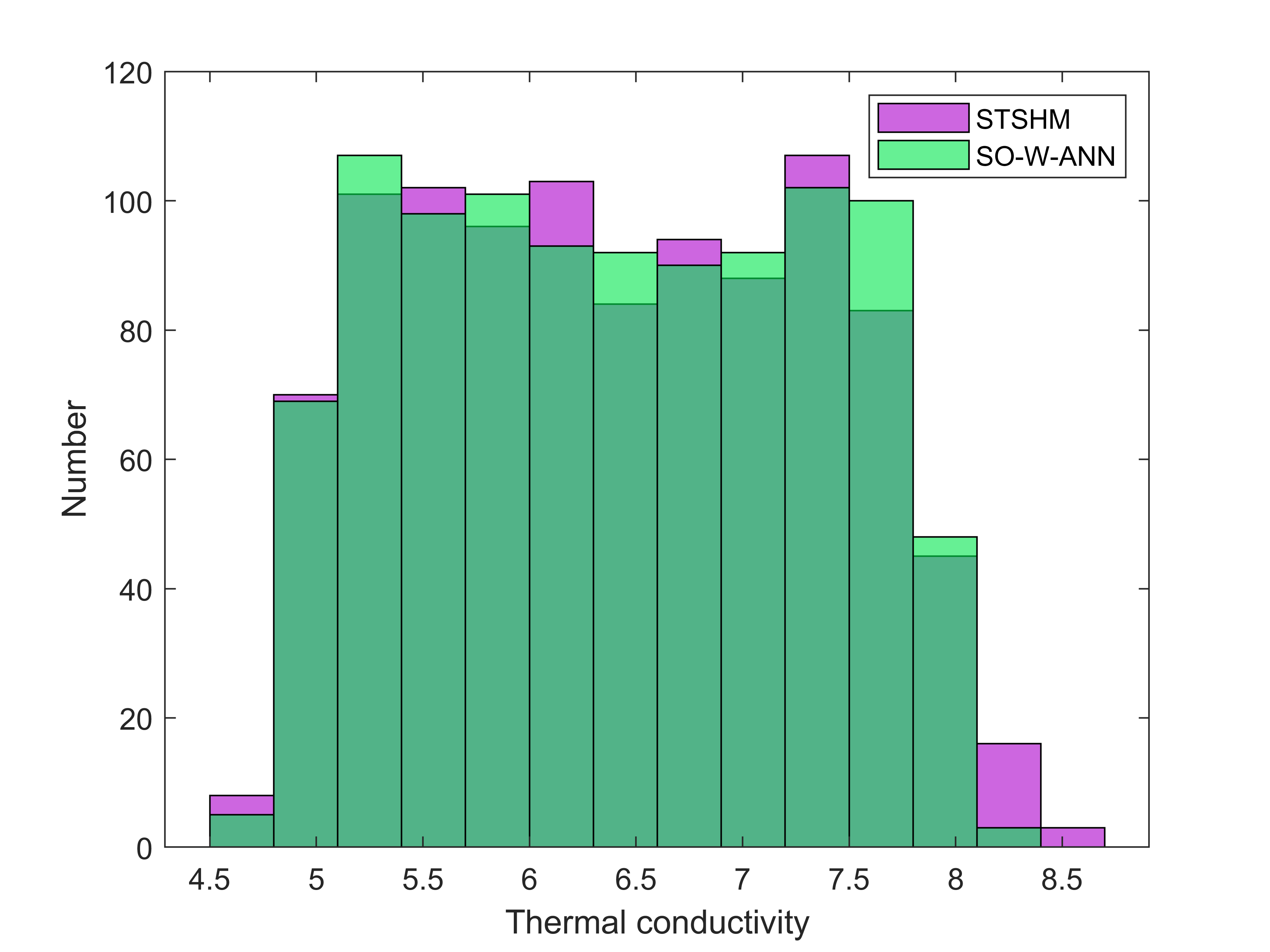}
		\end{minipage}%
	}%
	\subfigure[]{
		\begin{minipage}[t]{0.5\linewidth}
			\centering
			\includegraphics[width=60mm]{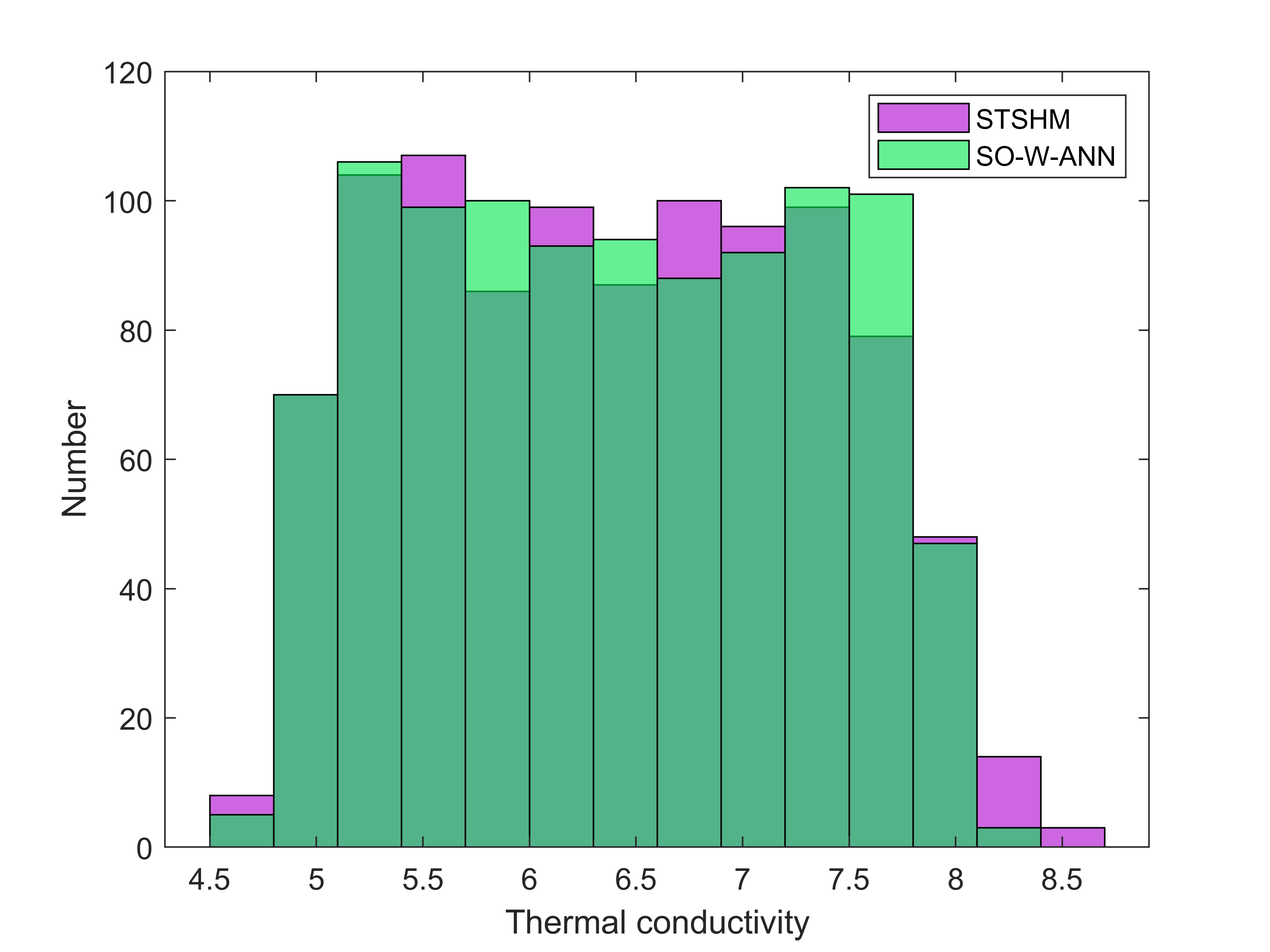}
		\end{minipage}%
	}
	\centering
	\caption{The predictive results by STSHM and self-optimization wavelet-learning
method: (a) $\kappa_{11}^*$; (b) $\kappa_{22}^*$.}
	\label{fig:7}
\end{figure}

After dynamic self-optimization, the optimal wavelet-ANN model is determined as neural structure 1251-176-6-324-145-150-2 with learning rate 0.00049. For the presented self-optimization wavelet-ANN model (SO-W-ANN), the training time is 16.53s, and the test time is 0.00037s. The training error of $\kappa_{11}^*$ is 2.11\% and the test error is 2.36\%, and the training error of $\kappa_{22}^*$ is 2.12\% and the test error is 2.40\%. The overall training error is 2.11\% and the test error is 2.38\%. Afterwards, Fig.\hspace{1mm}\ref{fig:7} shows the histogram of the predicted material parameters $\kappa_{11}^*$ and $\kappa_{22}^*$ compared to the predicted values calculated by the STSHM.

In addition, five samples are randomly generated to conduct the comparison analysis of the self-optimization wavelet-learning method and the traditional theoretical methods (Hashin-Shtrikman method and Lees method \cite{R55}). Table \ref{tab:3} provides the detailed comparison of predictive results.
\begin{table}[]
	\footnotesize
	\centering
	\caption{The comparison of self-optimization wavelet-learning method and traditional theoretical methods.}
	\begin{tabular}{|c|c|c|cl|c|}
		\hline
		Sample   & $\kappa_{11}^*$/$\kappa_{22}^*$ by STSHM & $\kappa_{11}^*$/$\kappa_{22}^*$ by SO-W-ANN & \multicolumn{2}{c|}{\begin{tabular}[c]{@{}c@{}}Upper/Lower Hashin-\\ Shtrikman bounds\end{tabular}} & \begin{tabular}[c]{@{}c@{}}Lees \\ method\end{tabular} \\ \hline
		Sample 1 & 7.050/6.962 & 7.137/7.136 & \multicolumn{2}{c|}{7.955/5.207}                 & 5.708          \\ \hline
		Sample 2 & 5.825/5.893 & 5.684/5.683 & \multicolumn{2}{c|}{6.511/4.752}                 & 5.038          \\ \hline
		Sample 3 & 7.696/7.554 & 7.956/7.955 & \multicolumn{2}{c|}{8.684/5.627}                 & 6.172          \\ \hline
		Sample 4 & 5.772/5.833 & 5.674/5.672 & \multicolumn{2}{c|}{6.447/4.755}                 & 5.016          \\ \hline
		Sample 5 & 4.684/4.770 & 4.866/4.865 & \multicolumn{2}{c|}{5.158/4.170}                 & 4.252          \\ \hline
	\end{tabular}
\label{tab:3}
\end{table}

\begin{figure}[htbp]
	\centering
	\subfigure[]{
		\begin{minipage}[t]{0.5\linewidth}
			\centering
			\includegraphics[width=60mm]{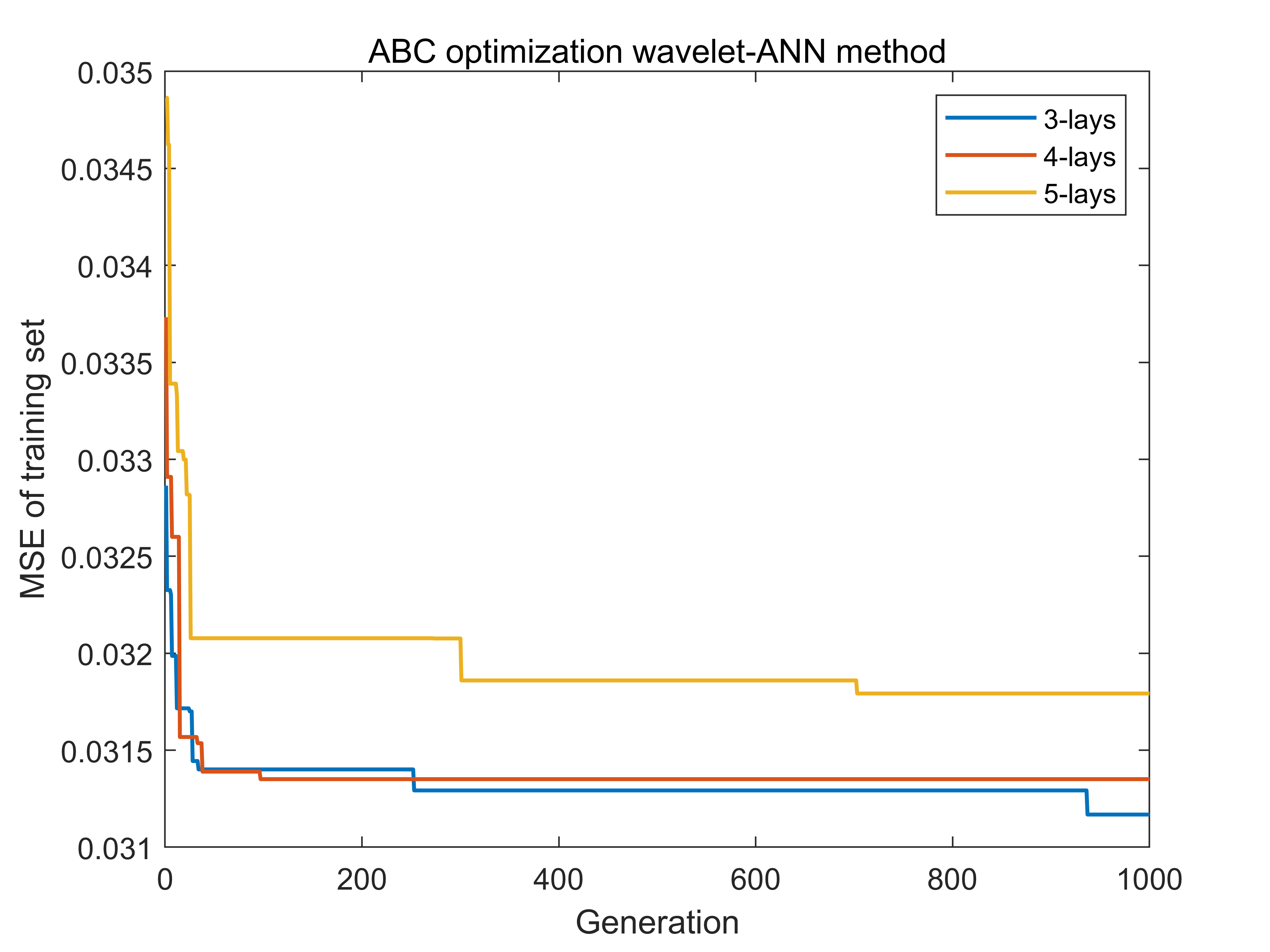}
		\end{minipage}%
	}%
	\subfigure[]{
		\begin{minipage}[t]{0.5\linewidth}
			\centering
			\includegraphics[width=60mm]{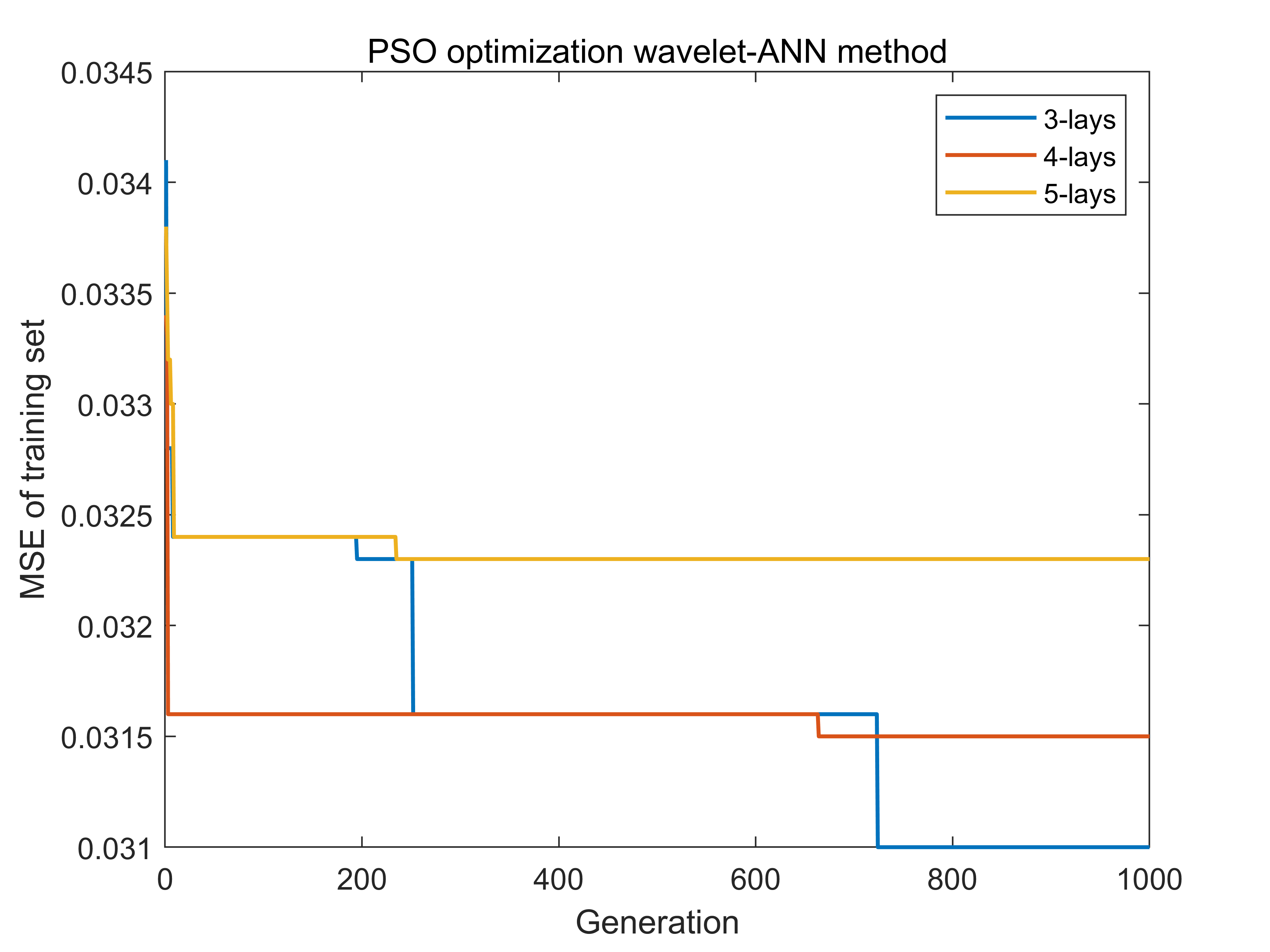}
		\end{minipage}%
	}
	\centering
	\caption{Decreasing curves of loss function of wavelet-learning predictive models: (a) ABC optimization; (b) PSO optimization.}
	\label{fig:8}
\end{figure}

For 3D particulate composites, raw 27000 material features are preprocessed by a three-layer wavelet decomposition. Then the new 3375 features $cA_3$ and temperature feature $\theta$ are adopted as neurons of input layer, $\kappa_{11}^*$, $\kappa_{22}^*$ and $\kappa_{33}^*$ were calculated by the stochastic three-scale homogenized method as outputs to establish the self-optimization wavelet-learning predictive models. In purpose of establishing a optimal predictive
model, ABC and PSO algorithms is introduced to self-optimize the adjustable parameters, which is the same as the 2D example. Simultaneously, the wavelet-learning method is maximum iterated 300 epochs. Then, the dynamic self-optimization processes of ABC and PSO algorithms are presented in Fig.\hspace{1mm}\ref{fig:8}.

\begin{figure}[htbp]
	\centering
	\subfigure{
		\begin{minipage}[t]{0.3\linewidth}
			\centering
			\includegraphics[width=45mm]{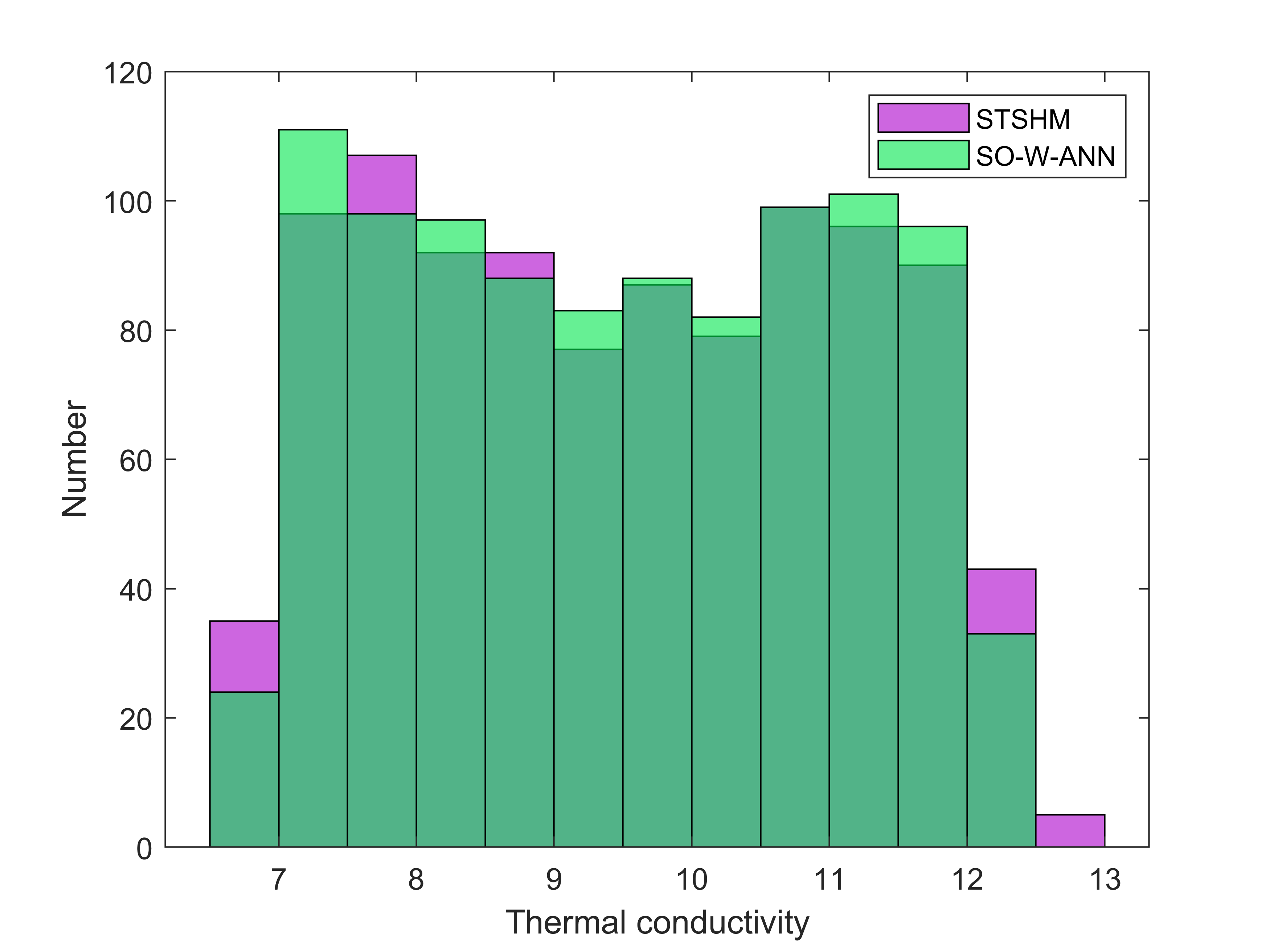}
			(a)
		\end{minipage}%
	}%
	\subfigure{
		\begin{minipage}[t]{0.3\linewidth}
			\centering
			\includegraphics[width=45mm]{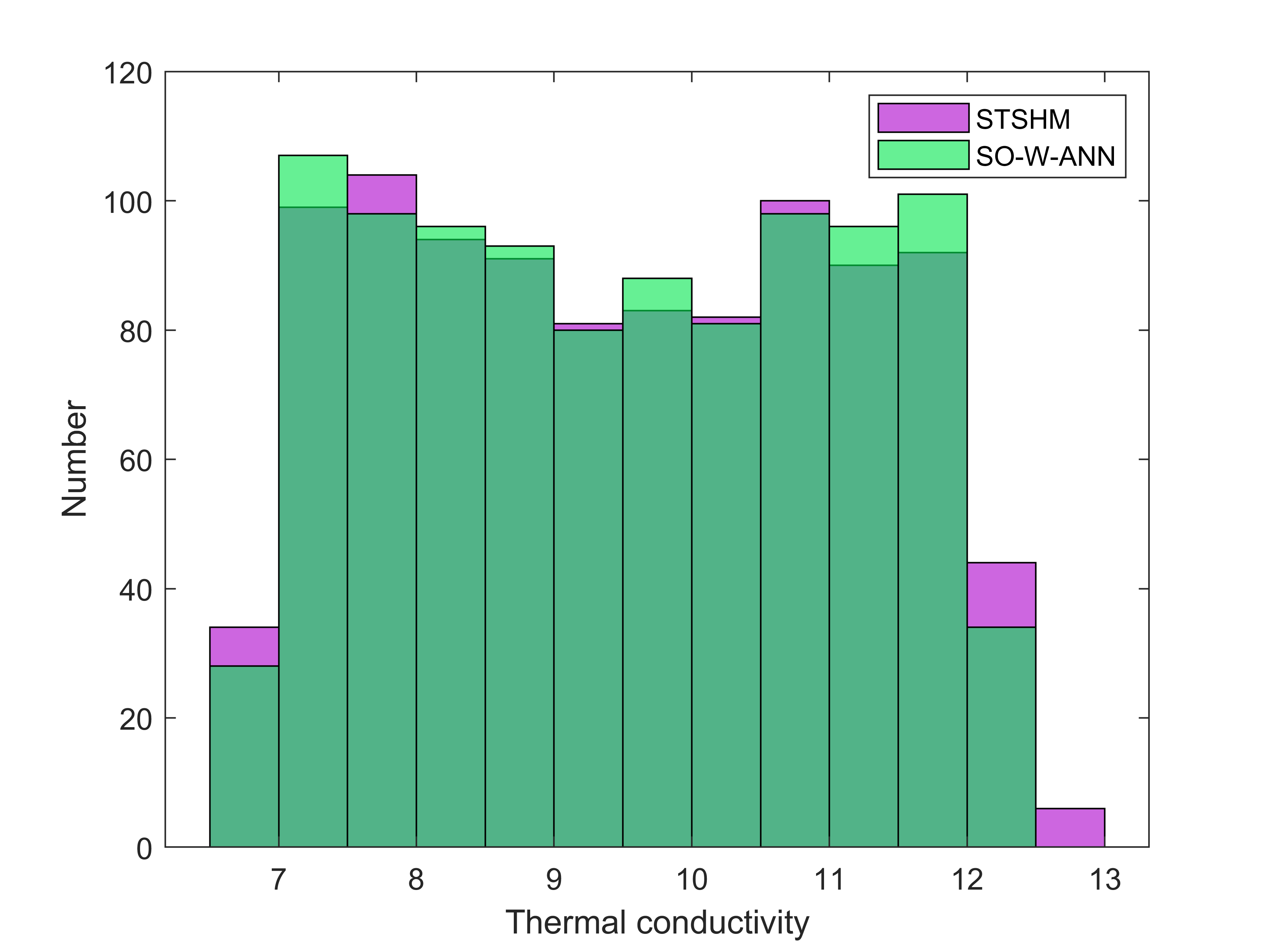}
			(b)
		\end{minipage}%
	}
	\subfigure{
		\begin{minipage}[t]{0.3\linewidth}
			\centering
			\includegraphics[width=45mm]{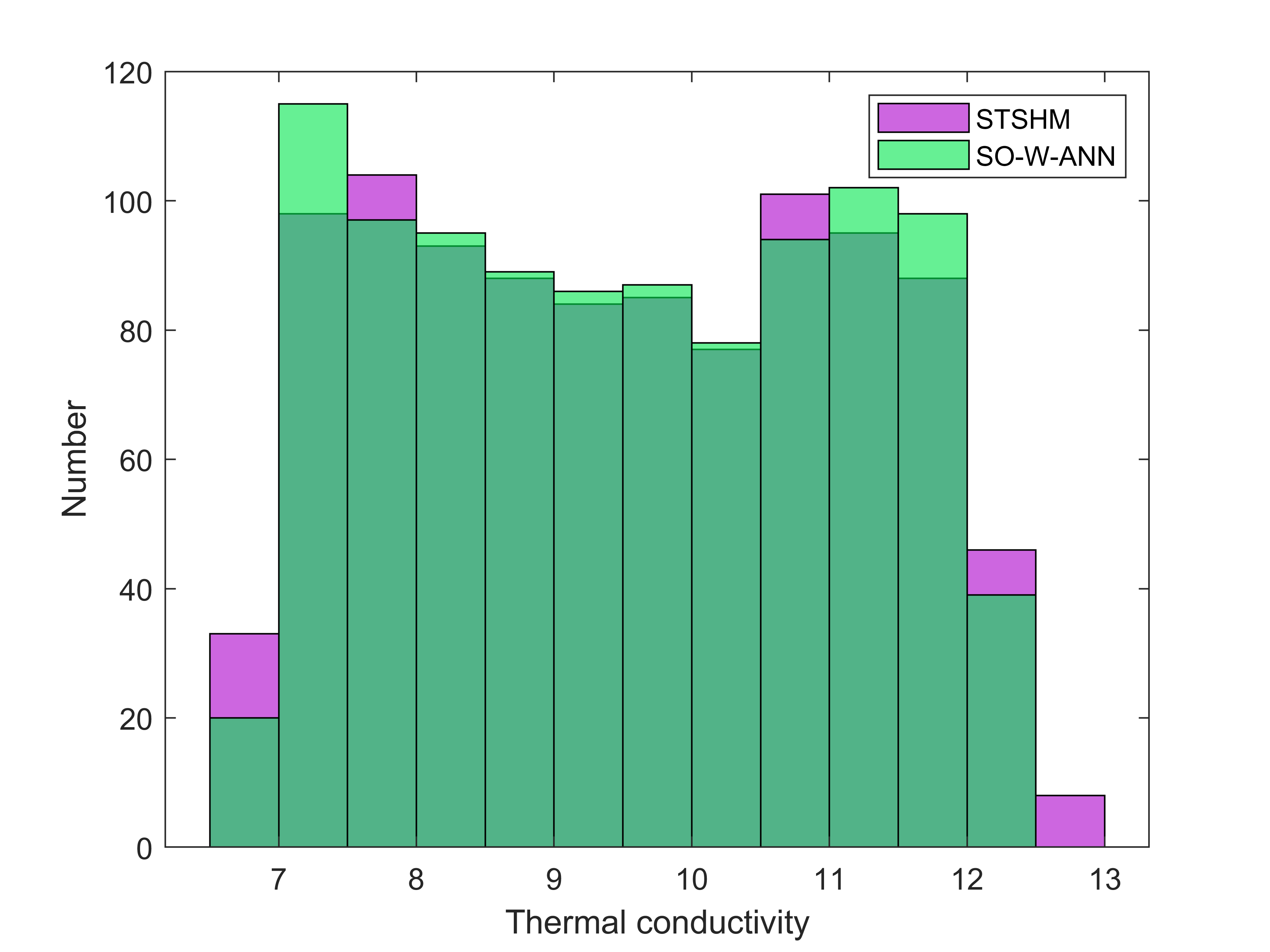}
			(c)
		\end{minipage}%
	}
	\centering
	\caption{The predictive results by STSHM and self-optimization wavelet-learning method: (a) $\kappa_{11}^*$; (b) $\kappa_{22}^*$; (c) $\kappa_{33}^*$.}
	\label{fig:9}
\end{figure}

After dynamic self-optimization, the optimized wavelet-ANN model is determined with neural architecture is 3376-104-72-163-3 with a learning rate of 0.000298. The training time of predictive model is 13.39s, and the test time is 0.00028s. The training errors of $\kappa_{11}^*$, $\kappa_{22}^*$ and $\kappa_{33}^*$ are 1.417\%, 1.416\% and 1.429\% respectively, and the test errors are 1.515\%, 1.506\% and 1.537\% respectively. The overall training error is 1.421\% and the test error is 1.519\%. Next, Fig.\hspace{1mm}\ref{fig:9} shows the histogram of the predicted material parameters $\kappa_{11}^*$, $\kappa_{22}^*$ and $\kappa_{33}^*$ compared to the predicted values calculated by the STSHM.

Furthermore, five samples are randomly generated to compare the prediction results of the self-optimization wavelet-learning method with the traditional theoretical methods (Hashin-Shtrikman method and Lees method). The detailed comparison of predictive results is displayed in Table \ref{tab:4}.
\begin{table}[]
	\footnotesize
	\centering
	\caption{The comparison of self-optimization wavelet-learning method and traditional theoretical methods.}
	\begin{tabular}{|c|c|c|cl|c|}
		\hline
		Sample   & $\kappa_{11}^*$/$\kappa_{22}^*$/$\kappa_{33}^*$ by STSHM & $\kappa_{11}^*$/$\kappa_{22}^*$/$\kappa_{33}^*$ by SO-W-ANN & \multicolumn{2}{c|}{\begin{tabular}[c]{@{}c@{}}Upper/Lower Hashin-\\ Shtrikman bounds\end{tabular}} & \begin{tabular}[c]{@{}c@{}}Lees \\ method\end{tabular} \\ \hline
		Sample 1 & 11.103/11.055/11.099 & 10.981/10.996/11.030 & \multicolumn{2}{c|}{11.579/8.447} & 9.444 \\ \hline
		Sample 2 & 8.296/8.262/8.310    & 8.469/8.489/8.503    & \multicolumn{2}{c|}{8.643/6.974}  & 7.378  \\ \hline
		Sample 3 & 9.677/9.684/9.671    & 9.821/9.808/9.833    & \multicolumn{2}{c|}{10.100/7.819} & 8.442  \\ \hline
		Sample 4 & 8.568/8.597/8.583    & 8.579/8.589/8.599    & \multicolumn{2}{c|}{8.953/7.080}  & 7.585  \\ \hline
		Sample 5 & 9.086/9.099/9.062    & 9.291/9.277/9.303    & \multicolumn{2}{c|}{9.482/7.264}  & 7.862  \\ \hline
	\end{tabular}
	\label{tab:4}
\end{table}

As can be seen in Tables \ref{tab:3} and \ref{tab:4}, the predictive results by the established SO-W-ANN are high-precision approximation to the computational results by STSHM. Additionally, the predictive results by the established SO-W-ANN fall between the upper and lower bounds of Hashin-Shtrikman method and are very close to the predictive values of Lees method. However, Hashin-Shtrikman method only gives an estimation interval for the lower and upper bounds of inhomogeneous material, which can not be directly utilized for engineering applications. Besides, Lees method can only obtain the isotropic prediction results and it is not applicative for anisotropic materials. From the perspective of computational efficiency, the presented SO-W-ANN can avert repetitive numerical computation by STSHM and can be utilized to efficient prediction for particulate composites with random configurations.
\subsection{Example 2. nonlinear fibrous composite C/SiC with two-scale random configurations}
\begin{figure}[htbp]
	\centering
	\subfigure{
		\begin{minipage}[t]{0.25\linewidth}
			\centering
			\includegraphics[width=30mm]{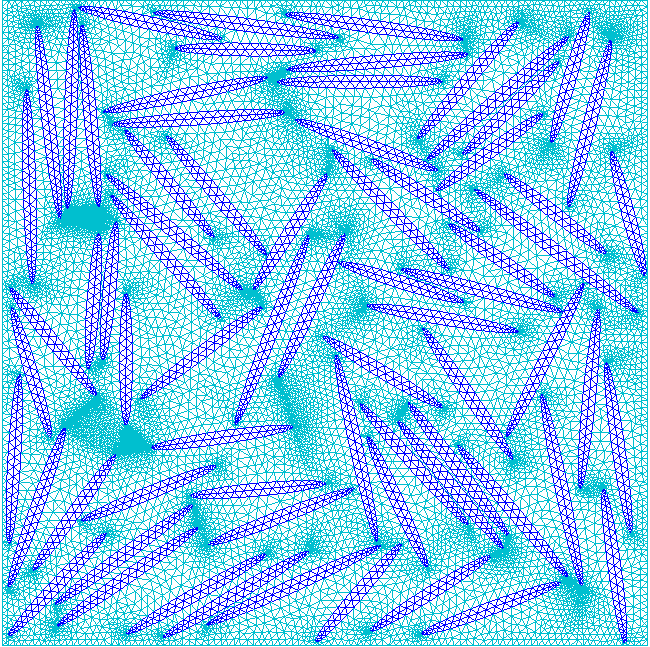}
			(a)
		\end{minipage}%
	}%
	\subfigure{
		\begin{minipage}[t]{0.25\linewidth}
			\centering
			\includegraphics[width=30mm]{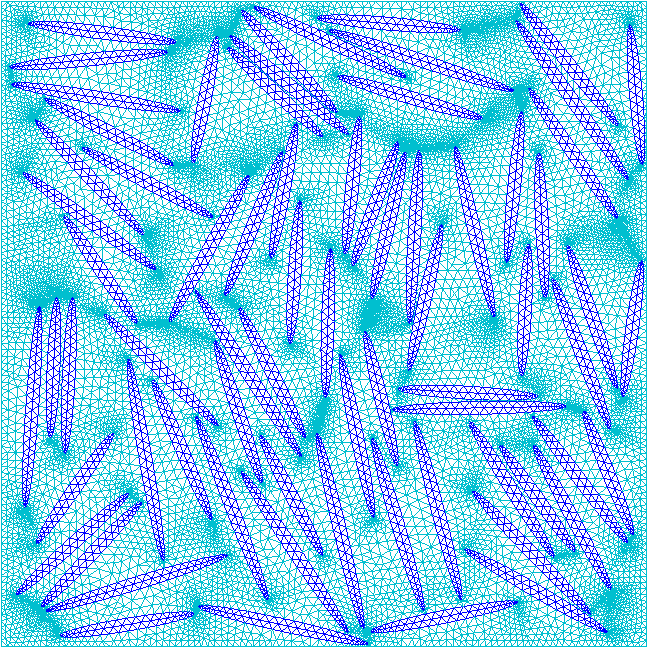}
			(b)
		\end{minipage}%
	}%
	\subfigure{
		\begin{minipage}[t]{0.25\linewidth}
			\centering
			\includegraphics[width=31mm]{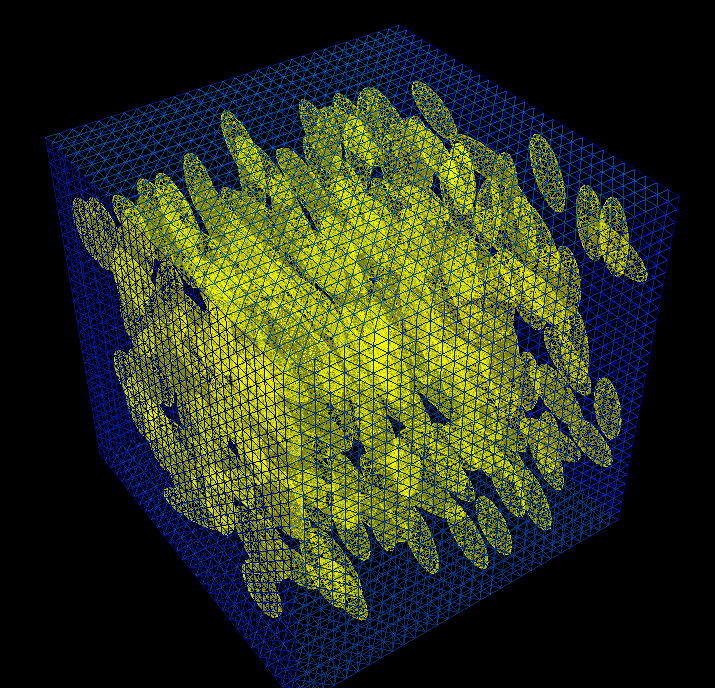}
			(c)
		\end{minipage}
	}%
	\subfigure{
		\begin{minipage}[t]{0.25\linewidth}
			\centering
			\includegraphics[width=31mm]{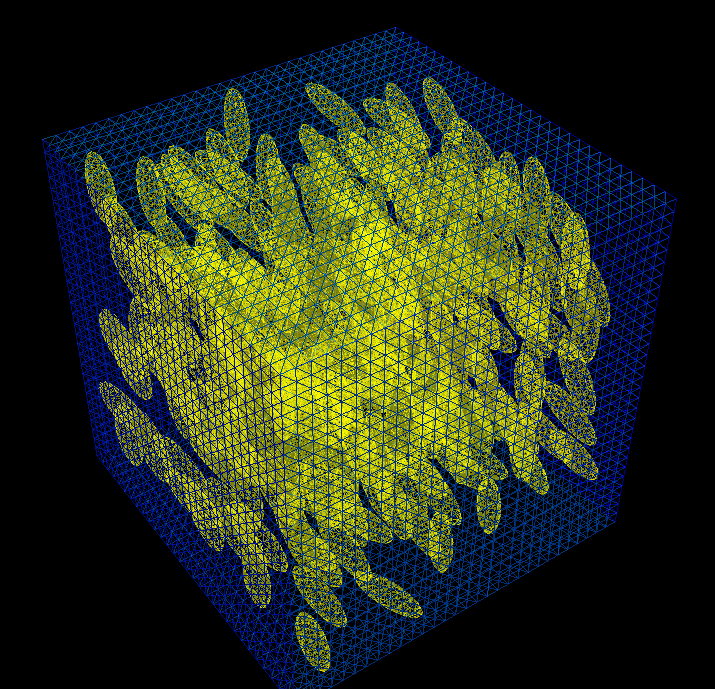}
			(d)
		\end{minipage}
	}%
	\centering
	\caption{Several microscopic RVEs of 2D and 3D C/SiC.}
	\label{fig:10}
\end{figure}

The composite material C/SiC is a typical high-contrast two-scale heterogeneous material with fiber-reinforced phase C and ceramic matrix SiC. Some microscopic RVEs of these composites in 2D and 3D cases are exhibited in Fig.\hspace{1mm}\ref{fig:10}. The nonlinear thermal conductivities with temperature-dependent properties for components C and SiC are ${\kappa_1} = 8 + 0.02535\theta$ and ${\kappa_2} = 250 + 0.02728\theta$, respectively \cite{R51}.

\begin{figure}[htbp]
	\centering
	\subfigure[]{
		\begin{minipage}[t]{0.5\linewidth}
			\centering
			\includegraphics[width=60mm]{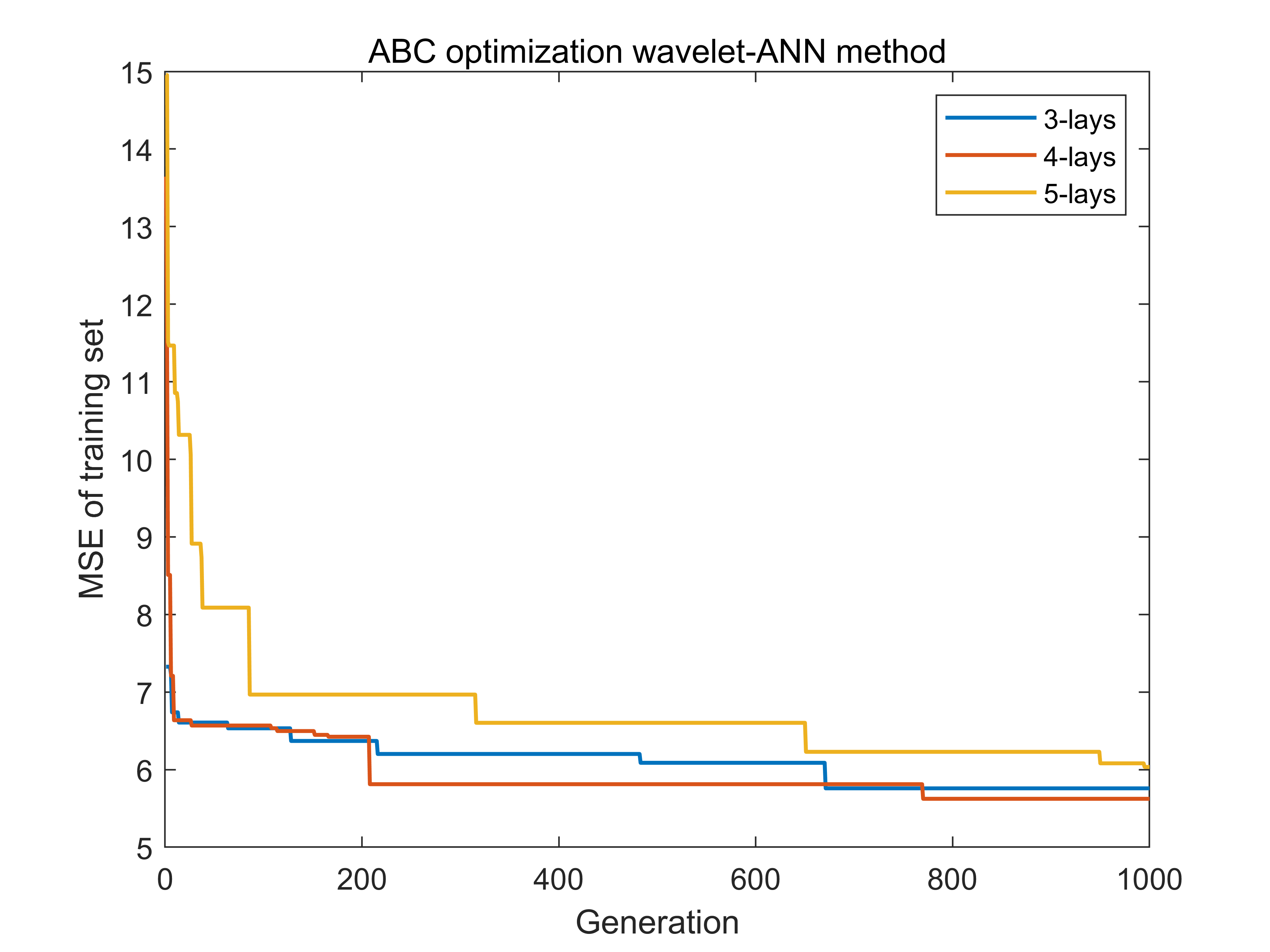}
		\end{minipage}%
	}%
	\subfigure[]{
		\begin{minipage}[t]{0.5\linewidth}
			\centering
			\includegraphics[width=60mm]{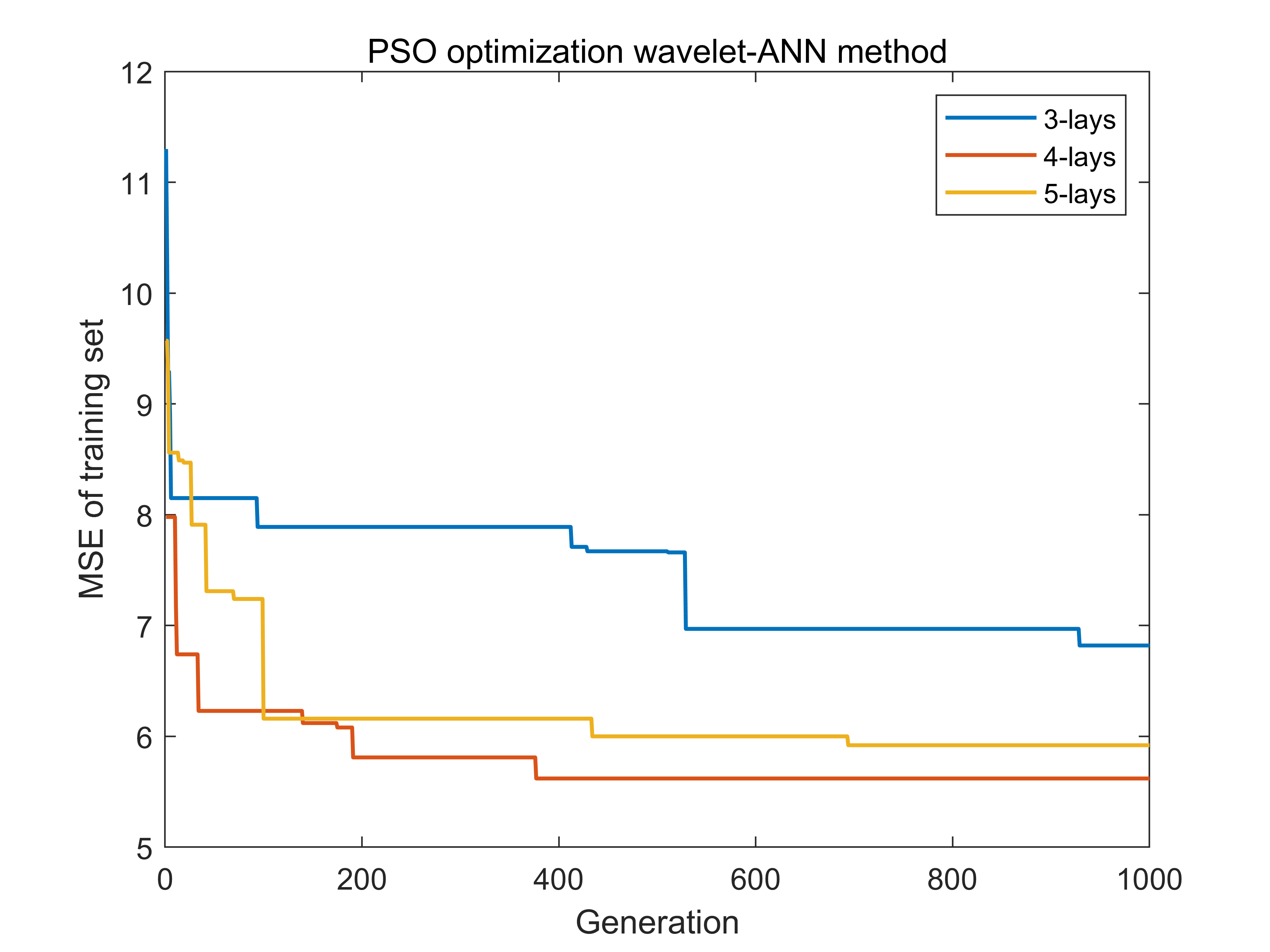}
		\end{minipage}%
	}
	\centering
	\caption{Decreasing curves of loss function of wavelet-learning predictive models: (a) ABC optimization; (b) PSO optimization.}
	\label{fig:11}
\end{figure}

In this example, the changeable range of service temperature of constituent materials is defined in the interval [500,1000)$K$. In order to ensure the stochastic diversity of the material samples, 50 samples are randomly generated per 5$K$, and a total of 5000 samples are generated in this way. 4000 samples among 5000 generated samples are randomly selected as training set and the rest 1000 samples are set aside for test set. For 2D fibrous composites, the material features in the raw database are preprocessed using three-level wavelet decomposition. The data scale of material features is compressed from 10000 to 1250. The 1250 wavelet coefficients $cA_3$ and temperature feature $\theta$ are used as inputs, material parameters $\kappa_{11}^*$ and $\kappa_{22}^*$ were calculated by the STSHM as outputs to build the self-optimization wavelet-learning predictive models. Then, ABC and PSO algorithms are employed to optimize the number of neurons in each hidden layer, the learning rate, and the depth of hidden layer of the wavelet-learning predictive models. The search scope for adjustable parameters of the wavelet-learning predictive models is the same as that of example 1 in Section 4.1. The wavelet-ANN is maximum iterated 500 epochs. Furthermore, we depict the descent results of loss function in Fig.\hspace{1mm}\ref{fig:11}.

\begin{figure}[htbp]
	\centering
	\subfigure[]{
		\begin{minipage}[t]{0.5\linewidth}
			\centering
			\includegraphics[width=60mm]{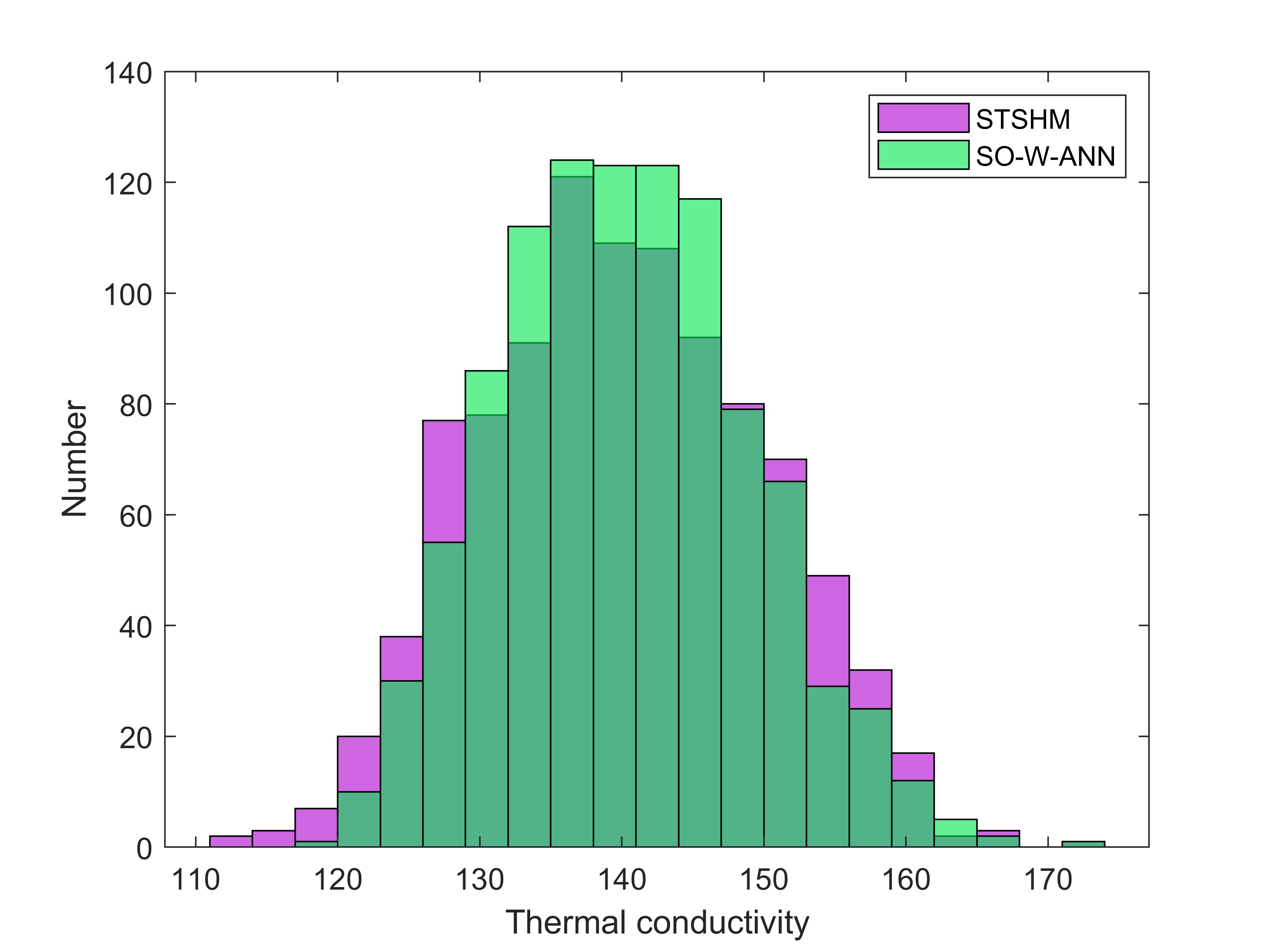}
		\end{minipage}%
	}%
	\subfigure[]{
		\begin{minipage}[t]{0.5\linewidth}
			\centering
			\includegraphics[width=60mm]{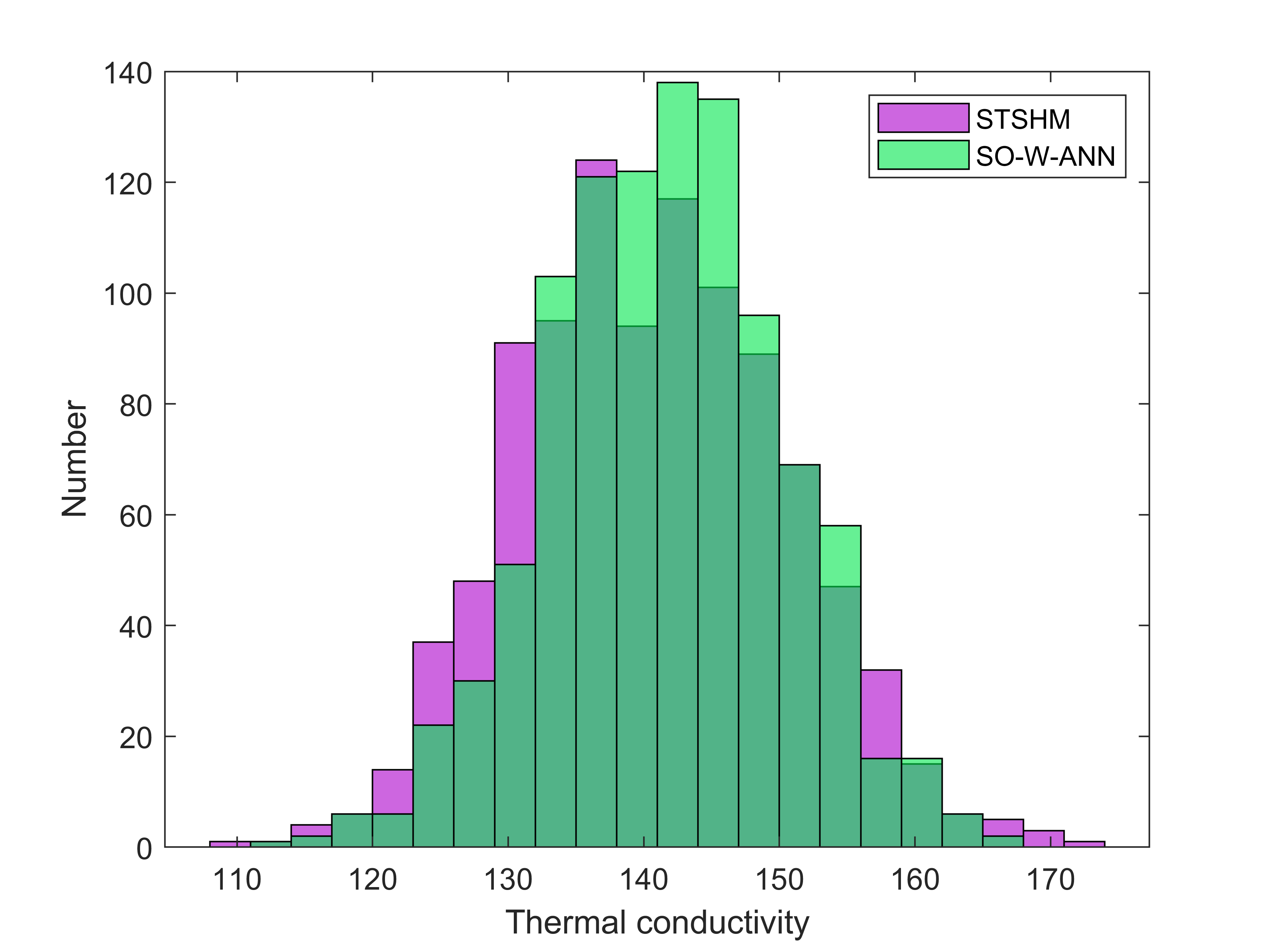}
		\end{minipage}%
	}
	\centering
	\caption{The predictive results by STSHM and self-optimization wavelet-learning method: (a) $\kappa_{11}^*$; (b) $\kappa_{22}^*$.}
	\label{fig:12}
\end{figure}

After dynamic self-optimization, the optimum wavelet-ANN model is obtained as neural structure 1251-386-334-292-203-2 with a learning rate of 0.000116. For this self-optimization wavelet-ANN model (SO-W-ANN), the training time is 33.13s, and the test time is 0.00034s. The training error of $\kappa_{11}^*$ is 1.01\% and the test error is 2.88\%, and the training error of $\kappa_{22}^*$ is 1.14\% and the test error is 2.86\%. The overall training error is 1.07\% and the test error is 2.87\%. Subsequently, the histogram of the predicted material parameters $\kappa_{11}^*$ and $\kappa_{22}^*$ compared to the predicted values computed by the STSHM is presented Fig.\hspace{1mm}\ref{fig:12}.

Besides, the predictive results of some detailed samples by the proposed self-optimization wavelet-learning method and two theoretical approaches (Hashin-Shtrikman method and Lees method) are exhibited in Table \ref{tab:5}.
\begin{table}[]
	\footnotesize
	\centering
	\caption{The comparison of self-optimization wavelet-learning method and traditional theoretical methods.}
	\begin{tabular}{|c|c|c|cl|c|}
		\hline
		Sample   & $\kappa_{11}^*$/$\kappa_{22}^*$ by STSHM & $\kappa_{11}^*$/$\kappa_{22}^*$ by SO-W-ANN & \multicolumn{2}{c|}{\begin{tabular}[c]{@{}c@{}}Upper/Lower Hashin-\\ Shtrikman bounds\end{tabular}} & \begin{tabular}[c]{@{}c@{}}Lees \\ method\end{tabular} \\ \hline
		Sample 1 & 119.183/141.746 & 126.472/141.487 & \multicolumn{2}{c|}{180.035/112.606} & 135.570 \\ \hline
		Sample 2 & 147.637/143.176 & 151.502/137.529 & \multicolumn{2}{c|}{188.693/131.539} & 149.427 \\ \hline
		Sample 3 & 135.315/137.763 & 132.422/136.678 & \multicolumn{2}{c|}{183.122/117.985} & 140.059 \\ \hline
		Sample 4 & 134.015/157.029 & 143.283/152.080 & \multicolumn{2}{c|}{189.464/131.461} & 150.066 \\ \hline
		Sample 5 & 137.083/140.916 & 135.657/140.608 & \multicolumn{2}{c|}{184.333/120.368} & 141.854 \\ \hline
	\end{tabular}
	\label{tab:5}
\end{table}

\begin{figure}[htbp]
	\centering
	\subfigure[]{
		\begin{minipage}[t]{0.5\linewidth}
			\centering
			\includegraphics[width=60mm]{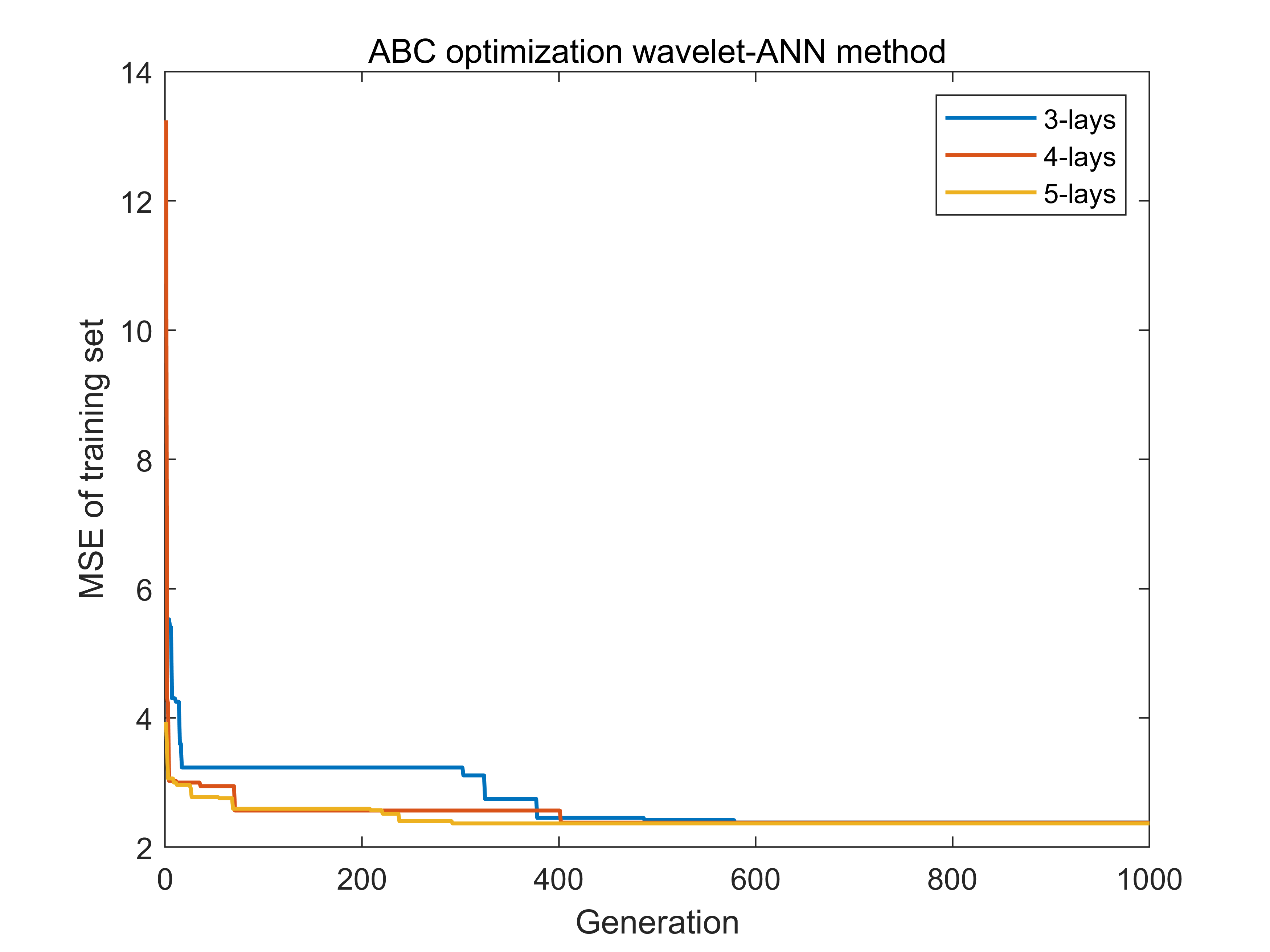}
		\end{minipage}%
	}%
	\subfigure[]{
		\begin{minipage}[t]{0.5\linewidth}
			\centering
			\includegraphics[width=60mm]{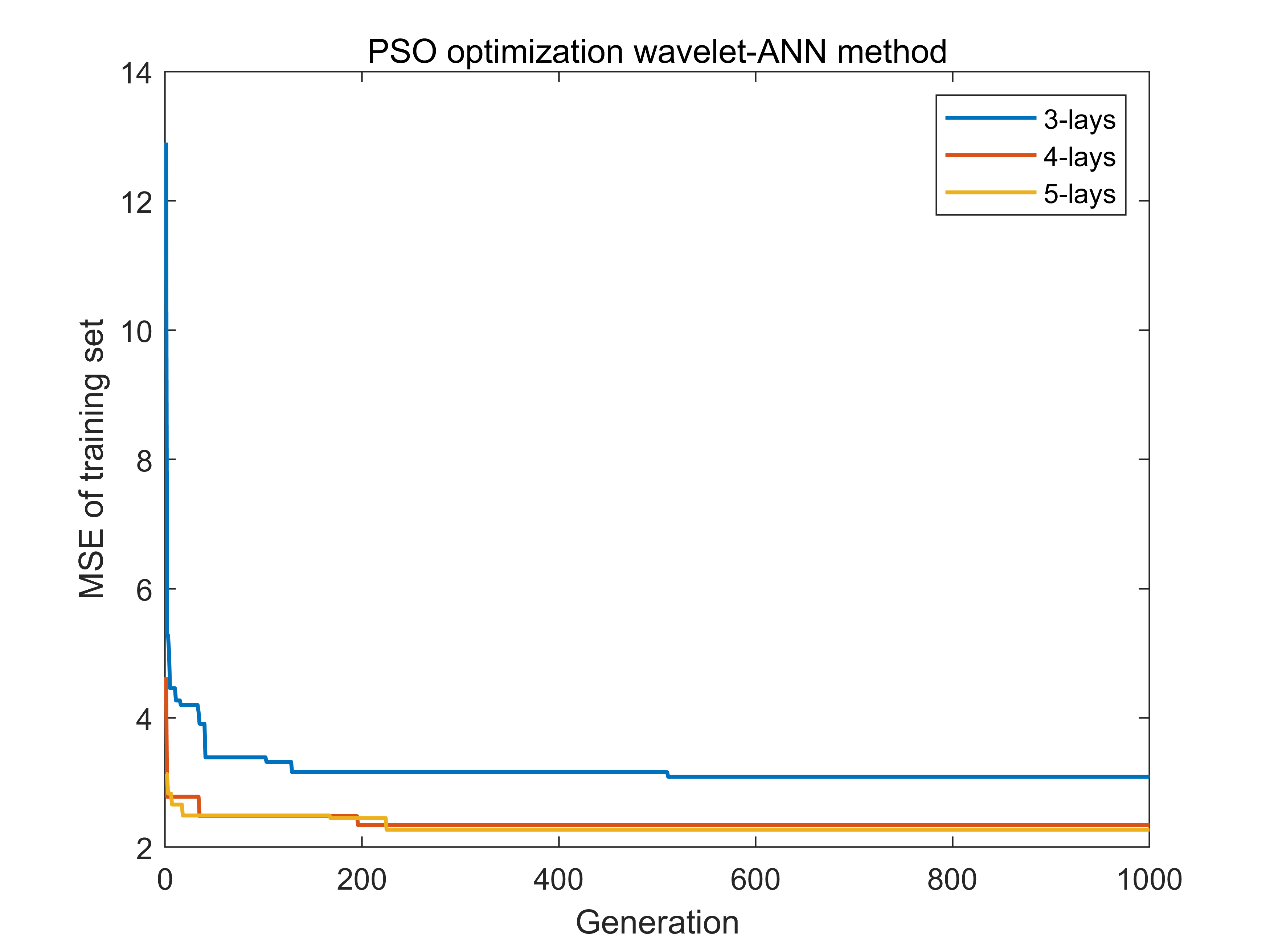}
		\end{minipage}%
	}
	\centering
	\caption{Decreasing curves of loss function of wavelet-learning predictive models: (a)
ABC optimization; (b) PSO optimization.}
	\label{fig:13}
\end{figure}

For 3D fibrous composites, 27000 original material features are decomposed into 3375 $cA_3$ by using three-layer wavelet decomposition. Then, the 3375 $cA_3$ coefficients and temperature feature $\theta$ are used as inputs, $\kappa_{11}^*$, $\kappa_{22}^*$ and $\kappa_{33}^*$ were calculated by STSHM as outputs to establish the self-optimization wavelet-learning predictive models. For the 3D case, the range of optimized parameters of the wavelet-learning predictive
models is the same as the 2D example. Furthermore, each wavelet-learning predictive model is iterated 300 times. Then the adjustable parameters of the predictive models are optimized by using ABC and PSO algorithms, the descent results of loss function are presented in Fig.\hspace{1mm}\ref{fig:13}.

\begin{figure}[htbp]
	\centering
	\subfigure{
		\begin{minipage}[t]{0.3\linewidth}
			\centering
			\includegraphics[width=45mm]{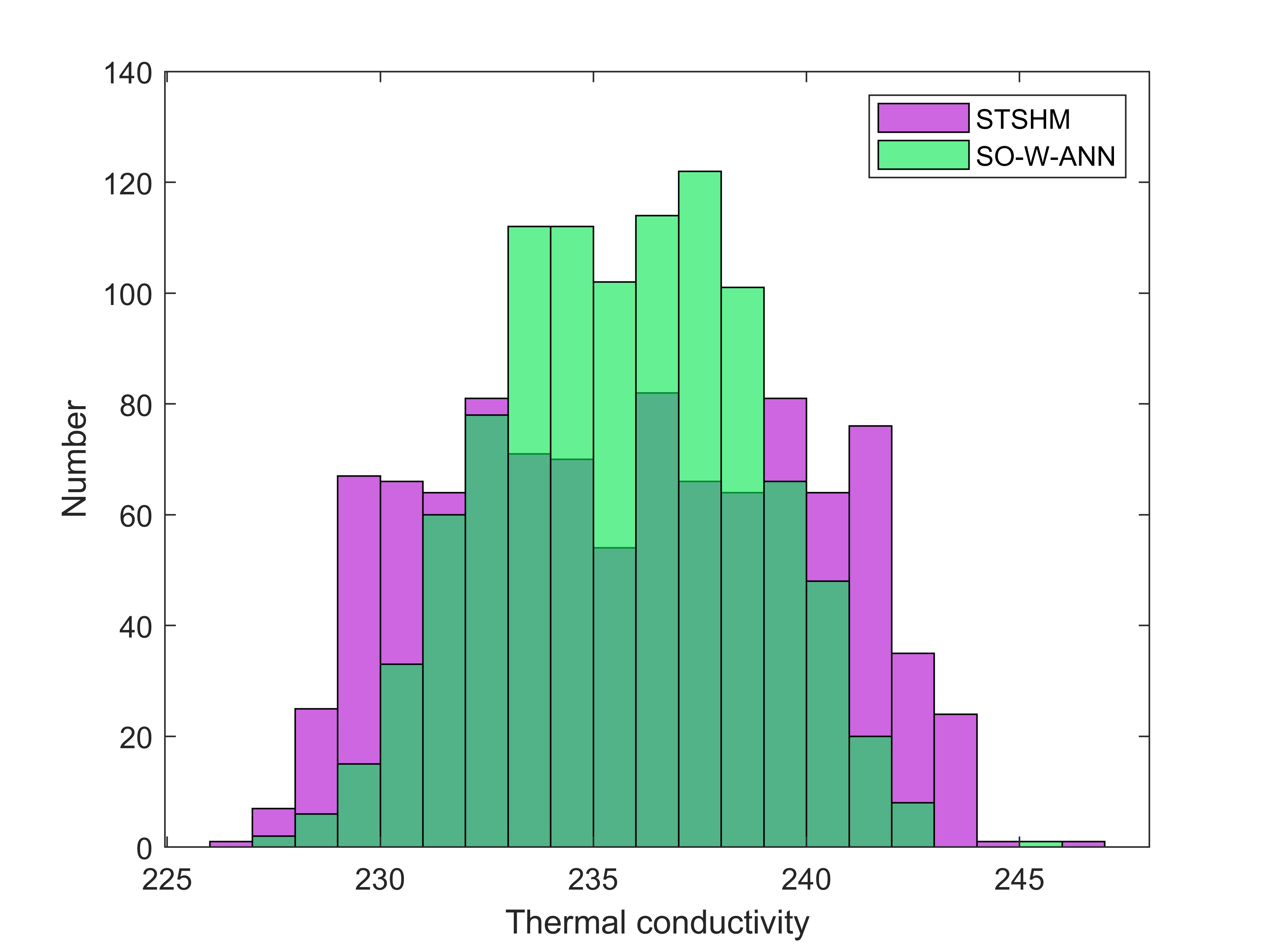}
			(a)
		\end{minipage}%
	}%
	\subfigure{
		\begin{minipage}[t]{0.3\linewidth}
			\centering
			\includegraphics[width=45mm]{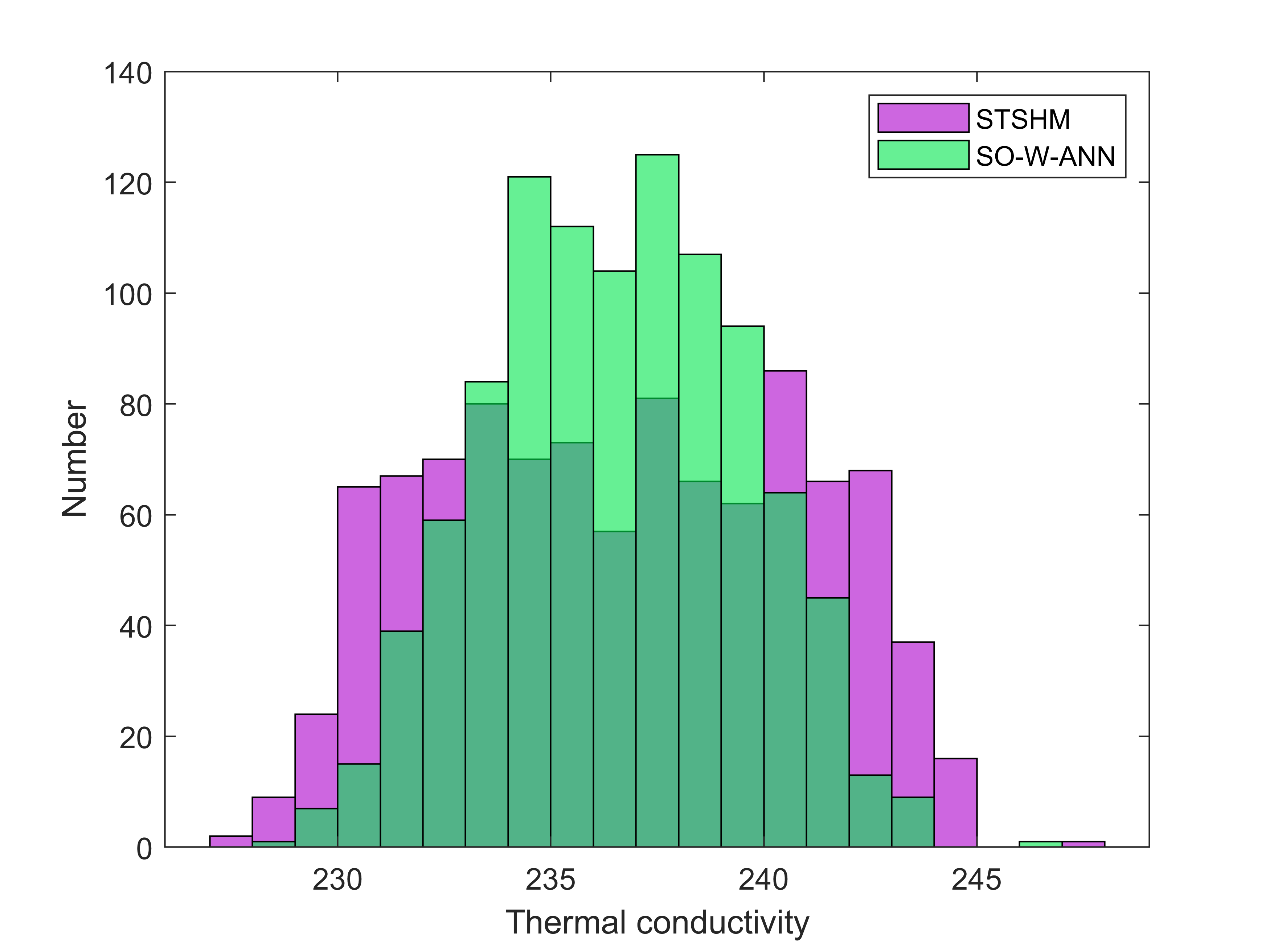}
			(b)
		\end{minipage}%
	}
	\subfigure{
		\begin{minipage}[t]{0.3\linewidth}
			\centering
			\includegraphics[width=45mm]{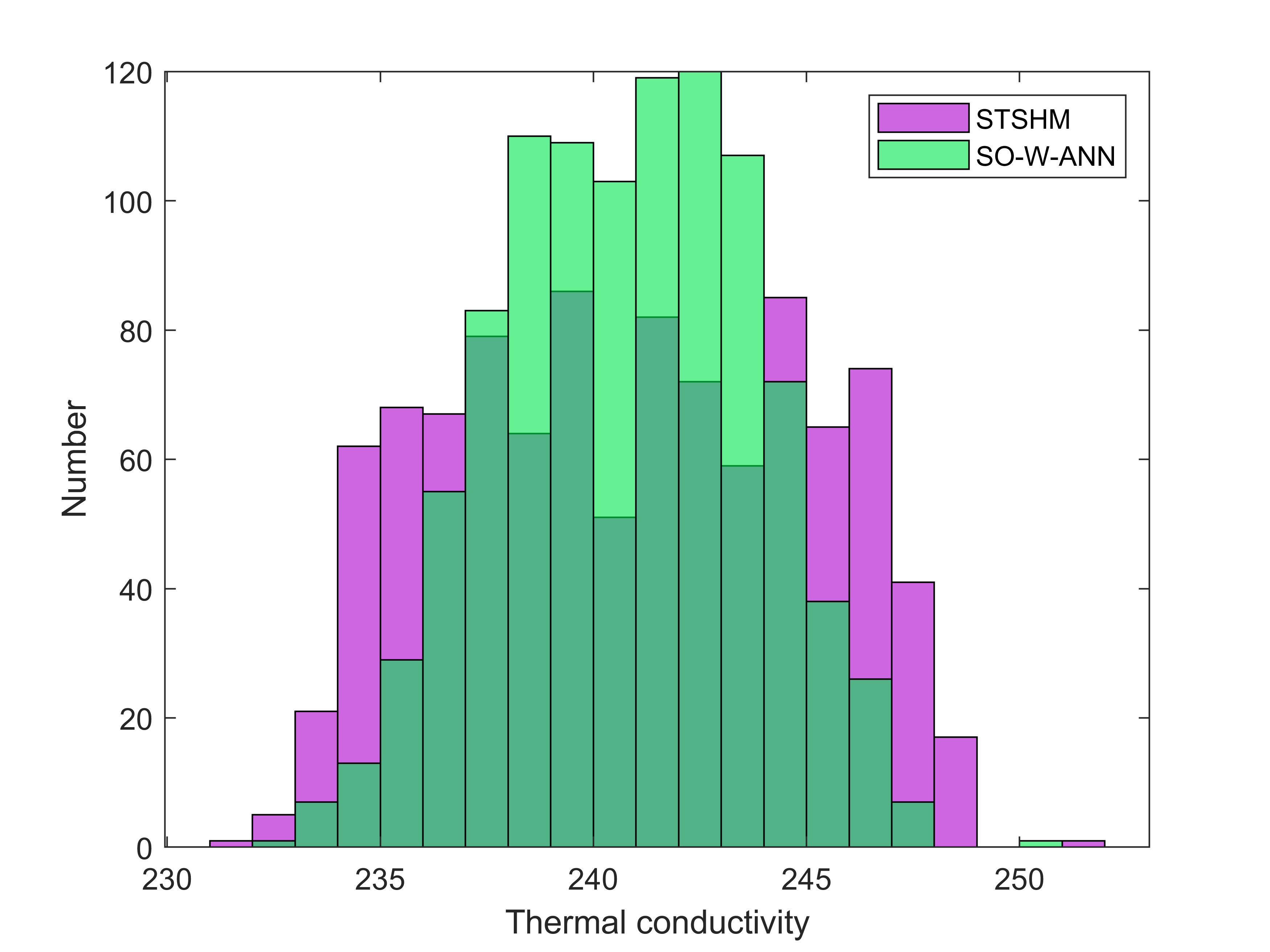}
			(c)
		\end{minipage}%
	}
	\centering
	\caption{The predictive results by STSHM and self-optimization wavelet-learning
method: (a) $\kappa_{11}^*$; (b) $\kappa_{22}^*$; (c) $\kappa_{33}^*$.}
	\label{fig:14}
\end{figure}

After real-time self-optimization, the optimal wavelet-ANN model is gained with network structure 3376-186-141-276-185-197-3 with a learning rate of 0.000231. For this self-optimization wavelet-ANN model (SO-W-ANN), the training time is 19.29s, and the test time is 0.00058s. The training errors of $\kappa_{11}^*$, $\kappa_{22}^*$ and $\kappa_{33}^*$ are 0.49\%, 0.50\% and 0.48\% respectively, and the test errors are 0.80\%, 0.81\% and 0.77\% respectively. The overall training error is 0.49\% and the test error is 0.79\%. And then, the predicted results of thermal conductivities $\kappa_{11}^*$, $\kappa_{22}^*$ and $\kappa_{33}^*$ by STSHM and SO-W-ANN are depicted Fig.\hspace{1mm}\ref{fig:14}.

Moreover, the predictive results of five samples by the self-optimization wavelet-learning method and two theoretical approaches (Hashin-Shtrikman method and Lees method) are exhibited in Table \ref{tab:6}.
\begin{table}[]
	\footnotesize
	\centering
	\caption{The comparison of self-optimization wavelet-learning method and traditional theoretical methods.}
	\begin{tabular}{|c|c|c|cl|c|}
		\hline
		Sample   & $\kappa_{11}^*$/$\kappa_{22}^*$/$\kappa_{33}^*$ by STSHM & $\kappa_{11}^*$/$\kappa_{22}^*$/$\kappa_{33}^*$ by SO-W-ANN & \multicolumn{2}{c|}{\begin{tabular}[c]{@{}c@{}}Upper/Lower Hashin-\\ Shtrikman bounds\end{tabular}} & \begin{tabular}[c]{@{}c@{}}Lees \\ method\end{tabular} \\ \hline
		Sample 1 & 233.581/234.394/238.761 & 235.229/235.771/240.290 & \multicolumn{2}{c|}{242.818/206.631} & 224.611 \\ \hline
		Sample 2 & 231.973/232.706/237.035 & 233.637/234.408/238.038 & \multicolumn{2}{c|}{241.104/204.473} & 222.801 \\ \hline
		Sample 3 & 231.922/232.510/236.852 & 235.131/235.736/240.725 & \multicolumn{2}{c|}{240.967/202.170} & 221.984 \\ \hline
		Sample 4 & 232.200/232.922/237.347 & 231.629/232.821/237.536 & \multicolumn{2}{c|}{241.449/200.075} & 221.659 \\ \hline
		Sample 5 & 236.820/237.718/241.946 & 237.433/238.158/241.990 & \multicolumn{2}{c|}{246.046/212.355} & 228.574 \\ \hline
	\end{tabular}
	\label{tab:6}
\end{table}

As illustrated in Tables \ref{tab:5} and \ref{tab:6}, the predicted results by the established SO-W-ANN accurately approximate the computational results by STSHM. Additionally, the predicted results by the established SO-W-ANN fall between the upper and lower bounds of Hashin-Shtrikman method and are very close to the predicted values obtained by Lees method. However, Hashin-Shtrikman method offers a relatively large estimation interval for the lower and upper bounds of high-contrast inhomogeneous material, that can not be applied in engineering applications. Besides, Lees approach can merely obtain the isotropic prediction results and it is not applicative for anisotropic materials. From the viewpoint of computing efficiency, the proposed SO-W-ANN can avoid repetitive numerical computation by STSHM and can be utilized to efficient prediction for fibrous composites with random configurations.
\subsection{Example 3. nonlinear concrete composite with three-scale random configurations}
The multi-level concrete material is a representative three-scale heterogeneous material with steel fiber-reinforced phase at meso-scale, limestone aggregate inclusion and cement mortar matrix at micro-scale. The nonlinear thermal conductivities with temperature-dependent properties for cement mortar, limestone aggregate and steel fiber are ${\kappa_1} = 1.774 - 1.6714 \times {10^{ - 3}}\theta  + 5.7 \times {10^{ - 7}}{\theta ^2}$, ${\kappa_2} = 4.282 - 4.898 \times {10^{ - 3}}\theta  + 2.118 \times {10^{ - 6}}{\theta ^2}$ and ${\kappa_3} = 48.601 - 0.0022\theta$ respectively \cite{R52}.

\begin{figure}[htbp]
	\centering
	\subfigure{
		\begin{minipage}[t]{0.25\linewidth}
			\centering
			\includegraphics[width=30mm]{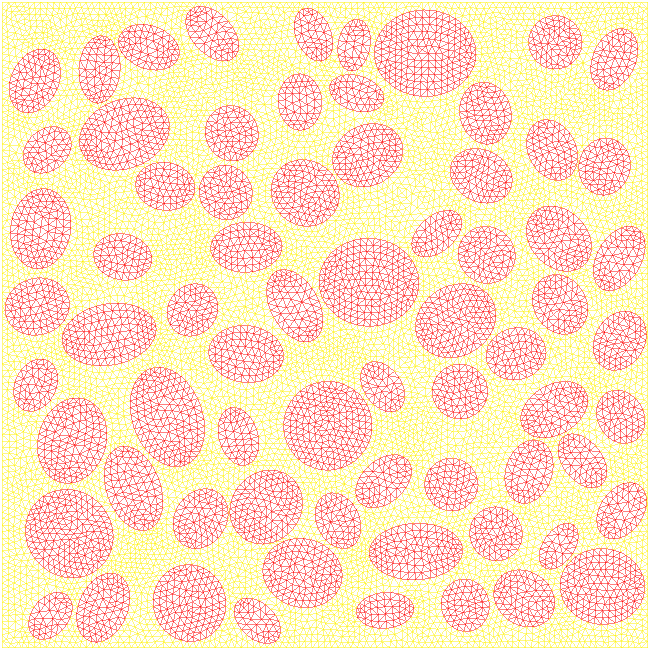}
			(a)
		\end{minipage}%
	}%
	\subfigure{
		\begin{minipage}[t]{0.25\linewidth}
			\centering
			\includegraphics[width=30mm]{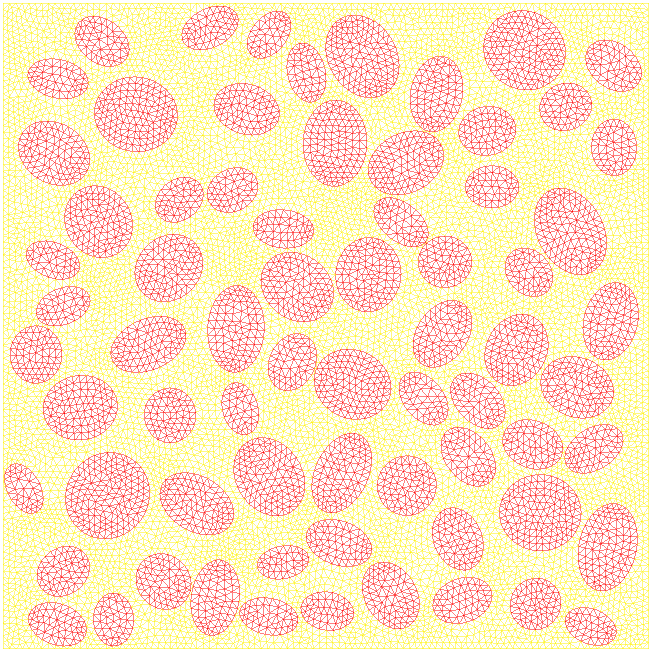}
			(b)
		\end{minipage}%
	}%
	\subfigure{
		\begin{minipage}[t]{0.25\linewidth}
			\centering
			\includegraphics[width=30mm]{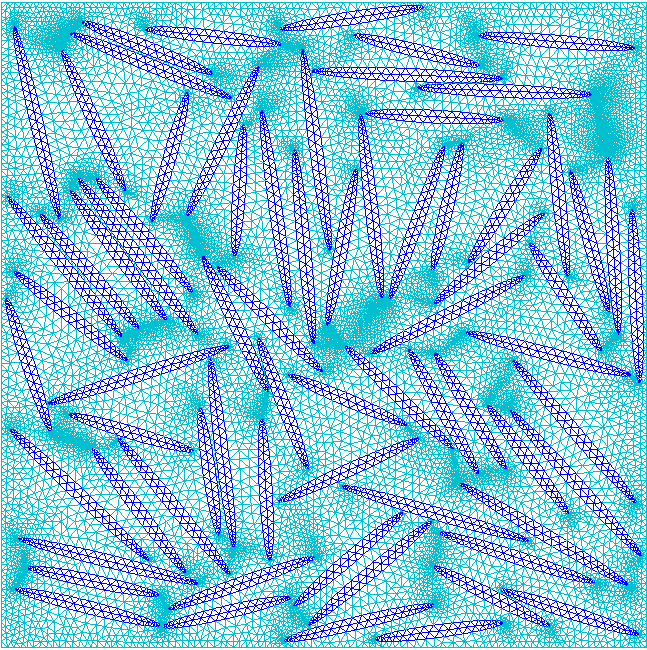}
			(c)
		\end{minipage}
	}%
	\subfigure{
		\begin{minipage}[t]{0.25\linewidth}
			\centering
			\includegraphics[width=30mm]{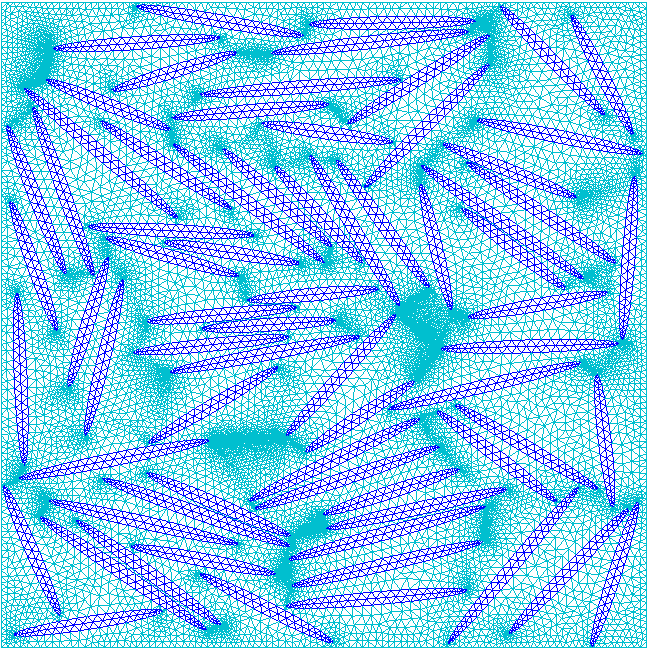}
			(d)
		\end{minipage}
	}%
	\centering
	\caption{Several microscopic and mesoscopic RVEs in 2D case: (a) and (b) are microscopic RVEs; (c) and (d) are mesoscopic RVEs.}
	\label{fig:15}
\end{figure}

\begin{figure}[htbp]
	\centering
	\subfigure[]{
		\begin{minipage}[t]{0.5\linewidth}
			\centering
			\includegraphics[width=60mm]{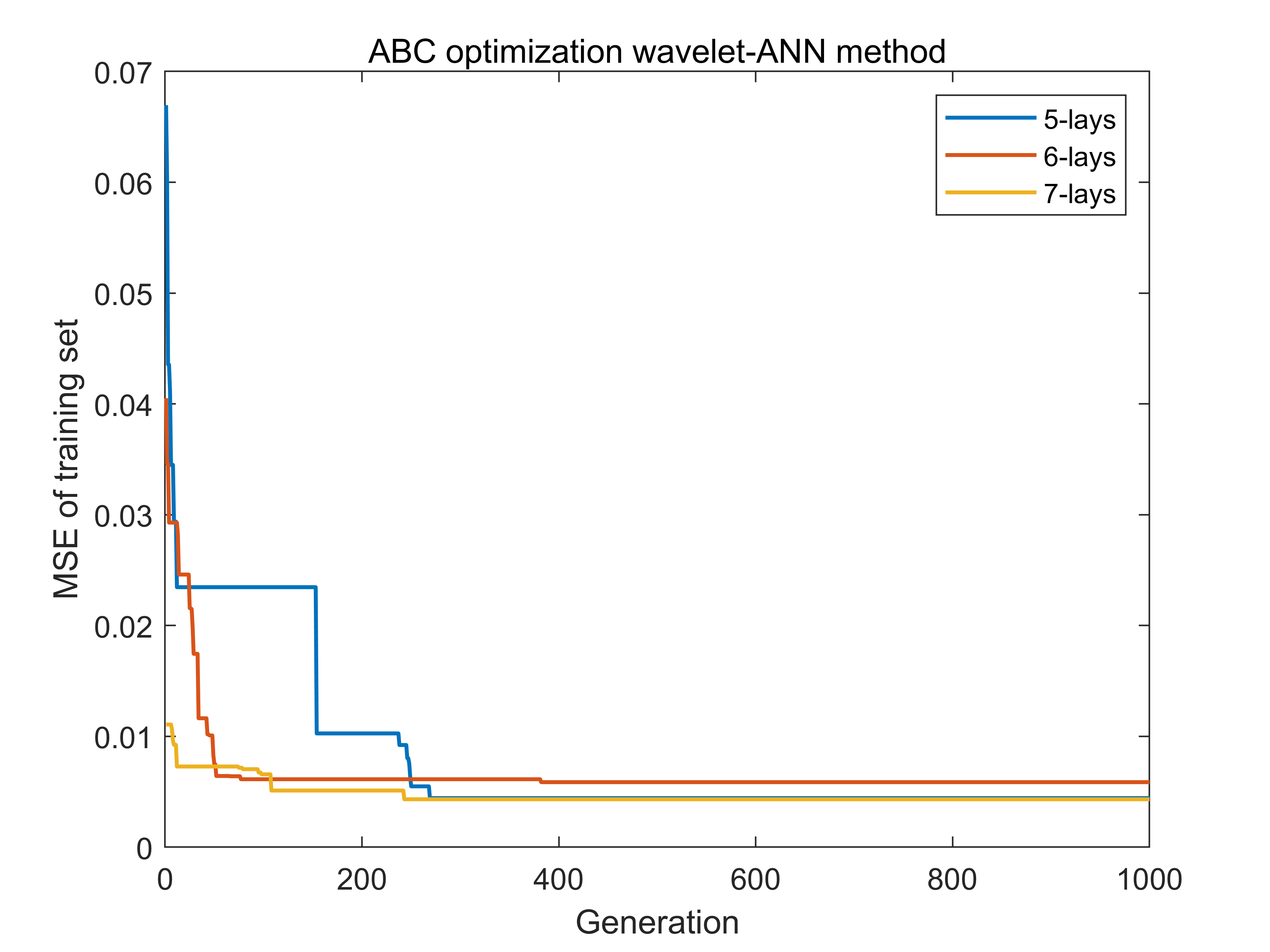}
		\end{minipage}%
	}%
	\subfigure[]{
		\begin{minipage}[t]{0.5\linewidth}
			\centering
			\includegraphics[width=60mm]{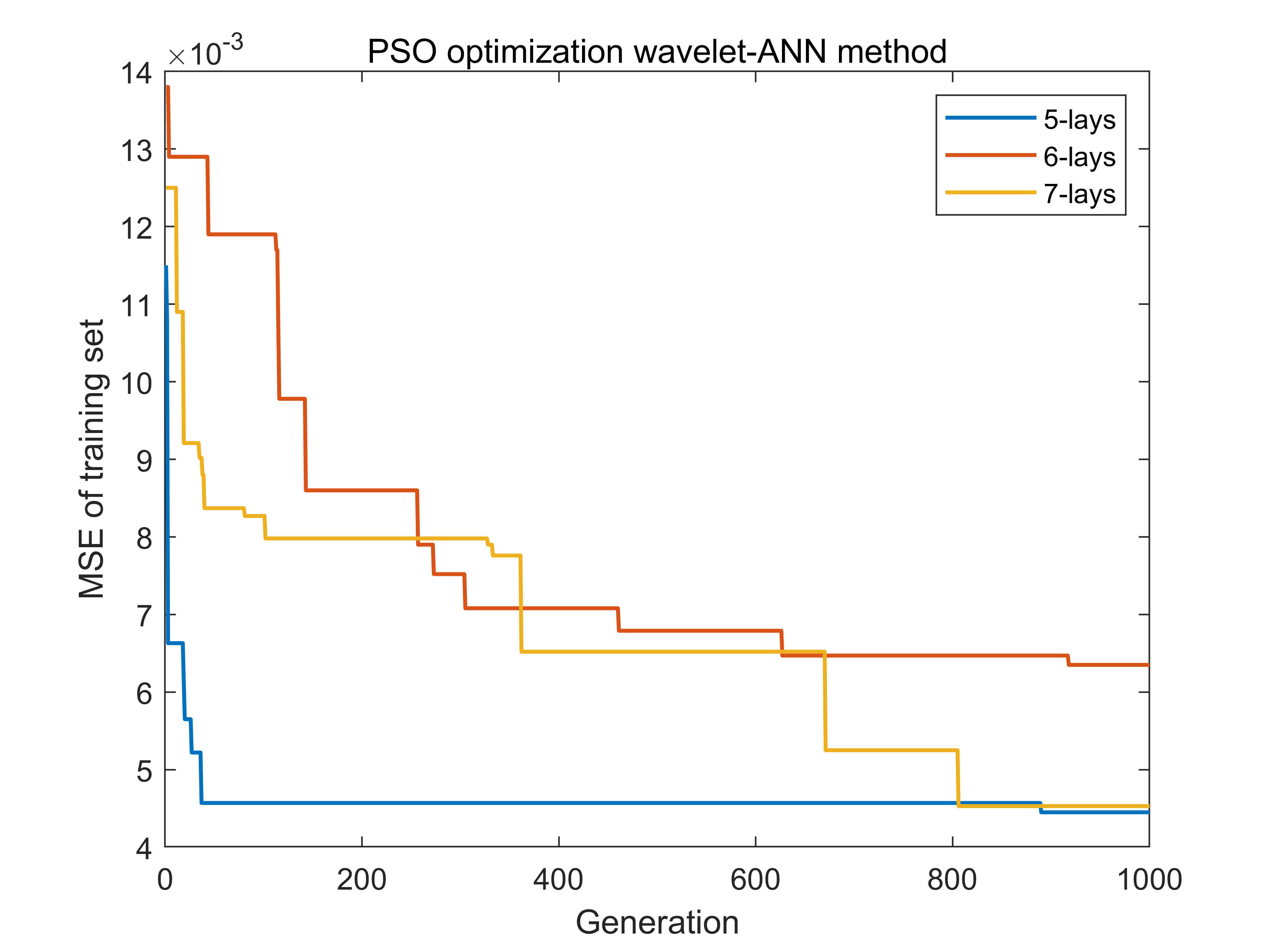}
		\end{minipage}%
	}
	\centering
	\caption{Decreasing curves of loss function of wavelet-learning predictive models: (a) ABC optimization; (b) PSO optimization.}
	\label{fig:16}
\end{figure}

For 2D concrete composites, several micro-scale and meso-scale RVEs of 2D concrete materials are presented in Fig.\hspace{1mm}\ref{fig:15}. The temperature variation of component materials is investigated in the range of [500,1000)$K$. For the sake of ensuring the statistical diversity of material samples, 200 samples are randomly generated per 5$K$, and a total of 20000 samples are randomly generated. Among these 20000 samples, 80\% are randomly selected for training process and the rest 20\% are set aside for testing process. The raw 10000 microscopic and 10000 mesoscopic material features in the raw material database are preprocessed via three-level wavelet decomposition. Then, the 2500 wavelet coefficients $cA_3$ and temperature $\theta$ are used as inputs, $\kappa_{11}^*$ and $\kappa_{22}^*$ were calculated by the stochastic three-scale homogenized method as outputs to build the self-optimization wavelet-learning predictive models. Moreover, self-optimization mechanism are introduced to search a optimal predictive model to accurately predict the equivalent nonlinear thermal conductivity of heterogeneous concrete material. Next, we employed ABC and PSO algorithms to optimize the number of neurons in each hidden layer, the learning rate, and the depth of hidden layer of the wavelet-learning predictive models. The number of neurons in each hidden layer is searched in the range of 1-500, the learning rate is searched in the range of $\left[ {1 \times {{10}^{ - 6}},5 \times {{10}^{ - 4}}} \right]$, and the depth of hidden layer is searched in the range of 5-7 layers. Moreover, each wavelet-learning predictive model is maximum iterated 500 epochs. Then using ABC and PSO algorithms to dynamically optimize wavelet-ANN model, the descent results of loss function are exhibited in Fig.\hspace{1mm}\ref{fig:16}.

\begin{figure}[htbp]
	\centering
	\subfigure[]{
		\begin{minipage}[t]{0.5\linewidth}
			\centering
			\includegraphics[width=60mm]{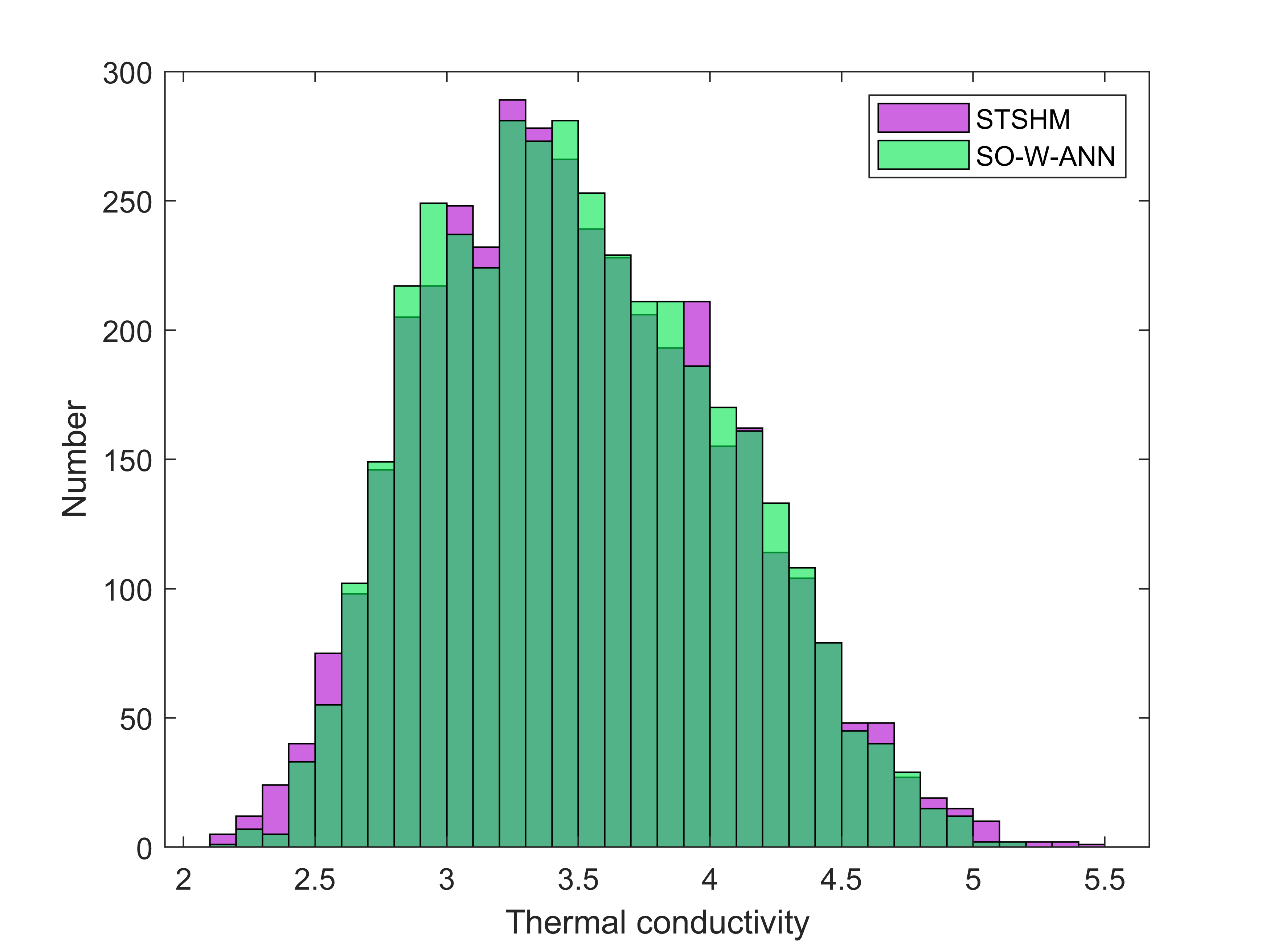}
		\end{minipage}%
	}%
	\subfigure[]{
		\begin{minipage}[t]{0.5\linewidth}
			\centering
			\includegraphics[width=60mm]{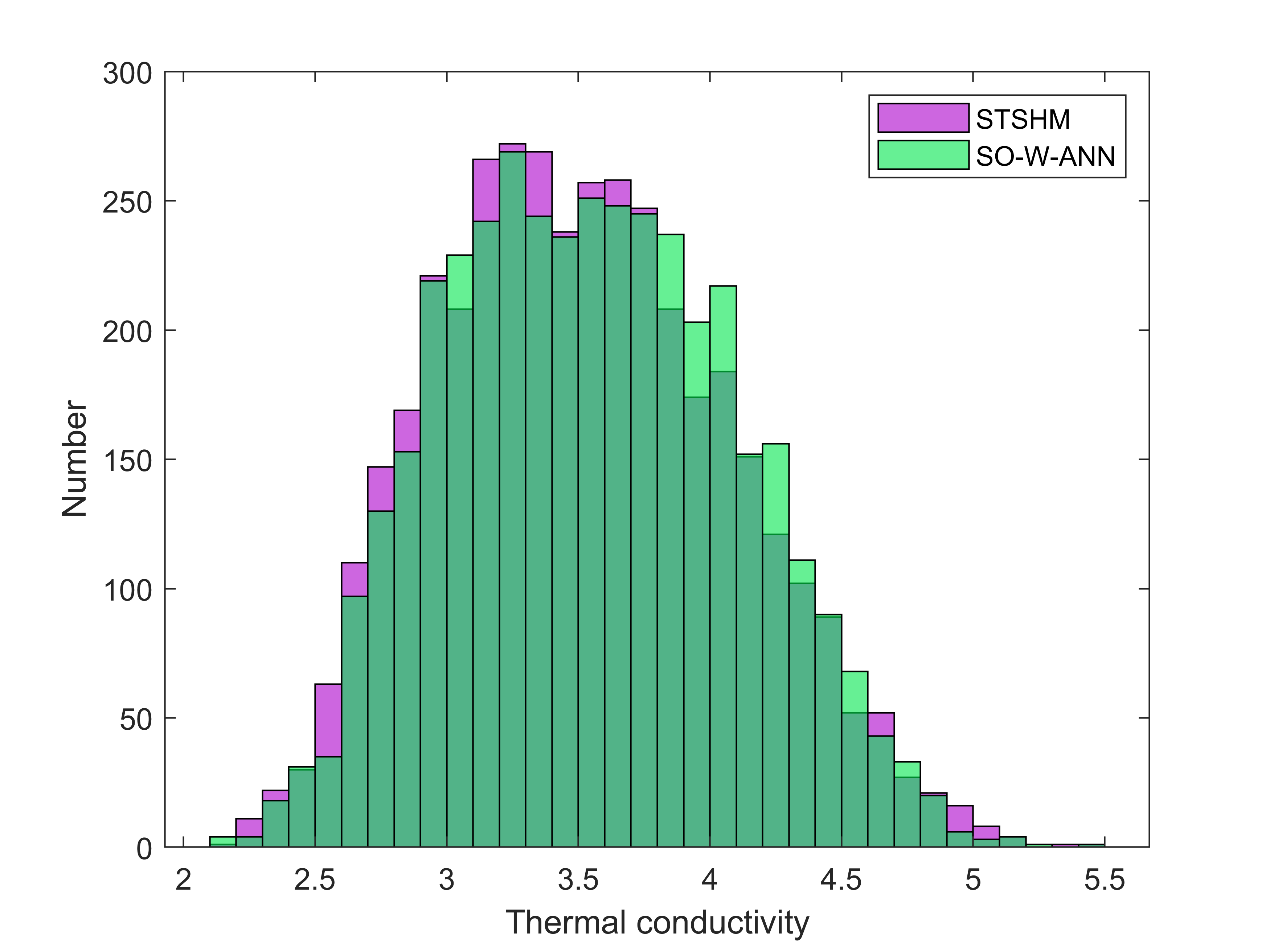}
		\end{minipage}%
	}
	\centering
	\caption{The predictive results by STSHM and self-optimization wavelet-learning method: (a) $\kappa_{11}^*$; (b) $\kappa_{22}^*$.}
	\label{fig:17}
\end{figure}

After dynamic self-optimization, the optimal wavelet-ANN model is determined as neural structure 2501-463-493-491-262-350-246-263-2 with learning rate 0.000033. For this self-optimization wavelet-ANN model (SO-W-ANN), the training time is 57.378s, and the test time is 0.000536s. The training error of material parameter $\kappa_{11}^*$ is 1.324\% and the test error is 4.561\%, and the training error of material parameter $\kappa_{22}^*$ is 1.469\% and the test error is 4.587\%. The overall training error is 1.397\% and the test error is 4.574\%. Next, Fig. \ref{fig:17} shows the histogram of the predicted material parameters $\kappa_{11}^*$ and $\kappa_{22}^*$ of SO-W-ANN compared to the predicted values calculated by the established STSHM.

In addition, five samples are randomly generated to conduct predictive analysis. The detailed comparison of the presented self-optimization wavelet-learning approach with traditional theoretical methods (Hashin-Shtrikman method and Lees method) is displayed in Table \ref{tab:7}.
\begin{table}[]
	\footnotesize
	\centering
	\caption{The comparison of self-optimization wavelet-learning method and traditional theoretical methods.}
	\begin{tabular}{|c|c|c|cl|c|}
		\hline
		Sample   & $\kappa_{11}^*$/$\kappa_{22}^*$ by STSHM & $\kappa_{11}^*$/$\kappa_{22}^*$ by SO-W-ANN & \multicolumn{2}{c|}{\begin{tabular}[c]{@{}c@{}}Upper/Lower Hashin-\\ Shtrikman bounds\end{tabular}} & \begin{tabular}[c]{@{}c@{}}Lees \\ method\end{tabular} \\ \hline
		Sample 1 & 4.253/3.671 & 4.180/3.971  & \multicolumn{2}{c|}{9.044/2.895} & 3.537 \\ \hline
		Sample 2 & 3.917/3.525 & 3.763/3.485 & \multicolumn{2}{c|}{8.620/2.587} & 3.235 \\ \hline
		Sample 3 & 3.578/4.216 & 3.774/4.057  & \multicolumn{2}{c|}{9.820/2.583} & 3.349  \\ \hline
		Sample 4 & 3.677/3.606  & 4.217/3.658  & \multicolumn{2}{c|}{9.299/2.590} & 3.274 \\ \hline
		Sample 5 & 4.164/3.661   & 4.601/3.805  & \multicolumn{2}{c|}{9.863/2.742} & 3.473 \\ \hline
	\end{tabular}
	\label{tab:7}
\end{table}

\begin{figure}[htbp]
	\centering
	\subfigure{
		\begin{minipage}[t]{0.25\linewidth}
			\centering
			\includegraphics[width=30mm]{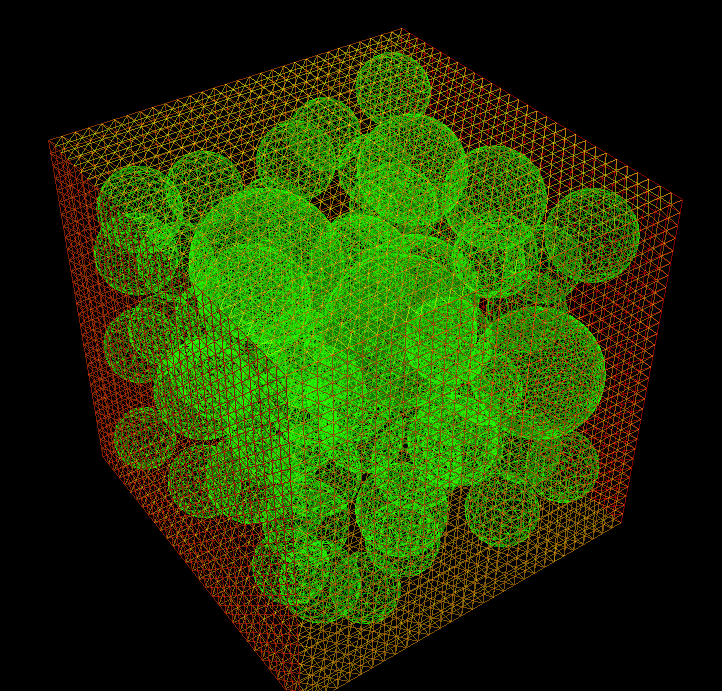}
			(a)
		\end{minipage}%
	}%
	\subfigure{
		\begin{minipage}[t]{0.25\linewidth}
			\centering
			\includegraphics[width=30mm]{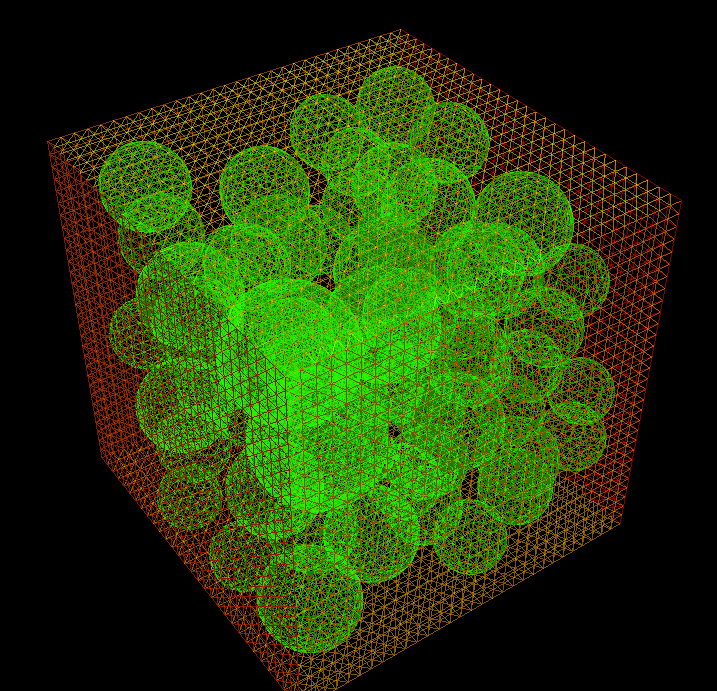}
			(b)
		\end{minipage}%
	}%
	\subfigure{
		\begin{minipage}[t]{0.25\linewidth}
			\centering
			\includegraphics[width=30mm]{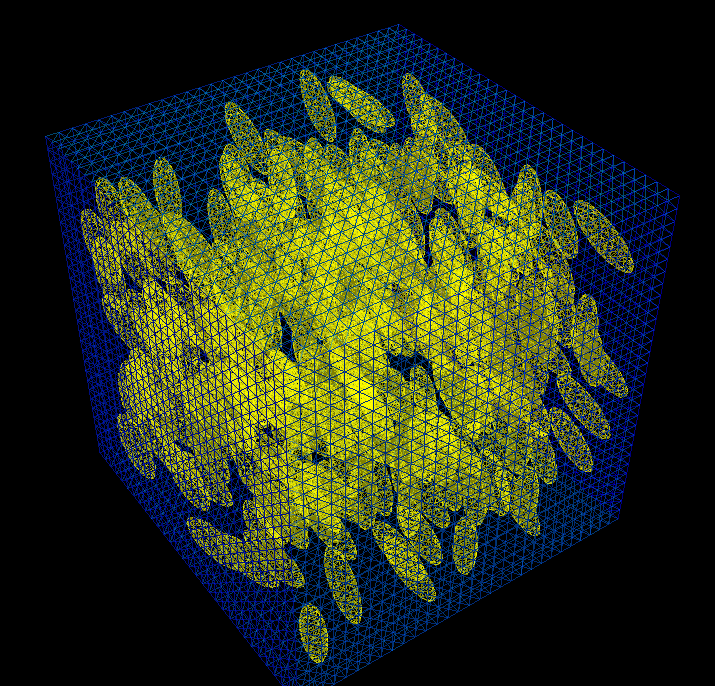}
			(c)
		\end{minipage}
	}%
	\subfigure{
		\begin{minipage}[t]{0.25\linewidth}
			\centering
			\includegraphics[width=30mm]{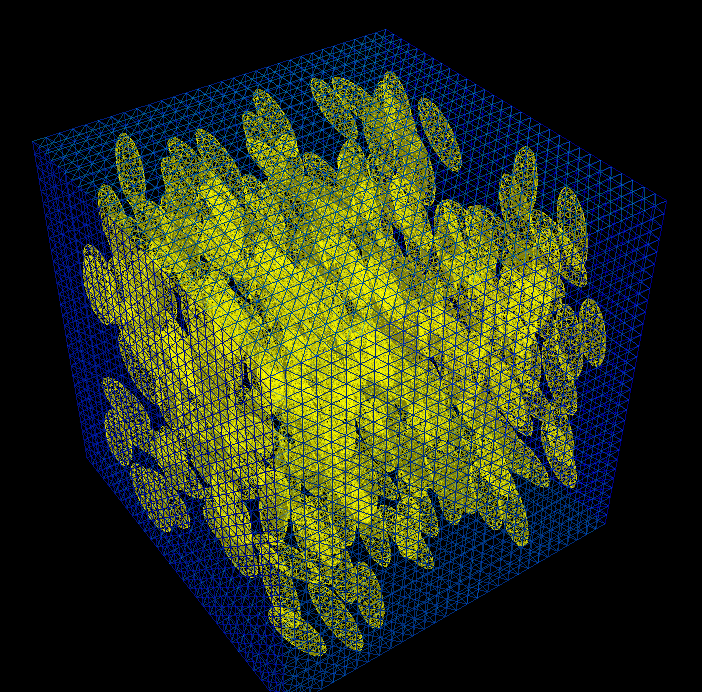}
			(d)
		\end{minipage}
	}%
	\centering
	\caption{Several microscopic and mesoscopic RVEs in 3D case: (a) and (b) are microscopic RVEs; (c) and (d) are mesoscopic RVEs.}
	\label{fig:18}
\end{figure}

\begin{figure}[htbp]
	\centering
	\subfigure[]{
		\begin{minipage}[t]{0.5\linewidth}
			\centering
			\includegraphics[width=60mm]{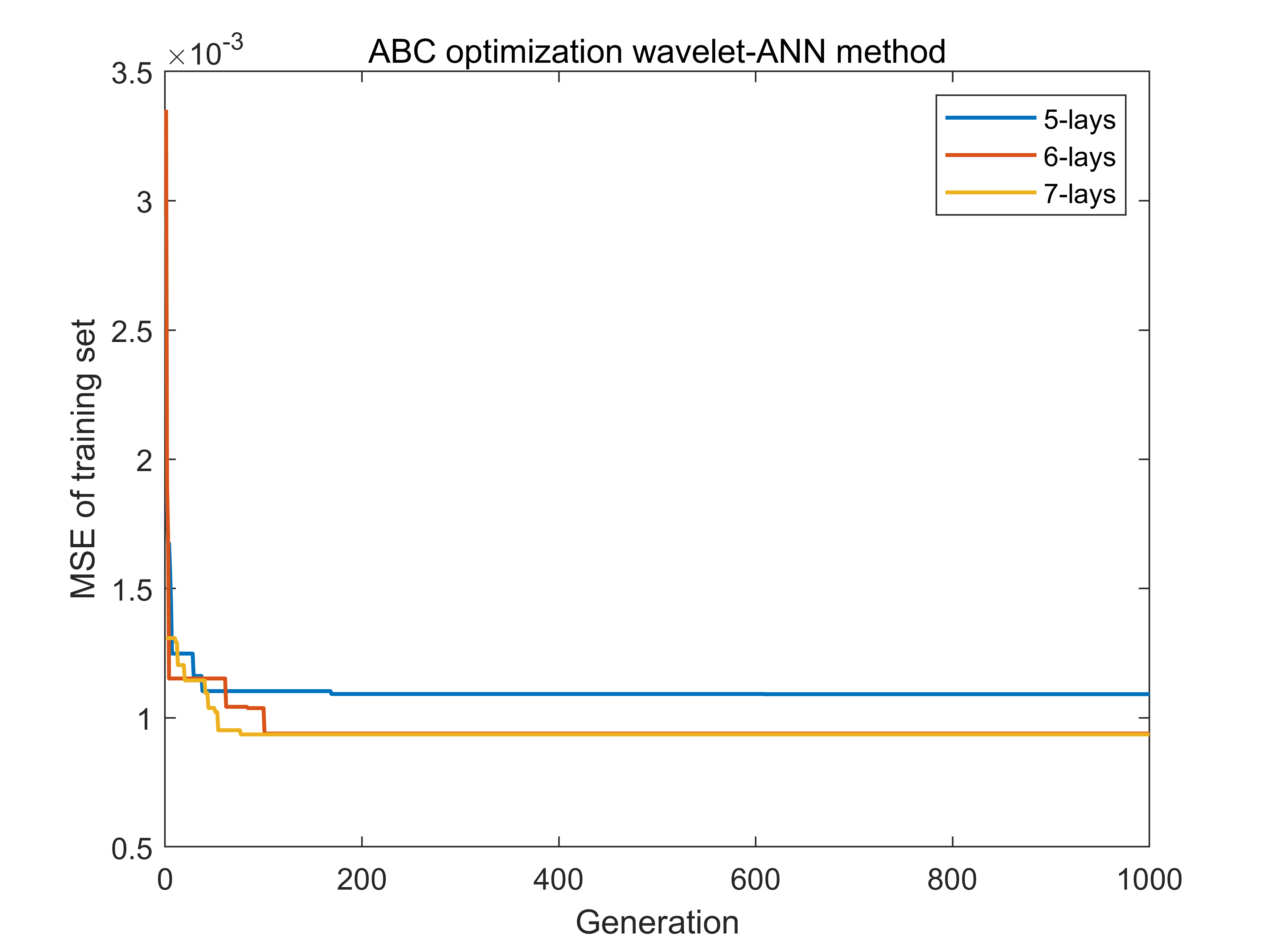}
		\end{minipage}%
	}%
	\subfigure[]{
		\begin{minipage}[t]{0.5\linewidth}
			\centering
			\includegraphics[width=60mm]{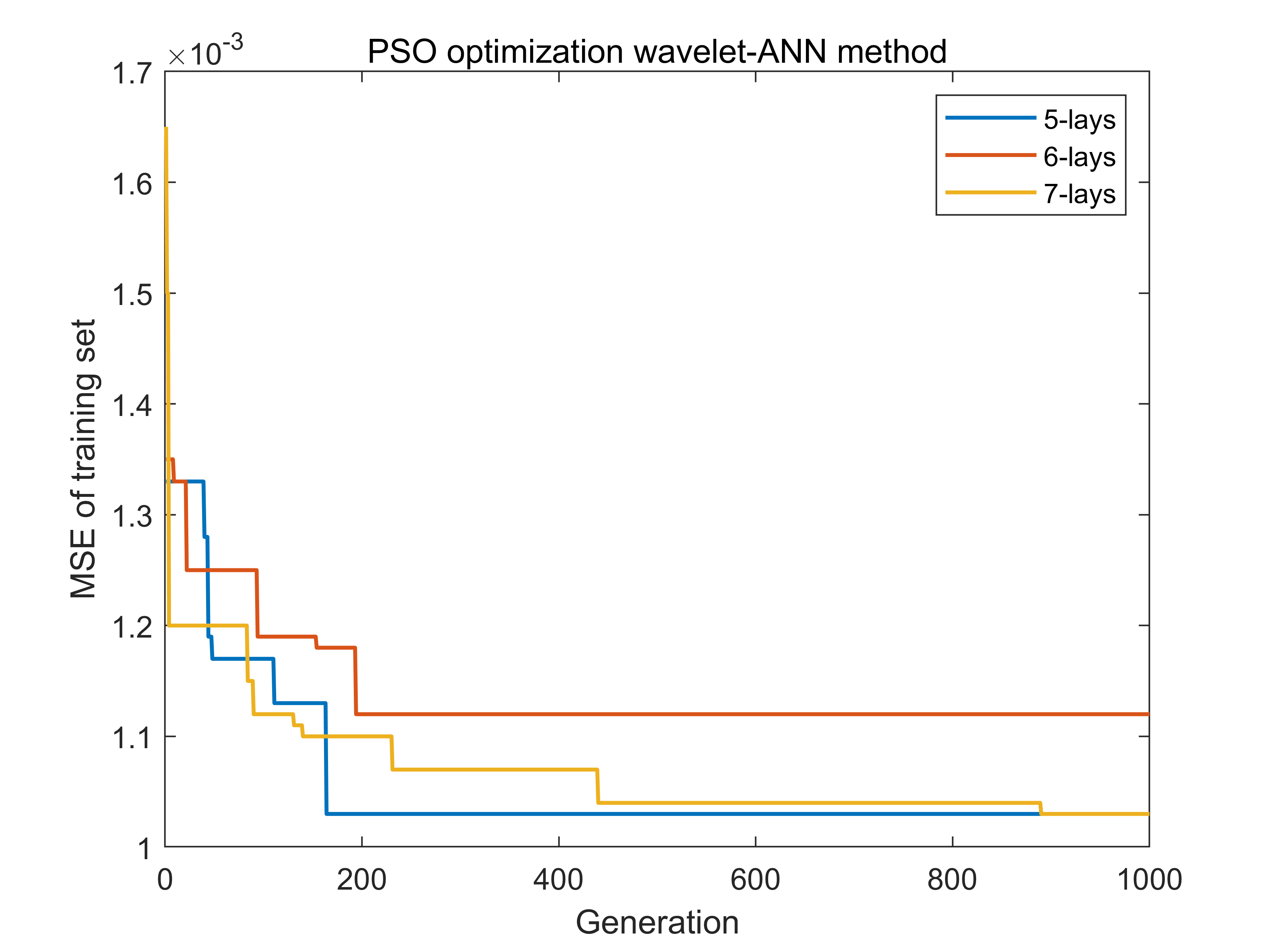}
		\end{minipage}%
	}
	\centering
	\caption{Decreasing curves of loss function of wavelet-learning predictive models: (a) ABC optimization; (b) PSO optimization.}
	\label{fig:19}
\end{figure}

For 3D concrete composites, some micro-scale and meso-scale RVEs of 3D concrete materials are presented in Fig.\hspace{1mm}\ref{fig:18}. The temperature variation of component materials is considered in the range of [500,1000)$K$. In an effort to pledge the stochastic diversity of material samples, 50 samples are randomly generated per 5$K$, and a total of 5000 samples are randomly generated. Among the 5000 samples, 80\% are randomly chosen as training set and the remaining 20\% are denoted as test set. After employing three-level wavelet decomposition to the raw 27000 microscopic and 27000 mesoscopic material features, the 6750 wavelet coefficients $cA_3$ and temperature $\theta$ are utilized as new inputs, and $\kappa_{11}^*$, $\kappa_{22}^*$ and $\kappa_{33}^*$ were calculated by the stochastic three-scale homogenized method as outputs to build the self-optimization wavelet-learning predictive models. For the 3D case, the range of optimized parameters of the wavelet-learning predictive models is the same as the 2D example. Furthermore, each wavelet-learning predictive model is iterated 300 epochs. Then the adjustable parameters of the predictive models are optimized by using ABC and PSO algorithms, the descent results of loss function are exhibited in Fig.\hspace{1mm}\ref{fig:19}.

\begin{figure}[htbp]
	\centering
	\subfigure{
		\begin{minipage}[t]{0.3\linewidth}
			\centering
			\includegraphics[width=45mm]{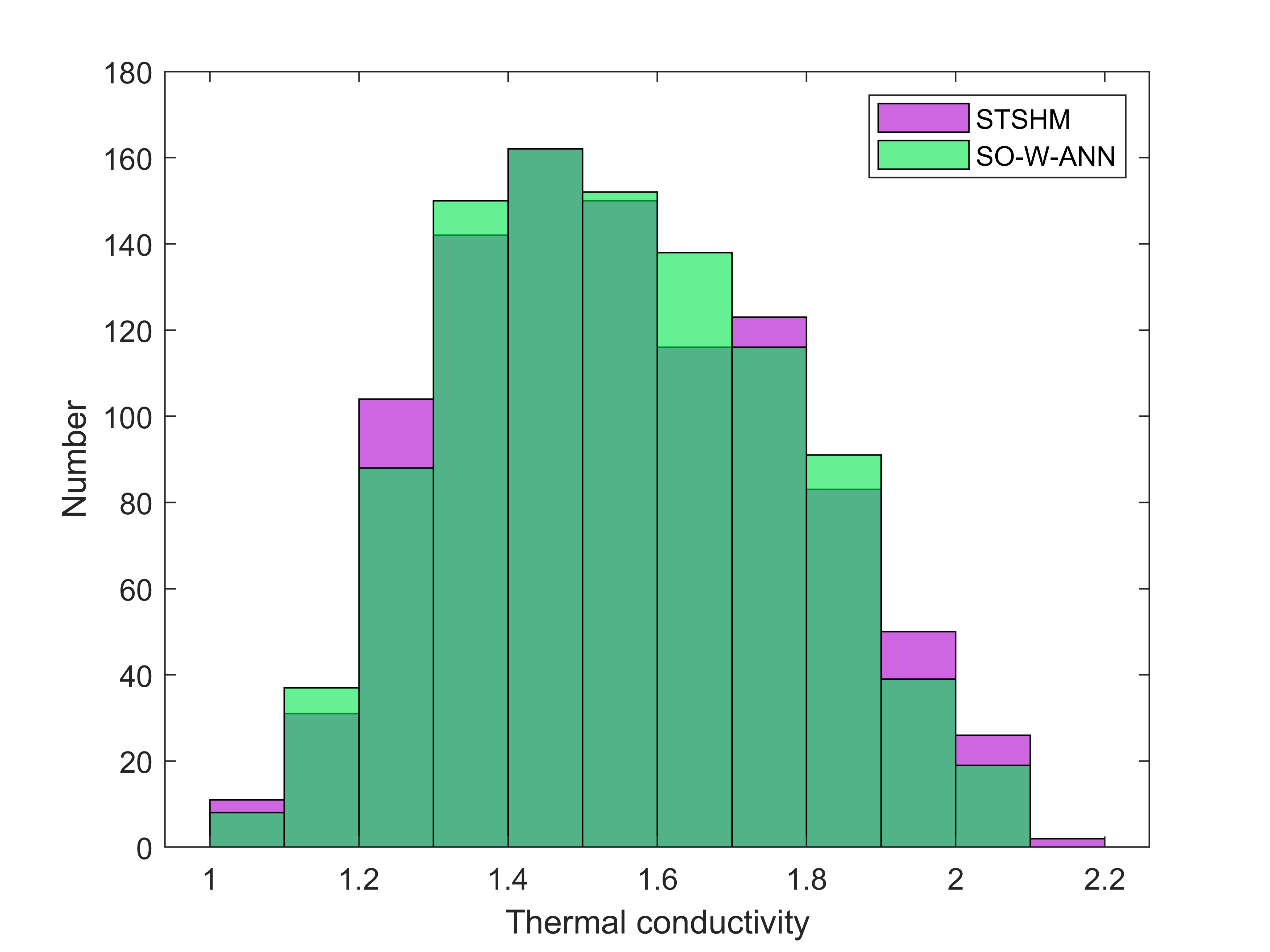}
			(a)
		\end{minipage}%
	}%
	\subfigure{
		\begin{minipage}[t]{0.3\linewidth}
			\centering
			\includegraphics[width=45mm]{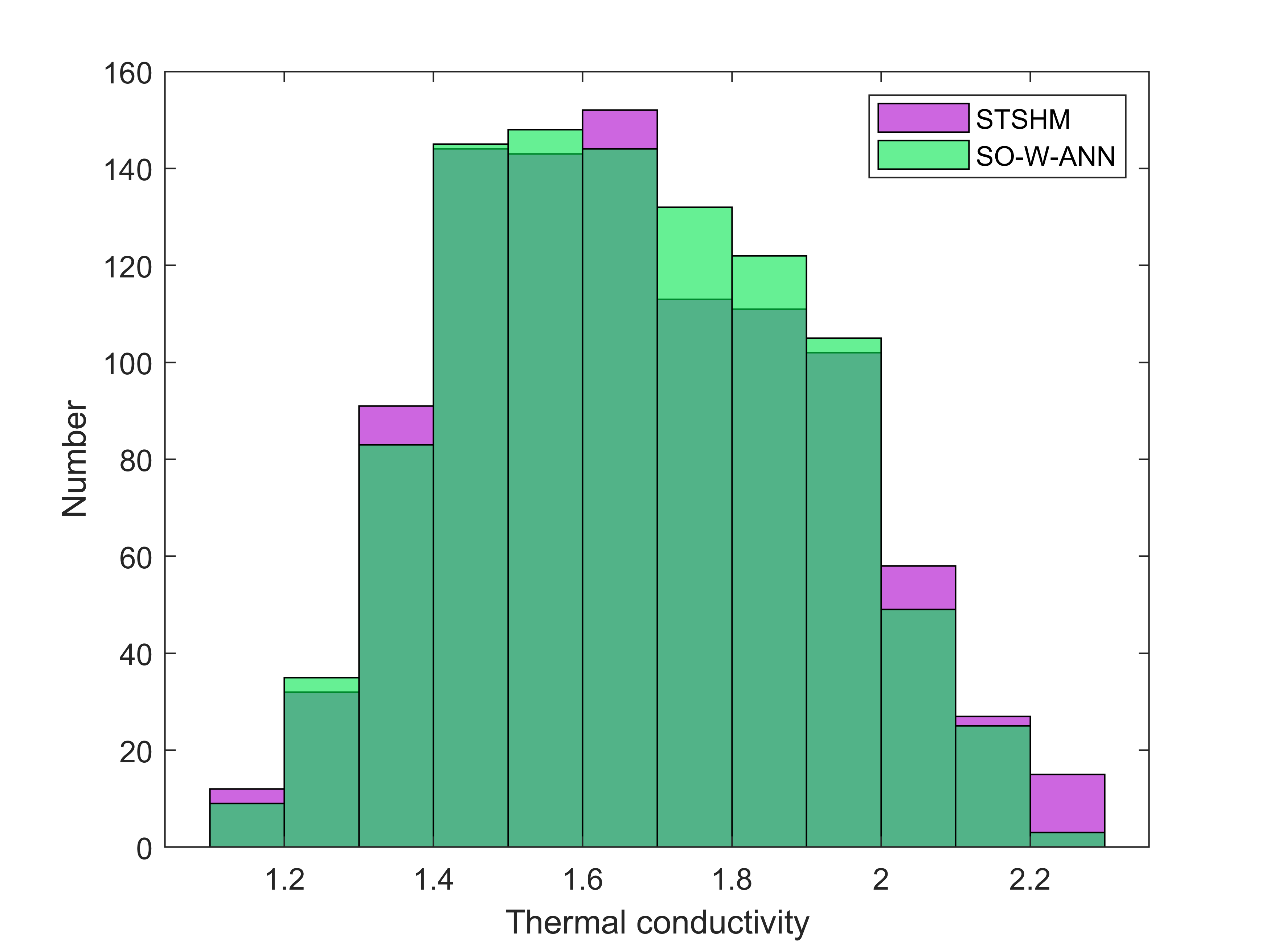}
			(b)
		\end{minipage}%
	}
	\subfigure{
		\begin{minipage}[t]{0.3\linewidth}
			\centering
			\includegraphics[width=45mm]{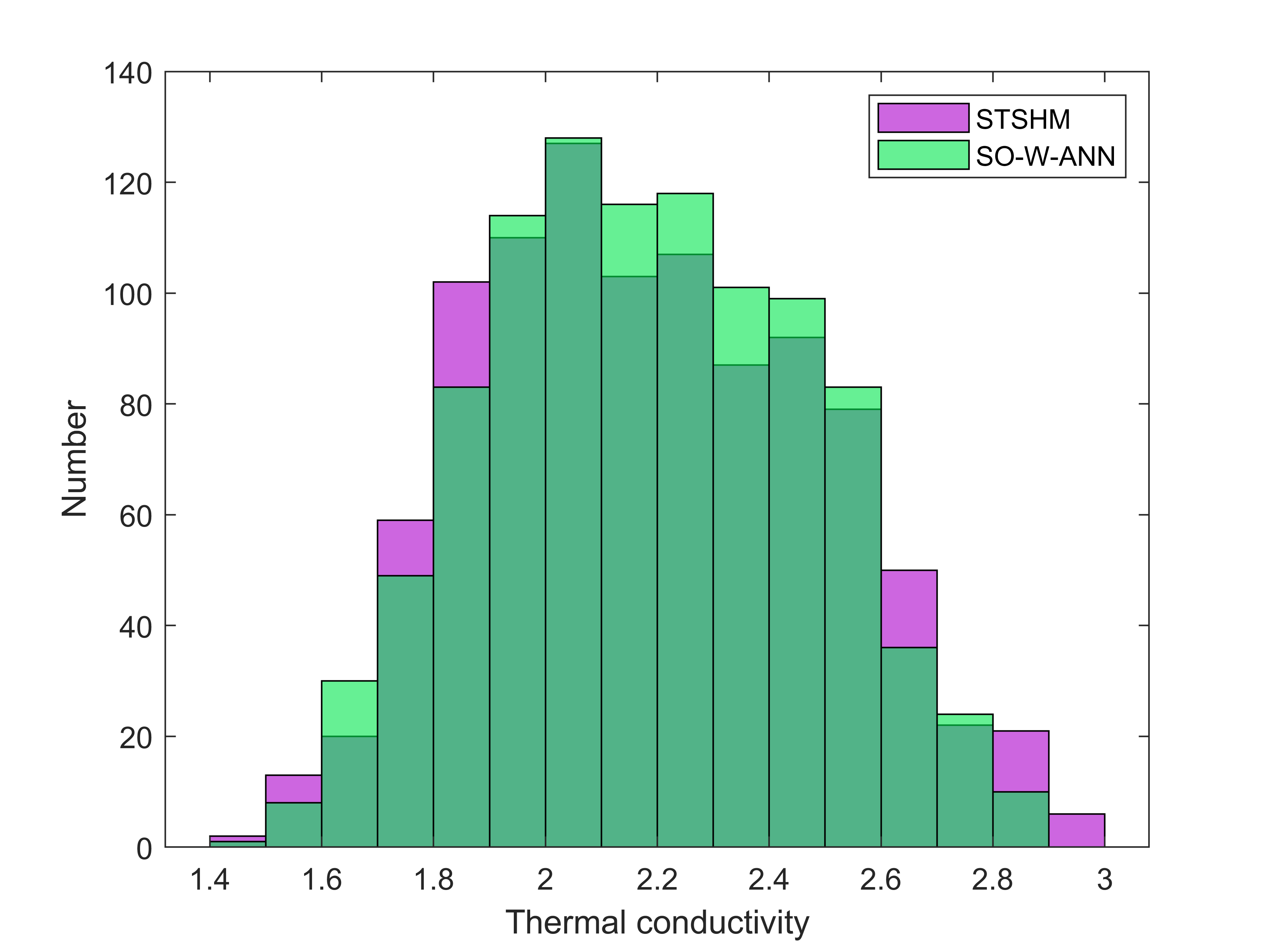}
			(c)
		\end{minipage}%
	}
	\centering
	\caption{The predictive results by STSHM and self-optimization wavelet-learning method: (a) $\kappa_{11}^*$; (b) $\kappa_{22}^*$; (c) $\kappa_{33}^*$.}
	\label{fig:20}
\end{figure}

After real-time self-optimization, the optimal wavelet-ANN model is determined as neural structure 6751-367-410-481-363-346-13-418-3 with learning rate 0.000011. For this self-optimization wavelet-ANN model (SO-W-
ANN), the training time is 27.926s, and the test time is 0.00055s. The training errors of $\kappa_{11}^*$, $\kappa_{22}^*$ and $\kappa_{33}^*$ are 1.131\%, 1.099\% and 1.031\% respectively, and the test errors are 3.199\%, 3.064\% and 2.855\% respectively. The overall training error is 1.087\% and the test error is 3.039\%. The predicted material
parameters $\kappa_{11}^*$, $\kappa_{22}^*$ and $\kappa_{33}^*$ of SO-W-ANN and the computational values by the STSHM are compared in Fig.\hspace{1mm}\ref{fig:20}.

Furthermore, the predictive results of five samples by the self-optimization wavelet-learning method and two theoretical approaches (Hashin-Shtrikman method and Lees method) are displayed in Table \ref{tab:8}.
\begin{table}[]
	\footnotesize
	\centering
	\caption{The comparison of self-optimization wavelet-learning method and traditional theoretical methods.}
	\begin{tabular}{|c|c|c|cl|c|}
		\hline
		Sample   & $\kappa_{11}^*$/$\kappa_{22}^*$/$\kappa_{33}^*$ by STSHM & $\kappa_{11}^*$/$\kappa_{22}^*$/$\kappa_{33}^*$ by SO-W-ANN & \multicolumn{2}{c|}{\begin{tabular}[c]{@{}c@{}}Upper/Lower Hashin-\\ Shtrikman bounds\end{tabular}} & \begin{tabular}[c]{@{}c@{}}Lees \\ method\end{tabular} \\ \hline
		Sample 1 & 1.335/1.443/1.839 & 1.252/1.352/1.797 & \multicolumn{2}{c|}{3.068/1.225} & 1.331 \\ \hline
		Sample 2 & 1.339/1.437/1.913 & 1.428/1.554/2.041 & \multicolumn{2}{c|}{3.415/1.222} & 1.339 \\ \hline
		Sample 3 & 1.655/1.765/2.328 & 1.577/1.701/2.220 & \multicolumn{2}{c|}{3.915/1.501} & 1.630 \\ \hline
		Sample 4 & 1.396/1.496/1.918 & 1.348/1.461/1.929 & \multicolumn{2}{c|}{3.202/1.274} & 1.385 \\ \hline
		Sample 5 & 1.752/1.900/2.458 & 1.788/1.930/2.507 & \multicolumn{2}{c|}{4.241/1.602} & 1.742 \\ \hline
	\end{tabular}
	\label{tab:8}
\end{table}

As displayed in Tables \ref{tab:7} and \ref{tab:8}, the predictive results by the proposed SO-W-ANN highly approximate the computational results by STSHM. Furthermore, the predictive results by the proposed SO-W-ANN fall between the upper and lower bounds of Hashin-Shtrikman method and are very close to the predicted values obtained by Lees method. However, Hashin-Shtrikman method gives a relatively large estimation interval for the lower and upper bounds of high-contrast concrete composites, which is inaccurate for direct practical applications. In addition, Lees method can only obtain the isotropic prediction results and it is not applicative for anisotropic materials. From the perspective of computational efficiency, the established SO-W-ANN can avert repetitive numerical computation by STSHM and can be applied to efficient prediction for concrete materials with randomly multi-level configurations.
\section{Conclusions and prospects}
\label{sec:5}
This work has developed a self-optimization wavelet-learning framework by combining respective advantages of stochastic three-scale homogenization method, wavelet transform, neural networks and intelligent optimization algorithm, which provides an accurately and efficiently predictive tool of estimating effective nonlinear thermal conductivity of random composites with complicated hierarchical configurations. The proposed self-optimization wavelet-learning approach simultaneously allows for random structural heterogeneities, temperature-dependent nonlinearity and material property uncertainties of heterogeneous materials to be analyzed. The present study contains the principal innovations as follows. (1) A computer generation algorithm for fibrous composites with high volume ratio inclusions is developed. The detailed volume ratio for short-fiber inclusions can be improved to 26.7\%. The microscopic and mesoscopic models of the composites are developed secondarily based on Freefem++ software. (2) A novel stochastic three-scale homogenized approach is developed to calculate the equivalent nonlinear thermal conductivity of random inhomogeneous materials, which is used as the data label for the material database. In addition, the structural information and material parameters of inhomogeneous materials are extracted by using a background mesh method and a filling technique, which are regarded as data features of the material database. It is worth emphasizing that the filling technique tackles the missing information problem of matrix material at the meso-scale. (3) The high-dimensional material features are processed by using wavelet decomposition, and the dimension-reduced low-dimensional features are utilized as the input data of neural networks. Wavelet decomposition can distill the principal features from the material database, compress the raw data features, and reduce the input data of the neural networks, which can benefit to raise the training efficiency and reduce over-fitting of predictive neural network models. (4) The network structure and learning rate of the wavelet-learning approach are optimized by combining intelligent optimization algorithms to obtain the self-optimization wavelet-learning methodology with the optimal predictive performance.

The computational results of numerical experiments on the basis of experimental material parameters of realistic random composites demonstrate that the self-optimization wavelet-learning method established in this paper can effectively estimate the equivalent nonlinear thermal conductivity of highly inhomogeneous materials with random hierarchical configurations. Compared with classical analytical and numerical methods, the proposed self-optimization wavelet-learning approach can avert repetitive numerical simulation and can be utilized to quickly predict the equivalent nonlinear thermal conductivity of random inhomogeneous materials. Further, in the future, this work can be improved and expanded in the following areas: (1) The efficiency of computer generation algorithm for random inhomogeneous materials with high volume ratio inclusions should be further upgraded and employed to simulate the multi-scale nonlinear problems. (2) The self-optimization wavelet-learning framework can be potentially employed to estimate any other nonlinear mechanical and multi-physical coupling properties of random heterogeneous materials. (3) To more precise prediction of effective physical properties of multi-level concrete materials, the physical size effect of concrete materials should be taken into account in overall self-optimization wavelet-learning framework. (4) More advanced optimization algorithms should be employed in our self-optimization wavelet-learning framework for obtaining the globally optimal predictive neural networks.

\section*{Acknowledgments}
This work was financially supported by the National Natural Science Foundation of China (Nos.\hspace{1mm}12001414, 62276202, and 62106186), the Natural Science Basic Research Plan in Shaanxi Province of China under Grant (No.\hspace{1mm}2022JQ-670), the Young Talent Fund of Association for Science and Technology in Xi'an, China (No.\hspace{1mm}095920221338), the Young Talent Fund of Association for Science and Technology in Shaanxi, China (No.\hspace{1mm}20220506), the Fundamental Research Funds for the Central Universities under Grant (No.\hspace{1mm}QTZX22047), the National Natural Science Foundation of China (No.\hspace{1mm}11971386). We also acknowledged the technical support by Xi'an Key Laboratory of Scientific Computation and Applied Statistics and the Key Technology Research of FRP-Concrete Composite Structure.




\bibliographystyle{model1a-num-names}
\bibliography{paper}







\end{document}